\documentclass[11pt]{article}
\usepackage{jheppub}
\usepackage{amsmath,amssymb,amsfonts,amsthm,mathtools}
\usepackage[dvipsnames]{xcolor}
\usepackage{comment}
\usepackage{hyperref}
\usepackage[utf8]{inputenc}
\usepackage[titletoc]{appendix}
\usepackage{tikz}
\usetikzlibrary{shapes,arrows,shadows}

\makeatletter
\def\@fpheader{\relax}
\makeatother

\newcommand\be{\begin{equation}}
\newcommand\ee{\end{equation}}
\newcommand\bea{\begin{eqnarray}}
\newcommand\eea{\end{eqnarray}}
\newcommand\ba{\begin{array}}
\newcommand\ea{\end{array}}
\newcommand\nn{\nonumber}
\newcommand\eref[1]{(\ref{#1})}
\newcommand\bc{\begin{center}}
\newcommand\ec{\end{center}}
\newcommand\pa{\partial}
\renewcommand\comment[1]{}

\numberwithin{equation}{section}
\allowdisplaybreaks

\title{Neural Network Approximations for Calabi--Yau Metrics}

\author[a]{Vishnu Jejjala}
\author[a,b]{\!, Dami\'an Kaloni Mayorga Pe\~na}
\author[c]{\!, Challenger Mishra}

\affiliation[\,a]{Mandelstam Institute for Theoretical Physics, School of Physics, NITheP, and CoE-MaSS,\\
University of the Witwatersrand, Johannesburg, WITS 2050, South Africa}
\affiliation[\,b]{Data Laboratory, Universidad de Guanajuato,\\
Loma del Bosque No.\ 103 Col.\ Lomas del Campestre C.P.\ 37150 Leon, Guanajuato, Mexico}
\affiliation[\,c]{Department of Computer Science \& Technology, University of Cambridge,\\ 15 J.J.\ Thomson Ave., Cambridge CB3 0FD, United Kingdom}

\emailAdd{vishnu@neo.phys.wits.ac.za}
\emailAdd{damian.mayorgapena@wits.ac.za}
\emailAdd{cm2099@cam.ac.uk}

\abstract{
Ricci flat metrics for Calabi--Yau threefolds are not known analytically.
In this work, we employ techniques from machine learning to deduce numerical flat metrics for the Fermat quintic, for the Dwork quintic, and for the Tian--Yau manifold.
This investigation employs a single neural network architecture that is capable of approximating Ricci flat K\"ahler metrics for several Calabi--Yau manifolds of dimensions two and three.
We show that measures that assess the Ricci flatness of the geometry decrease after training by three orders of magnitude.
This is corroborated on the validation set, where the improvement is more modest.
Finally, we demonstrate that discrete symmetries of manifolds can be learned in the process of learning the metric. 
}

\begin{document}

\maketitle
\parskip=10pt

\section{Introduction}
Superstring theory supplies an architectural framework for obtaining the real world from a consistent theory of quantum gravity.
In the critical setting, string theory makes the remarkable prediction that spacetime is ten dimensional.
Commensurate with $\mathcal{N}=1$ supersymmetry in four dimensions, we may use a compact Calabi--Yau threefold to reduce the theory down from ten dimensions~\cite{Candelas:1985en,Green:1987mn}.

A Calabi--Yau space is an $n$-dimensional complex manifold $\mathcal{M}$ with a K\"ahler metric with local holonomy in $SU(n)$.
Mathematicians were already interested in these manifolds in the 1950s.
Calabi conjectured~\cite{Calabi:1954} that given a closed $(1,1)$-form $\frac{1}{2\pi}C_1(\mathcal{M})$ representing the first Chern class of a K\"ahler manifold $\mathcal{M}$, there is a unique K\"ahler metric in the same K\"ahler class whose Ricci tensor is the closed $(1,1)$-form $C_1(\mathcal{M})$.
This result implies the existence of a unique Ricci flat K\"ahler metric within each K\"ahler class.
Unfortunately, the proof of the conjecture, \textit{viz.}, Yau's theorem~\cite{yau1977calabi, Yau:1978}, does not explicitly construct the flat metric.
Except for the trivial Calabi--Yau manifold, which is the even dimensional torus, we did not until recently have analytic closed form expressions for flat metrics on Calabi--Yau spaces.
This situation is changing as~\cite{Gaiotto:2009hg, Kachru:2018van, Kachru:2020tat} have tackled the problem for K3.

Luckily, topology on its own is a sufficiently powerful tool to enable string model building.
Starting from the work of~\cite{Braun:2005ux} and~\cite{Bouchard:2005ag}, numerous semi-realistic models of particle physics have been constructed on appropriately chosen Calabi--Yau spaces.
These models are semi-realistic because, while the visible sector recapitulates the matter content and the interactions of the minimal supersymmetric Standard Model (MSSM), we do not have a detailed understanding of the Yukawa couplings or the mass hierarchies among the generations.
To go beyond an analysis of the spectrum and break supersymmetry in a controlled manner, we must fix the $\mathcal{N}=1$ K\"ahler potential.
Doing this demands knowledge of the Ricci flat metric on the base manifold.
Determining the mass of the electron from first principles therefore requires an understanding of the geometry as well as the topology of Calabi--Yau spaces.

Significant progress has been made in obtaining numerical metrics on Calabi--Yau geometries.
Using the Gauss--Seidel method, Headrick and Wiseman~\cite{Headrick:2005ch} constructed numerical Ricci flat metrics on a one parameter family of K3 surfaces obtained as blowups of $T^4/\mathbb{Z}_2$.
Donaldson~\cite{donaldson2001,donaldson2005some} subsequently developed an algorithm to solve for the metrics numerically.
The essence of this approach is to consider generalizations of the Fubini--Study metric induced from the embedding of a hypersurface (or a complete intersection) in an ambient space.
In particular, weighted projective spaces are endowed with a simple choice for a K\"ahler metric: the Fubini--Study (FS) metric.
For Calabi--Yau manifolds constructed as hypersurfaces or complete intersections in (products of) weighted projective spaces, a K\"ahler metric can be obtained from the pullback of the ambient Fubini--Study metric onto the hypersurface or complete intersection defining the embedding.
Generalizing the expression for the K\"ahler potential of this metric, Donaldson provides a family of so called ``balanced'' metrics that in a particular $k\to\infty$ limit, recovers the Ricci flat Calabi--Yau metric.
This idea was then applied to finding numerical metrics on the Fermat quintic~\cite{Douglas:2006rr}.
Since then, there have been a number of related numerical approaches, notably involving
energy functionals~\cite{Headrick:2009jz}, the Hermitian Yang--Mills equations~\cite{Douglas:2006hz, Braun:2007sn, Braun:2008jp, Anderson:2010ke, Anderson:2011ed, Ashmore:2020ujw}, and general scaling properties~\cite{Cui:2019uhy}.

Beginning with~\cite{He:2017set, Krefl:2017yox, Ruehle:2017mzq, Carifio:2017bov}, modern methods in machine learning have successfully been applied to various problems in high energy theoretical physics and mathematics; see~\cite{Ruehle:2020jrk} for a review.
For example, topological invariants in knot theory~\cite{Jejjala:2019kio, Gukov:2020qaj, Craven:2020bdz} are machine learnable.
Topological properties of complete intersection Calabi--Yau threefolds have also been successfully reproduced~\cite{Bull:2018uow, Bull:2019cij, Erbin:2020srm, Erbin:2020tks}.
Less work has been done in purely geometric directions.
Ashmore, He, and Ovrut~\cite{Ashmore:2019wzb} used machine learning to improve the performance of Donaldson's algorithm to obtain approximate Ricci flat metrics on the Fermat quintic.
In this work, we apply neural networks to obtain numerical metrics on the quintic and Tian--Yau Calabi--Yau threefolds.

The organization of this paper is as follows.
In Section~\ref{sec:general}, we make some general remarks on Calabi--Yau spaces.
In Section~\ref{sec:donaldson}, we briefly review Donaldson's algorithm and discuss measures of flatness.
In Section~\ref{sec:ricci}, we survey Ricci flow.
In Section~\ref{sec:nn}, we introduce neural networks.
In Section~\ref{sec:methodology}, we describe our methodology.
In Section~\ref{sec:results}, we present our results for the torus, the K3 surface, the Fermat quintic, the Dwork family, and the Tian--Yau manifold.
In Section~\ref{sec:symms}, we assess how well the neural network that yields the Ricci flat metric learns discrete symmetries of the Calabi--Yau.
Finally, in Section~\ref{sec:disc}, we discuss the results and provide a prospectus for our future work. 

\paragraph{Note:}
As this work was nearing completion, two other papers by Anderson, Gerdes, Gray, Krippendorf, Raghuram, and Ruehle~\cite{Anderson:2020hux} and Douglas, Lakshminarasimhan, and Qi~\cite{Douglas:2020hpv} appeared that also found numerical metrics on Calabi--Yau threefolds using machine learning techniques.
Many of our methods and conclusions overlap with theirs.
See also~\cite{Raghuram:2020sgds, Ruehle:2020bc, Douglas:2020sd, Krippendorf:2020sd}.

\section{General remarks}\label{sec:general}
Recall that a complex $n$-dimensional compact \textit{Calabi--Yau manifold} $\mathcal{M}$ is determined by any of the following equivalent statements: 
\begin{itemize}
\item The first real Chern class of $\mathcal{M}$ is zero.
\item A positive power of the canonical bundle of $\mathcal{M}$ is trivial.
\item $\mathcal{M}$ has a K\"ahler metric with local holonomy in $SU(n)$.
\item $\mathcal{M}$ admits a metric with vanishing Ricci curvature. This metric is unique in each K\"ahler class. 
\end{itemize}
In cases where the manifold is simply connected, this is equivalent to saying that $\mathcal{M}$ possess a unique, nowhere vanishing $(n,0)$-form, that the positive power of the canonical bundle is the first power, and the local holonomy is global.
As we are interested in finding the Ricci flat metric, we adopt the weaker definition.

As a complex K\"ahler manifold, the metric of $\mathcal{M}$ is a Hermitian matrix that can be derived from a K\"ahler potential $K(z,\bar{z})$:
\begin{equation}
g_{a\bar{b}}=\partial_a \partial_{\bar{b}} K(z,\bar{z}) ~.
\end{equation}
The metric can be used to construct the K\"ahler form as 
\begin{equation}
J=\frac{\rm i}{2} g_{a\bar{b}}\, dz^a \wedge d\bar{z}^{\bar{b}} ~.
\end{equation}
This is a closed $(1,1)$-form.
The corresponding Ricci tensor is given by: 
\begin{equation}
R_{a\bar{b}}=\partial_a \partial_{\bar{b}}\log \det g ~.
\end{equation}
For further details, we direct the reader to the classic references~\cite{candelas1988lectures, hubsch1992calabi}.

The Calabi--Yau spaces we will study are realized as hypersurfaces in products of projective space.
The trivial example of a Calabi--Yau manifold is the torus $T^{2n}$.
The torus $T^2$ can be embedded in projective space $\mathbb{P}^2$ as a cubic equation
\be
T^2: \quad z_1^3 + z_2^3 + z_3^3 = 0 \subset \mathbb{P}^2 ~, \label{eq:cubic}
\ee
where $[z_1:z_2:z_3]$ identifies a point in $\mathbb{P}^2$.
Similarly, we can generalize this type of embedding to construct K3 as a quartic hypersurface in $\mathbb{P}^3$ and as well write quintic hypersurfaces in $\mathbb{P}^4$:
\bea
\text{K3}: &\quad&  z_1^4 + z_2^4 + z_3^4 + z_4^4 = 0 \subset \mathbb{P}^3 ~, \label{eq:quartic} \\
\text{Fermat quintic}: &\quad& z_1^5 + z_2^5 + z_3^5 + z_4^5 + z_5^5 = 0 \subset \mathbb{P}^4 ~, \label{eq:quintic} \\
\text{Dwork family}: &\quad& z_1^5 + z_2^5 + z_3^5 + z_4^5 + z_5^5 - 5\psi z_1 z_2 z_3 z_4 z_5 = 0 \subset \mathbb{P}^4 ~, \quad \psi^5 \ne 1 ~. \label{eq:dwork}
\eea
Any pair of complex analytic K3 surfaces are diffeomorphic as smooth four real dimensional manifolds.
The Hodge number $h^{1,1} = 20$ for K3.
The quintic hypersurface has $h^{1,1} = 1$ and $h^{1,2} = 101$.
We will also study the complete intersection Calabi--Yau manifold given by the configuration matrix
\be
\text{Tian--Yau}: \quad \left[ \ba{c||ccc} \mathbb{P}^3 & \,3   & 0 & 1 \cr \mathbb{P}^3 & \,0 & 3 & 1 \ea \right]_{\chi=-18} \qquad \Longleftrightarrow \qquad \left\{ \ba{ccc} \alpha^{ijk} z_i z_j z_k &=& 0 ~, \cr  \beta^{ijk} w_i w_j w_k &=& 0 ~, \cr  \gamma^{ij} z_i w_j &=& 0 ~, \ea \right.
\label{eq:ty}
\ee
where $w_i$ are coordinates on the first $\mathbb{P}^3$ and $z_i$ are coordinates on the second $\mathbb{P}^3$.
The Tian--Yau manifold has $h^{1,1} = 14$ and $h^{1,2} = 23$.
A freely acting $\mathbb{Z}_3$ quotient yields a Calabi--Yau manifold with $\chi=-6$~\cite{tian1987three}, corresponding to three generations of elementary particles; this was one of the initial testbeds for string phenomenology~\cite{Greene:1986bm}.
We will study numerical metrics on these geometries.

\begin{figure}[t!]
\centering
\includegraphics[scale=.33]{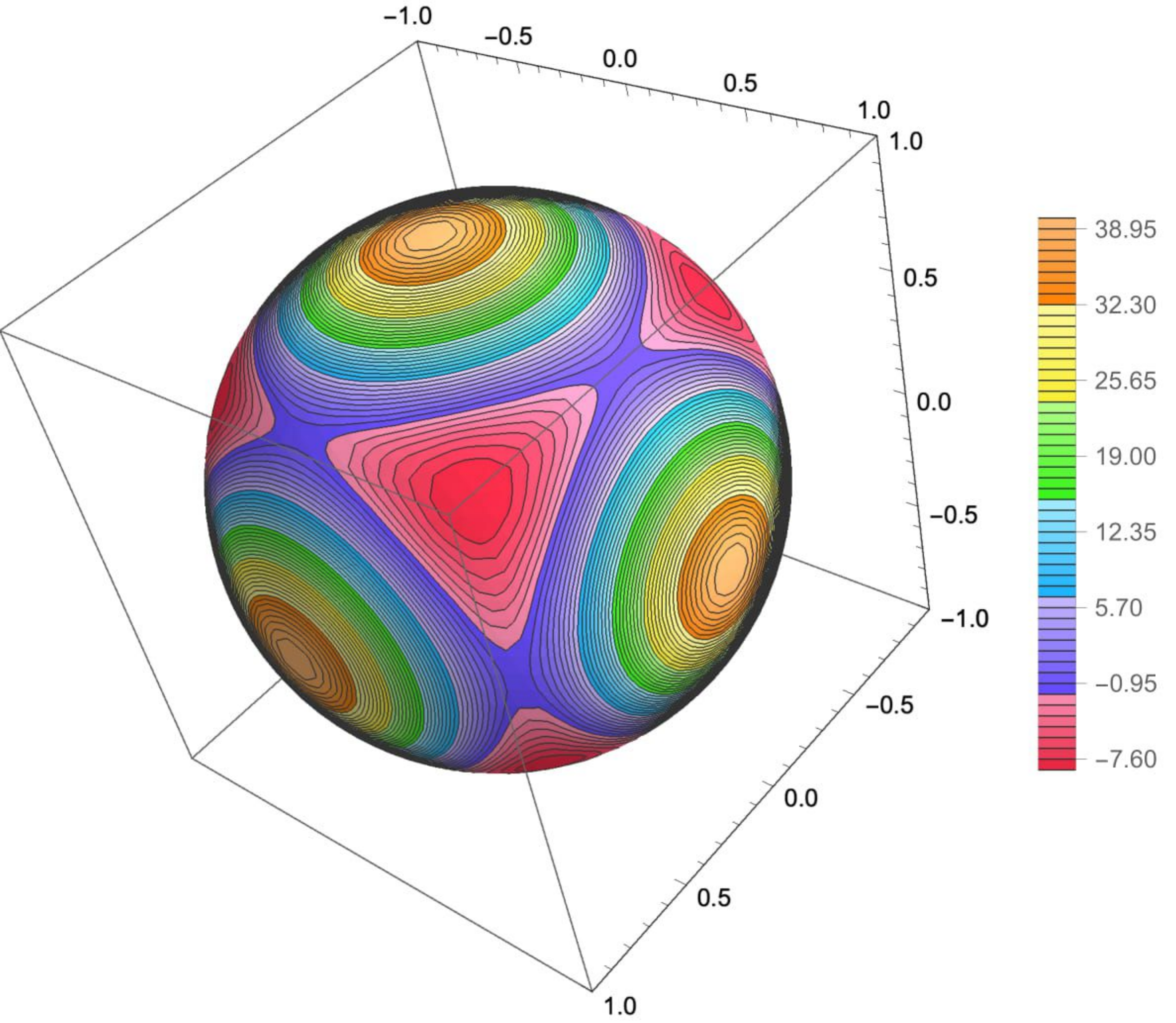}\\[20pt]	
\caption{\small\textit{Visualizing the scalar curvature of the metric obtained by restricting the Fubini--Study metric of  $\mathbb{P}^4$ to the quintic hypersurface. The three axes represent real parts of the three independent co-ordinates of the Fermat quintic \eqref{eq:quintic} in a patch $z_i=0$ for some $i$. The contour of scalar curvature is then presented atop a three dimensional sphere, and is evidently not uniform.}}
  \label{fig:FS_Quintic}
  \vspace{10pt}
\end{figure}

On the manifold $\mathbb{P}^n$, an atlas is given by $n+1$ coordinate charts on which each of the $z_i\ne 0$. In homogeneous coordinates, a possible K\"ahler potential for the projective space $\mathbb{P}^n$ has the form 
\be
K(z,\bar{z})=\frac{1}{\pi} {\rm log}(|z|^2) ~, \qquad |z|^2 = \sum_{a=1}^{n+1} z^a \bar{z}^{\bar{a}} ~.
\label{eq:kpot}
\ee
This potential leads to the following metric:
\be
g_{a\bar{b}} (z,\bar{z})= \frac{|z|^2\delta^{a\bar{b}} - \bar{z}^{a}\bar{z}^{\bar{b}}}{\pi |z|^4} ~.
\ee
This is the \textit{Fubini--Study} (FS) metric on $\mathbb{P}^n$.
It is induced from the definition of $\mathbb{P}^n$ as a quotient space $S^{2n+1}/S^1$, where $S^{2n+1}$ carries the round metric. The corresponding metric on each of the $n+1$ charts can be obtained via matrix multiplication with the Jacobian of the transformation from homogeneous to affine coordinates. In a similar fashion, the pullback of the Fubini--Study metric on $\mathbb{P}^n$ to the hypersurface induced by the embedding yields the Fubini--Study metric on the Calabi--Yau. This Fubini--Study metric is not the Ricci flat K\"ahler metric. In Figure~\ref{fig:FS_Quintic}, we have plotted the Ricci scalar for the Fubini--Study metric on a section of the Fermat quintic. 

For $\mathbb{P}^n$, one naturally has the $n+1$ affine charts, each labeled by an index $i$ corresponding to the coordinate that is set to one (\textit{i.e.}, $z^{i}=1$). Additionally, given the structure of the embedding equations for $T^2$, K3, and the quintic, we have a natural collection of charts labeled by $z_i\ne 0$ and $z_{j\ne i}$, the latter of which is obtained from solving the equation defining the hypersurface. Therefore, each chart for the hypersurface will be labelled by two indices $(i,j)$. 

The Tian--Yau manifold is given as the intersection of three transversely intersecting loci on $\mathbb{P}^3\times\mathbb{P}^3$. One has then two $\mathbb{C}^*$ actions and therefore two coordinates that can be set to one. For simplicity we define $z^{k+4}=w^k$, such that the index in the coordinate $z$ runs from $1,\ldots,8$. Taking the affine patch resulting from setting the coordinates $z^l=1$ and $z^m=w^{m-4}=1$, with $l\leq 4$ and $m>4$, we can use the two cubic equations to solve for coordinates $z^{s}$ and $z^t$ with $s\leq 4$, $s\neq l$ and $t>4$ $t \neq m$. Finally we use the bilinear equation to obtain a third dependent coordinate which can belong to either of the $\mathbb{P}^3$'s, say $z^r$, $r\neq l\,,\,m\,,\,s\,,\,t$. We thus have a chart in the Tian--Yau manifold defined by five indices $(l\,,\,m\,;\,s\,,\,t\,,\,r)$

\section{Donaldson's algorithm}\label{sec:donaldson}
Donaldson~\cite{donaldson2001} supplied a constructive algorithm for obtaining the Ricci flat metric starting from the Fubini--Study metric.
Weighted projective spaces are endowed with a simple choice of K\"ahler metric derived from the following K\"ahler potential given in Equation \eqref{eq:kpot}. Another valid K\"ahler potential is
\begin{equation}
K^{(k)}(z,\bar{z})=\frac{1}{k\pi} \log\left(h^{\alpha \bar{\beta}}s_\alpha \bar{s}_{\bar{\beta}}\right) ~,
\end{equation}
where $s_\alpha$ is an element of a basis for degree $k$ holomorphic polynomials over $\mathcal{M}$.
Take $N_k$ to be the dimension of such a basis and define 
\begin{equation}
H_{\alpha\bar{\beta}}=\frac{N_k}{\text{Vol}_\Omega}\int_\mathcal{M} d\text{Vol}_\Omega\ \left(\frac{s_\alpha \bar{s}_{\bar{\beta}}}{h^{\sigma \bar{\rho}}s_\sigma \bar{s}_{\bar{\rho}}}\right) ~.
\end{equation}
At level $k$, one can use $h^{\alpha\bar{\beta}}=(H_{\alpha\bar{\beta}})^{-1}$ and proceed iteratively until reaching a stable ``balanced'' metric $H_{\alpha\bar{\beta}}$.
This is guaranteed to exist by virtue of Donaldson's theorem.
Furthermore, as $k$ increases, the metric $g_{a\bar{b}}$ obtained from the balanced metric approaches the desired Ricci flat Calabi--Yau metric.
Donaldson's algorithm will provide a reference point to which we compare our results.

\subsection{Flatness measures}\label{sec:flatness}
As we are computing metrics numerically, we need some diagnostic that tells us how close we are to the flat metric.
A number of such diagnostics have appeared in the literature~\cite{Douglas:2006rr, Douglas:2006hz, Braun:2007sn, Anderson:2010ke, Ashmore:2019wzb}.
We examine the $\sigma$-measure: 
\begin{equation}
\sigma=\frac{1}{\text{Vol}_\Omega}\int_{\mathcal{M}} d\text{Vol}_\Omega\ \left|1-\frac{\text{Vol}_\Omega}{\text{Vol}_J}\cdot \frac{J^n}{\Omega\wedge \bar{\Omega}}\right| ~.
\end{equation}
This makes use of the fact the volume can be computed using either the holomorphic and anti-holomorphic top forms or the closed $(1,1)$ K\"ahler form $J$:
\begin{equation}
\text{Vol}_\Omega=\int_{\mathcal{M}} \Omega \wedge \bar{\Omega} ~, \qquad
\text{Vol}_J = \int_{\mathcal{M}} J^n ~,
\end{equation}
with $J^n$ denoting the $n$-fold wedge product of $J$.
Note that this measure becomes zero whenever $J^n=\alpha\Omega\wedge \bar{\Omega}$, with $\alpha$ an overall rescaling. 

Another measure comes from integrating the Ricci curvature scalar directly, \textit{i.e.}, 
\begin{equation}
||R||=\frac{\text{Vol}_J^{1/n}}{\text{Vol}_\Omega}\int_{\mathcal{M}} d\text{Vol}_J\ |R| ~.
\end{equation}
This is known as the $||R||$-measure and in some sense it is equivalent to the sigma measure.
Namely, it can be shown that as the $\sigma$-measure approaches zero, the $||R||$-measure goes to zero as well.

\section{Ricci flow}\label{sec:ricci}
Ricci flow gives a partial differential equation for a Riemannian metric $g$.
It was introduced by Hamilton~\cite{hamilton1982} and famously employed by Perelman to prove the Poincar\'e conjecture in three dimensions~\cite{perelman2002entropy, perelman303109ricci, perelman2003finite}.
See~\cite{chow2007ricci, kleiner2008notes} for surveys of the method.
In high energy theory, Ricci flow was used to find numerical black hole solutions~\cite{Headrick:2006ti} and a numerical K\"ahler--Einstein metric on $dP_3$~\cite{Doran:2007zn}, while~\cite{Jackson:2013eqa, Fonda:2016ine} discuss Ricci flow in a holographic context.
K\"ahler--Ricci flow applies the method to a K\"ahler manifold.
We specialize our notation to this case.
The differential equation tells us that
\be
\frac{\pa }{\pa \lambda} g_{a\bar{b}}(\lambda) = -\text{Ric}_{a\bar{b}}(\lambda) =  \frac{\pa^2}{\pa z^a \pa \bar{z}^{\bar{b}}} \log \det g ~,
\label{eq:rf}
\ee
where $\lambda$ is a parameter defining a family of metrics.
This differs by a factor of two from the Riemannian case~\cite{song2012lecture}.
We may take $g(0)$ to be the Fubini--Study metric and evolve this according to the differential equation.
A fixed point of this flow is the flat metric.
It turns out the Ricci flow preserves the K\"ahler class~\cite{cao1985deformation, chen2006ricci}.
The right hand side of~\eref{eq:rf} therefore provides another measure of how close we are to being Ricci flat, and the Ricci flow evolves the Fubini--Study metric to the flat metric on a Calabi--Yau.

For many purposes, it is convenient to augment the metric with a scalar function $f$.
When we do this, the Ricci flow equation~\eref{eq:rf} arises from a variational principle.
Ricci flow is then a gradient flow.
This is suggestive because neural networks often employ gradient flow in order to accomplish deep learning.
Define the Perelman functional
\bea
\mathcal{F}(g,f) &=& \int_{\mathcal{M}} d\mu\ e^{-f} \big( R + |\nabla f|^2 \big) \nn \\ &=& \int_{\mathcal{M}} dm\ \big( R + |\nabla f|^2 \big) ~,
\label{eq:act}
\eea
where $R$ is the scalar curvature.
In physics language, we have introduced a dilaton $f$.
The second equality in~\eref{eq:act} recasts $f$ in terms of the measure
\be
f := \log \frac{d\mu}{dm} ~.
\label{eq:dilaton}
\ee
Using this definition, variation of $\mathcal{F}$ gives an equation of motion for the modified Ricci flow:
\bea
\frac{\pa}{\pa \lambda} g_{a\bar{b}} &=& -(\text{Ric}_{a\bar{b}} + \nabla_a \nabla_{\bar{b}} f) ~.
\label{eq:mrf}
\eea
Mapping~\eref{eq:act} to a string action, the Ricci flow can be thought of as a beta function equation for the spacetime metric, which we treat as a coupling for the non-linear sigma model on the worldsheet.
In this case, the fixed points are conformal and correspond to Ricci flat target spaces that satisfy the vacuum Einstein equation.
Solutions to~\eref{eq:mrf} realize Perelman's energy monotonicity condition:
\be
\frac{d}{d\lambda} \mathcal{F} = 2\int_{\mathcal{M}} d\mu\ e^{-f} |\text{Ric}_{a\bar{b}} + \nabla_a \nabla_{\bar{b}} f|^2 = 2\int_{\mathcal{M}} dm\ |\text{Ric}_{a\bar{b}}|^2 ~.
\label{eq:frf}
\ee
This implies the existence of a coupled set of partial differential equations for the metric and dilaton:
\bea
\frac{\pa}{\pa \lambda} g_{a\bar{b}} &=& -(\text{Ric}_{a\bar{b}} + \nabla_a \nabla_{\bar{b}}f) ~, \label{eq:metricci} \\
\frac{\pa}{\pa \lambda} f &=& -\Delta f - R ~. \label{eq:bhe}
\eea
The previous two equations are the same as~\eref{eq:mrf} and~\eref{eq:dilaton}.
In particular, the second equation~\eref{eq:bhe} is a backward heat equation.
To efficiently address this coupled system, we must consistently solve the modified Ricci flow forward in time and the heat equation backward in time. The backward heat equation is not parabolic and therefore we have no guarantee of a solution to exist for a given initial condition $f(\lambda=0)$. In  order to solve this system, one has to bring~\eqref{eq:metricci} and~\eqref{eq:bhe} into the following form: 
\begin{align}
   \frac{\pa}{\pa \lambda} g_{a\bar{b}}\,&=\,-\text{Ric}_{a\bar{b}}\,,\\
      \frac{\pa}{\pa \lambda} f\,&=\,-\Delta f+|\nabla f|^2-R\,.
\end{align}
In this fashion, we can solve for $g$ in for $\lambda\in[0,T]$, with $T$ such that $g(\lambda)$ is smooth in $[0,T]$. One can then use the solution for $g$ and solve backwards in $\lambda$ starting from a boundary condition $f(\lambda=T)$. 

\section{Neural networks}\label{sec:nn}
A fully connected neural network correlates an input vector to an output vector in order to approximate a true result.
This correlation is highly non-linear.
We can write the $k$-layer neural network as a function
\be
\mathbf{v}_\text{out} = f_\theta(\mathbf{v}_\text{in}) =  L^{(k)}_\theta(\sigma^{(k-1)}( \cdots L_\theta^{(2)}(\sigma^{(1)} (L_\theta^{(1)}(\mathbf{v}_\text{in}))))) \approx \mathbf{v}_\text{true} ~,
\ee
where
\be
L_\theta^{(m)}(\mathbf{v}) = W^{(m)}_\theta \cdot \mathbf{v} + \mathbf{b}^{(m)}_\theta ~.
\ee
The $\mathbf{b}^{(m)}_\theta$ is called a bias vector, and $W^{(m)}_\theta$ is called a weight matrix.
The superscript denotes that we are in the $m$-th hidden layer of the neural network.
If $W^{(m)}_\theta$ is an $n_m\times n_{m-1}$ matrix, there are $n_m$ neurons in this layer.
In choosing the architecture, we specify $k$ and $n_m$ for $m=1,\ldots,k$ as well as the functions $\sigma^{(m)}$.
The elements of the bias vectors and the weight matrices are collectively termed hyperparameters, which we label by the subscript $\theta$.
These are fixed in training.
The non-linearity acts elementwise.
Three standard choices are
\bea
\text{logistic sigmoid}: &\quad& \sigma(x) = \frac{1}{1+e^{-x}} ~, \nn \\
\text{ReLU}: &\quad& \sigma(x) = x\, \Theta(x) ~, \\
\text{tanh}: &\quad& \sigma(x) = \tanh x ~. \nn
\eea
The training is executed by minimizing a specified loss function on the training set.
Optimization of the hyperparameters is accomplished, for example, by stochastic gradient descent (SGD) or adaptive moment estimation (Adam).
In fixing the hyperparameters, we pass the entire training set through the neural network multiple times; each time we do this is called an epoch.
Because the datasets are large, each epoch is reached by splitting the training set into several batches.
The number of epochs and the size of each batch are therefore parameters that enter the training.
Validation is performed by testing the trained neural network on inputs unseen during training.
The universal approximation theorem~\cite{Cybenko1989, Hornik1991} states that, with mild assumptions, a feedforward neural network with a single hidden layer and a finite number of neurons can approximate continuous functions on compact subsets of $\mathbb{R}^n$.
The performance of the neural network is gauged by whether a distance function $d(\mathbf{v}_\text{out}, \mathbf{v}_\text{true})$ is sufficiently small when this is suitably averaged over a dataset.

\section{Methodology for numerical Calabi--Yau metrics}\label{sec:methodology}

\subsection{Generation of points}\label{sec:points}
As inputs we take the affine coordinates describing various points in the manifold of interest.
For the computation of the numerical volumes and integrals in general, it is necessary to work with a uniform distribution of points.
For this purpose, the following method has been employed.
The same method was used in~\cite{Braun:2007sn,Ashmore:2019wzb}.
The general philosophy for Calabi--Yaus cut as degree $n+1$ hypersurfaces in $\mathbb{P}^n$ is based on the fact that lines in $\mathbb{P}^n$ are uniformly distributed with respect to the $SU(n+1)$ symmetry of the Fubini--Study metric.
Therefore, sampling the manifold with points at the intersection of each line with the hypersurface permits us to evaluate numerical integrations in a straightforward manner, taking the Fubini--Study metric as a measure of point distribution. 
The point selection process proceeds as follows. 
\begin{itemize}
\item First we generate a real vector $\mathbf{v}$ of random entries $-1\leq v_i\leq1$, $i=1\,\ldots\,2n$.
We only keep those vectors satisfying $|v|\leq1$.
\item Project the points to the hypersphere $S^{2n-1}$ by setting $\hat{v}=v/|v|$.
Furthermore, use $\hat{v}$ to build a point in $P\in\mathbb{P}^n$:
\begin{equation}
P=[\hat{v}_1+{\rm i}\hat{v}_2:\ldots\,:\hat{v}_{2n-1}+{\rm i}\hat{v}_{2n}] ~.
\label{eq:pp}\end{equation}
\item Build a line $L_{kj}=\{P_k+\lambda P_j\,|\, \lambda\in\mathbb{C}\}$ using two points $p_k$ and $p_j$ constructed in the manner highlighted above. 
\item For each line, one takes the points $\{p_l\}$ resulting from the intersection of the line with the hypersurface.
As they arise from random lines uniformly distributed with respect to the $SU(n+1)$ symmetry of the hypersphere, they are uniformly distributed with respect to the Fubini--Study metric on the hypersurface. 
\end{itemize}
A point $p=[z_1: \ldots :z_{n+1}]\in\mathbb{P}^n$ can be described in affine patches.
If the coordinate $z_m\neq 0$, then we can define the affine coordinates in the $m$-th patch as 
\begin{equation}
x^{m}_r=\begin{cases} z_r/z_m ~, \quad r<m ~, \\
z_{r+1}/z_m ~, \quad r>m ~. \end{cases}
\end{equation}
There is a preferred presentation for any given point.
Assume that for the point $p$ we have ${\rm max}_i(|z_i|^2)=|z_m|^2$.
We therefore know that in the $m$-th patch, the point $p$ is then described by the coordinates, 
\begin{equation}
x^{m}=(x^{m}_1\,,x^{m}_2\,\ldots\,,x^{m}_{n})\in D^n\,, \quad |x^{m}_{n}|\leq 1
\end{equation}
with $D$ a unitary disk in $\mathbb{C}$.
Note that in all other patches, the presentation of the same point will lie outside of the corresponding polydisk $D^n$ \cite{Cui:2019uhy}.
Furthermore, the hypersurface equation permits us to get rid of one coordinate that we denote the dependent coordinate.
In principle, one is allowed to choose which coordinate to take as the independent one.
For matters of numerical stability \cite{Ashmore:2019wzb}, it is recommended that we take as the dependent coordinate the one for which $|\partial Q/\partial x^{m}_r|$ is the maximum, with $Q$ being the corresponding hypersurface equation.
Assume that for the point $p$, this happens for the affine coordinate $x^{m}_l$.
For the entire manifold, we then split its points into different patches $S^{(m,k)}$, with $k=l$ if $l<m$ or $k=l+1$ if $l>m$.
Note that even though, there is a preferred patch for each point, in principle it has a presentation in all of the other patches, provided none of its homogeneous coordinates is zero. Recall that all the hypersurface equations considered in this work remain invariant under permutation of the coordinates. For this reason we expect the local metric expressions over the different patches to be identical. 

A slight modification of the point selection procedure has to be implemented for the complete intersection Tian--Yau manifold. For simplicity we take~\eref{eq:ty} to have the following form 
\begin{align}
Q_1&=w_1^3+w_2^3+w_3^3+w_4^3=0,\label{eq:q1}\\
Q_2&=z_1^3+z_2^3+z_3^3+z_4^3=0,\\
Q_3&=w_1 z_1+w_2 z_2+w_3 z_3+w_4 z_4=0\label{eq:q3}\,.
\end{align}
For the Tian--Yau the selection of points start with a random set of points in $\mathbb{P}^3$ exactly as done in~\eqref{eq:pp}. 
\begin{itemize}
\item We construct a line $L_{kj}=\{P_k+\lambda P_j\,|\, \lambda\in\mathbb{C}\}$ and a plane $P_{lmn}=\{P_l+\alpha P_m+\beta P_n\,|\, \alpha,\beta\in\mathbb{C}\}$. Take the line to belong to the first $\mathbb{P}^3$ and the plane to belong to the second $\mathbb{P}^3$. 
\item We then find all points in $L_{kj}\times P_{lmn}$ that satisfy the defining equations of the Tian--Yau manifold. Assume that $p\times p^\prime$ is one of these points. Note that $p^\prime \times p$ is also a solution of the system of equations. We keep both of these solutions as part of the set of sample points in the Tian--Yau complete intersection. 
\end{itemize}
Since lines and planes are both uniformly distributed over the hypersphere $S^7$ it is safe to assume that the points generated in this manner are uniformly distributed with respect to the symmetry of the Fubini--Study metric of the ambient space. Note that the set of transversely intersecting equations provides three dependent coordinates. Therefore the patches in this case will be described by two coordinates set to one using the $\mathbb{C}^*$ actions in the ambient space plus three dependent coordinates obtained from solving the Tian--Yau system of equations.

In contrast to the hypersurfaces, we can notice that arbitrary coordinate permutations do not necessarily leave \eqref{eq:q1}--\eqref{eq:q3} invariant. Therefore, not all local metrics are equivalent up to permutations. In order to work out the matching of patches, we need to follow a different approach. For the Tian--Yau manifold, we consider only training with the $\sigma$-measure as a loss; we defer the inclusion of the $\mu$- and $\kappa$-measures to forthcoming work. 

Now, let us briefly sketch how numerical integration proceeds. For simplicity, we focus on the hypersurfaces, but the same results can be straightforwardly extrapolated to the complete intersection case. Taking points in their corresponding preferred patch $S^{(m,l)}$ makes the numerical integration over $\mathcal{M}$ relatively feasible.
Picking the preferred presentation for each of the points makes the different patches to be almost disjoint.
(The choice is ambiguous for points at the boundary regions.
However, this does not affect the general integration process as the intersections constitute a set of measure zero.)
Assume that we want to obtain numerical estimates for the volume of $\mathcal{M}$, using the sample points generated with the method previously illustrated.
This volume is given by the expression 
\begin{equation}
{\rm Vol}_{\Omega}=\int_{\mathcal{M}} d{\rm Vol}_{\Omega} = \int_{\mathcal{M}} d{\rm Vol}_\text{FS}\ \left(\frac{d{\rm Vol}_{\Omega}}{d{\rm Vol}_\text{FS}}\right)\,.
\end{equation}
As the set of points of interest are uniformly distributed with respect to the Fubini--Study metric and $d{\rm Vol}_{\rm FS}$ is the Fubini--Study differential volume,
we can numerically approximate the volume as generated with the method previously illustrated.
This volume is given by the expression 
\begin{equation}
{\rm Vol}_{\Omega}=\frac{1}{N} \sum_{l=1}^N w_M (p_l)\,,\quad w_M (p_l)=\frac{d{\rm Vol}_{\Omega}}{d{\rm Vol}_{FS}} ~,
\end{equation}
where $N$ is the number of points under consideration.
Similarly, any integral over $\mathcal{M}$ can be evaluated numerically in the following manner:
\begin{equation}
\int_{\mathcal{M}} d{\rm Vol}_{\Omega}\ f(z) =\frac{1}{N} \sum_{l=1}^N f(p_l) w_M (p_l) ~.
\end{equation}

\subsection{Neural network architecture}
Our goal is to approximate the Hermitian Calabi--Yau metric from the output of a neural network.
As an $n\times n$ Hermitian matrix, the metric $g$ can be parameterized in terms of the following product
\begin{equation}
g=L\cdot D\cdot L^\dag ~,
\end{equation}
where $L$ is a lower triangular matrix with ones along the diagonal and $D$ is a diagonal matrix with real entries.
$L$ contains $n(n-1)$ real parameters, and $D$ must consist of $n$ real and positive numbers.
This is a variant of the classical Cholesky decomposition.
Since we want the metric to be generated from neural network outputs, we employ two neural networks for this process.
The first artificial neural network (ANN1) produces  $n$ outputs $o_1^{(1)}, \ldots, o_n^{(1)}$ that will serve to construct the matrix $D$.
\textit{A priori} these outputs need not to be positive, and for this reason we construct $D$ with the output exponentials or squares
\begin{equation}
D={\rm diag}(e^{o_1^{(1)}},\ldots,e^{o_n^{(1)}}), \text{~~~or,~~} D={\rm diag}({(o_1^{(1)})^2},\ldots,{(o_n^{(1)}})^2) ~.
\end{equation} 
The second artificial neural network (ANN2) outputs $o_i^{(2)}$ are combined into the entries of the matrix $L$.
Figure~\ref{fig:arch} displays the architecture of the neural network. For example, in the K3 case the metric $L$ takes the form 
\begin{equation}
L=\begin{pmatrix} 1 & 0 \\
o_1^{(2)}+{\rm i} o_2^{(2)} & 1 \end{pmatrix} 
\end{equation}
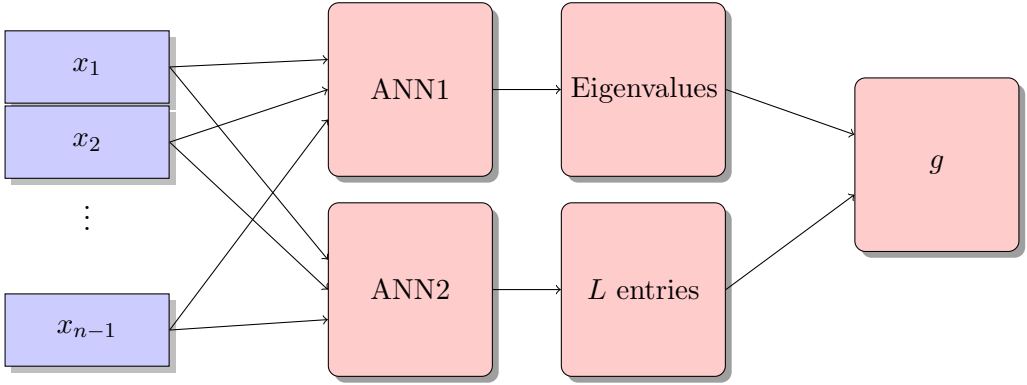
\begin{figure}[h!]
\centering
\tikzstyle{sensor}=[draw, fill=blue!20, text width=5em, 
    text centered, minimum height=2.5em,drop shadow]
\tikzstyle{ann} = [above, text width=5em, text centered]
\tikzstyle{wa} = [sensor, text width=5em, fill=red!20, 
    minimum height=6em, rounded corners, drop shadow]
\tikzstyle{sc} = [sensor, text width=13em, fill=red!20, 
    minimum height=10em, rounded corners, drop shadow]

\begin{tikzpicture}
    \node (wa) [wa]  {ANN1};
    \path (wa.west)+(-3.2,0.3) node (asr1) [sensor] {$x_1$};
    \path (wa.west)+(-3.2,-0.7) node (asr2)[sensor] {$x_2$};
    \path (wa.west)+(-3.2,-2.0) node (dots)[ann] {$\vdots$}; 
    \path (wa.west)+(-3.2,-3.2) node (asr3)[sensor] {$x_{n-1}$};    
   
    \path (wa.south)+(0,-1.5) node (vote0) [wa] {ANN2};
     \path (wa.east)+(2.0,0) node (vote) [wa] {Eigenvalues};
      \path (vote.south)+(0,-1.5) node (vote2) [wa] {$L$ entries};
   \path (vote.west)+(5.0,-1.0) node (vote1) [wa] {$g$};

    \path [draw, ->] (asr1.east) -- node [above] {} 
        (wa.160) ;
    \path [draw, ->] (asr2.east) -- node [above] {} 
        (wa.180);
    \path [draw, ->] (asr3.east) -- node [above] {} 
        (wa.200);
     \path [draw, ->] (asr1.east) -- node [above] {} 
        (vote0.160) ;
    \path [draw, ->] (asr2.east) -- node [above] {} 
        (vote0.180);
    \path [draw, ->] (asr3.east) -- node [above] {} 
        (vote0.200);
    \path [draw, ->] (wa.east) -- node [above] {} 
        (vote.west);       
    \path [draw, ->] (vote.east) -- node [above] {} 
        (vote1.160);
    \path [draw, ->] (vote2.east) -- node [above] {} 
        (vote1.200);
     \path [draw, ->] (vote0.east) -- node [above] {} 
        (vote2.west);
\end{tikzpicture}
\caption{\emph{Flow chart of the algorithm. Two separate neural networks provide the eigenvalues as well as the entries of the diagonal matrix $L$. They get combined into the metric $g$ that is used to minimize the combined loss function for both networks simultaneously.} 
}
   \label{fig:arch}
\end{figure}

The best architecture found for our purposes is to take ANN1 and ANN2 to be be multi--layer perceptron (MLP) neural networks with three hidden layers. Each hidden layer has $500$ neurons. No specific initialization was used. All experiments were implemented in \texttt{PyTorch}. The learning rate was taken to be $0.001$; the batch size was $500$ points. We use Adam optimization. The training and test set sizes were chosen to be $2000$ each, and the experiments were run for $20000$ epochs. The inputs to both neural networks are points in the affine patch and we input both dependent and independent coordinates.  

\subsection{Loss functions}
As we have already noted in Section~\ref{sec:flatness}, the $\sigma$- and $||R||$-measures are positive and bounded quantities that measure how close the metric approximation is to the Ricci flat metric.
The $\sigma$-measure has the advantage that we do not need to take derivatives of the neural network outputs. 

Additionally, we must take into account two other properties, in the case of the neural network the metric is not directly obtained from a K\"ahler potential.
The K\"ahler property of the metric has to be checked, and this is guaranteed when the K\"ahler form is closed:
\begin{equation}
dJ=0 \qquad \Longrightarrow \qquad \partial_a g_{b\bar{c}}-\partial_b g_{a\bar{c}}=0 ~.
\end{equation}
Let us then define the quantity 
\begin{equation}
k_{ab\bar{c}}= \partial_a g_{b\bar{c}}-\partial_b g_{a\bar{c}} ~.
\end{equation}
We can define the Frobenius norm of $k$ as follows 
\begin{equation}
|k|^2=\sum_{a,b,\bar{c}}|k_{ab\bar{c}}|^2 ~.
\end{equation}
From this, we define the $\kappa$-measure:
\begin{equation}
\kappa=\frac{{\rm Vol}_J^{1/n}}{{\rm Vol}_\Omega}\int_{\mathcal{M}}d{\rm Vol}_J\ |k|^2 ~.
\end{equation}

An additional property to check is that the boundary conditions for the metric are satisfied.
Take a point $p$ that has a presentation in the patch $S^{(m,l)}$ as well as in the patch $S^{(m^\prime,l^\prime)}$.
Assume that the metric in the first patch is described by $g^{(m,l)}$ and by $g^{(m^\prime,l^\prime)}$ on the second patch.
Then at point $p$, 
\begin{equation}
g^{(m,l)}=J_{(m^\prime,l^\prime)}^{(m,l)}g^{(m^\prime,l^\prime)}\bar{J}_{(m^\prime,l^\prime)}^{(m,l)} ~.
\end{equation}
where $J_{(m^\prime,l^\prime)}^{(m,l)}$ are the Jacobians of transformation between coordinate patches. Now define the matrix 
\begin{equation}
M(m^\prime,l^\prime;m,l)=g^{(m,l)}-J_{(m^\prime,l^\prime)}^{(m,l)}g^{(m^\prime,l^\prime)}\bar{J}_{(m^\prime,l^\prime)}^{(m,l)} ~,
\end{equation}
and define the $\mu$-measure
\begin{equation}
\mu=\frac{1}{N_p!}\sum_{m^\prime,l^\prime}\sum_{m.l\neq m^\prime,l^\prime}\frac{1}{{\rm Vol}_\Omega}\int_{\mathcal{M}}d{\rm Vol}_J\ |M(m^\prime,l^\prime;m,l)|^2 ~,
\end{equation}
where $|M(m^\prime,l^\prime;m,l)|$ denotes the Frobenius norm and $N_p$ is the number of patches. 

A good measure of how close the metric approximation is to the actual flat metric can be constructed in the following manner:
\begin{equation}
{\rm Loss}=\alpha_\sigma \sigma+\alpha_{\kappa}\kappa+\alpha_\mu \mu\,,
\label{eq:loss}
\end{equation} 
with $\alpha_\sigma$, $\alpha_{\kappa}$  and $\alpha_\mu$ are real positive coefficients that ensure the three quantities of interest make commensurate contributions to the loss function. 

\subsection{Ricci flow with a neural network}

One alternative approach to obtaining the Ricci flat metric of a Calabi--Yau manifold would be to follow the Ricci flow starting with the Fubini--Study metric (in case such manifold is embedded as a hypersurface, or complete intersection in a weighted projective space). This contrasts with the general approach of constructing a functional for which the flat metric can be obtained as a minimum for such a functional. This more direct implementation poses interesting challenges. In the first place on has to ensure that numerical errors do not propagate as one evolves the flow in the parameter $\lambda$, potentially driving us away from the desired flat metric. 

\begin{figure}[h!]
  \vspace{10pt}
\centering
\includegraphics[scale=.25]{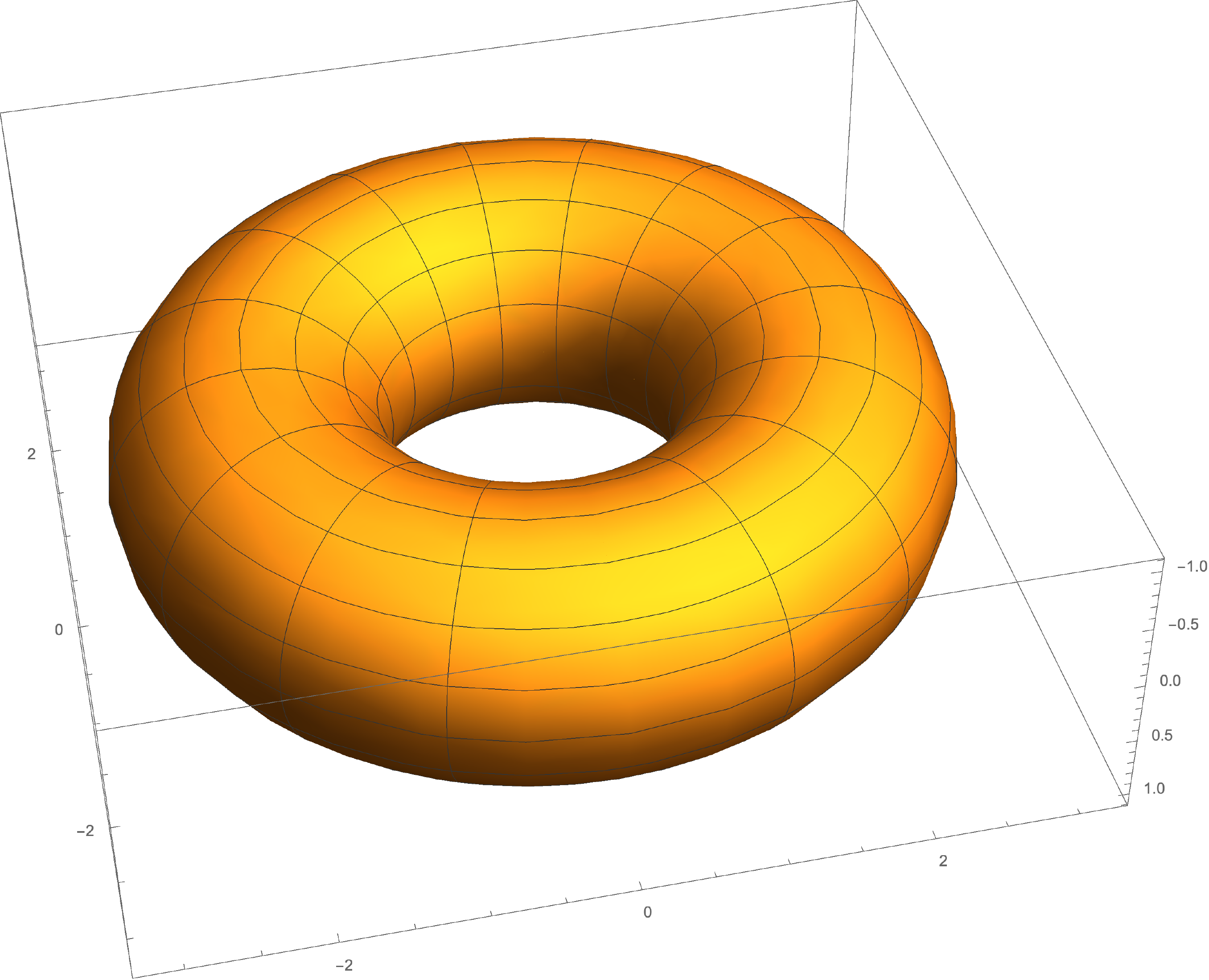}~~\includegraphics[scale=.4]{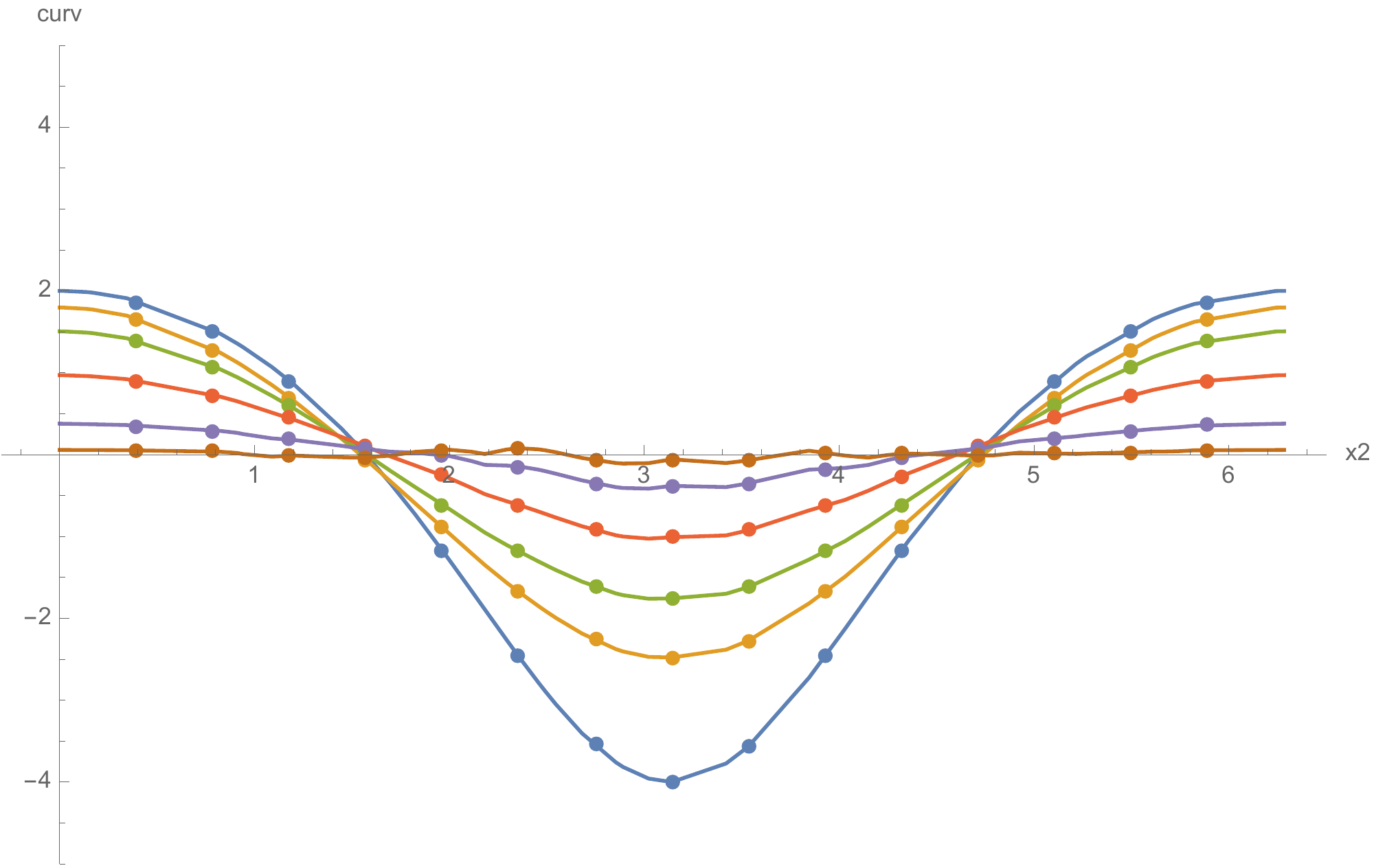}\\[20pt]
\caption{\small\textit{The scalar curvature of the real torus on the left is plotted as a function of the angular coordinate of the torus. By solving the Ricci flow equations we see a gradual shift to a metric of approximately zero scalar curvature. }}
  \label{fig:RealT2fig}
  \vspace{10pt}
\end{figure}

A a proof of principle, we have applied this method to learn the flat metric on the square torus treated as a real manifold. We take the domain $x\in [0,2\pi)$ and $y\in[0,2\pi)$. We take the metric coming from the embedding into $\mathbb{R}^3$ given by
\begin{equation}
    g(\lambda=0)=\begin{pmatrix}
    (c + a\, \cos y)^2 & 0\\
 0 & a^2 
\end{pmatrix}
\label{eq:init}
\end{equation}
with $a$ and $c$ being the corresponding torus radii. 
We present this in Figure \ref{fig:RealT2fig}. The approach here is slightly different as we employ Hamilton's Ricci flow formalism to flow the metric \eqref{eq:init} to the Ricci flat metric. The evolution of the Ricci scalar as function of the $y$ parameter is obtained using a numerical partial differential equation solver in \texttt{Mathematica}.

There are several challenges to overcome in the implementation of real Ricci flow on Riemannian manifolds. One such challenge is the likely appearance of curvature singularities at finite time as one evolves the flow. This is a typical feature due to the non linear nature of the Ricci flow equation. Additionally, due to gauge fixing, coordinate singularities might appear as well. Complex and especially K\"ahler manifolds permit us to deal with this problem in a simpler form, as the metric can be described in terms of a single function. In itself, K\"ahler--Ricci flow is more robust as convergence is proven~\cite{cao}. Also, for K\"ahler--Ricci flow on Calabi--Yau manifolds one generically does not expect the appearance of coordinate singularities.

A preliminary avenue to explore the flow would be to consider a neural network approximation of the metric. For simplicity, instead of considering gradient flow for the Perelman loss, we can consider solving the differential equation taking the neural network to approximate the solution for the metric $g$. In that case we do not need to worry about the dilaton and take the flow equation to be 
\begin{equation}\label{real:RicciFlow}
    \partial_\lambda g_{a\bar{b}}=-R_{a\bar{b}} ~.
\end{equation}
In order to set the boundary conditions, we start with a given known metric (\textit{e.g.}, the Fubini--Study metric) at $\lambda=0$:
\begin{equation}
g(\lambda=0)=g_{FS} ~.
\end{equation}
On the neural network side, this means that we start by training the network to reproduce the Fubini--Study metric.
We can then update the data at $\lambda=\Delta \lambda$ by computing the Ricci tensor $R_{a\bar{b}}$ on the neural network approximation and then train the network at $\lambda=\Delta \lambda$ with the following data:
\begin{equation}\label{stepflow}
g_{a\bar{b}}(\lambda=\Delta \lambda)=g_{a\bar{b}}(0)-R_{a\bar{b}}(0) \Delta \lambda ~,
\end{equation}
with $R_{a\bar{b}}$ being the neural network approximation for the Ricci tensor.
As such we are required to approximate not simply the metric, but also the corresponding Ricci tensor as accurately as possible. This requires a learning paradigm where derivative observations are available, synthetically generated or otherwise. Therefore, we would need to compute second derivatives of the network approximating the flat metric. Standard autograd tools in widely used machine learning packages (\textit{e.g.}, \texttt{PyTorch}, \texttt{TensorFlow}) allow fast gradient computations of neural networks.  It has been noted in the machine learning literature that  derivative observations can improve predictors as well as generalization. This has been demonstrated by the use of both Bayesian and non-Bayesian tools. In~\cite{wu2017exploiting}, for example, it was shown that derivative observations can improve the predictive power of Gaussian processes. It was recently shown in~\cite{czarnecki2017sobolev} that the same holds for neural networks. The authors proposed a new training paradigm: \textit{Sobolev training}, wherein the loss function is constructed out of observations of the function values as well as derivatives, up to some order. Additionally, there are theoretical guarantees for existence of networks (with ReLU or leaky ReLU activations) that can approximate a function, with the network's derivatives approximating the function's derivatives. A consequence of Sobolev training is that it has lower sample complexity than regular training. 

This learning paradigm is directly applicable to our situation, where both the metric and its derivatives (connections, Ricci tensor) have geometric meanings. The distinction from the method proposed in~\cite{czarnecki2017sobolev} is that the values of the metric or its derivatives are not readily available, except at the beginning of the flow. As such, we propose to generate this data at any step of the flow, from the network approximating the metric at the previous step, \`a la \eqref{stepflow}. We use the following dynamic loss function: 
\begin{equation}\label{threelosses}
{\rm Loss(\lambda)}=\alpha_0 {\rm MSE}(g_{NN}(\lambda),g(\lambda))+\alpha_1 {\rm MSE}(\nabla g_{NN}(\lambda),\nabla g(\lambda))+\alpha_2 {\rm MSE}(\nabla^2 g_{NN}(\lambda),\nabla^2 g(\lambda))\,,
\end{equation}
where $\lambda$ is the flow parameter, MSE denotes the mean square error, $g_{NN}(\lambda)$ is the approximation metric and $g(\lambda)$ is the target metric, when the flow parameter is $\lambda$. 
The parameters $\alpha_0$, $\alpha_1$, and $\alpha_2$ are weights set to ensure that the different mean squared error losses are of the same order. 

While we have preliminary results for the complex torus and K3, the conclusions we may draw from the implementation are so far only tentative. Part of the issue here is in designing a loss function that efficiently localizes to the solution of a coupled set of partial differential equations,~\eref{eq:metricci} and~\eref{eq:bhe}.
The results for $T^2$, K3, and Calabi--Yau threefolds will be reported in forthcoming work. 

\section{Results}\label{sec:results}
\subsection*{Overview}
In this section, we document the results of our machine learning experiments in modelling Ricci flat metrics on Calabi--Yau manifolds. We consider the complex torus in one dimension, the quartic K3 surface, the Fermat quintic, a second member of the Dwork family of quintics, and the Tian--Yau manifold. For these geometries, we model the flat metric by a neural network whose architecture was chosen keeping simplicity in mind. Since neural network are essentially black boxes, the simplicity of the network is beneficial in the analysis of the networks. We choose a network with three hidden layers with $500$ nodes each. We employ three distinct activation functions --- $\tanh$, ReLU, and logistic sigmoid, which as we will see lead to different training dynamics. We optimize over the network parameters initialized randomly, using either the full loss function in~\eqref{threelosses} or the partial $\sigma$-loss. We highlight the case of the Fermat quintic in greater detail. The longest it took to train our networks in any of the experiments was one hour on a laptop.\footnote{We use a $2.6$ GHz 6-Core Intel Core i7 processor.}

\subsection{The torus}\label{sec:torus}
Consider the torus $T^2=\mathbb{C}/\Lambda$ with $\Lambda=\langle1,\tau\rangle_\mathbb{Z}$.
A general point on the torus can be described by the complex coordinate $z\in\mathbb{C}$, with the identification $z\sim z+\lambda$ for any $\lambda\in\Lambda$.
In this way the metric on the torus inherited from $\mathbb{C}$ is
\begin{equation}\label{eq:rflat}
ds^2=dz\,d\bar{z}\,,
\end{equation}
which is obviously Ricci flat.
By contrast, treating $T^2 \simeq S^1\times S^1 \subset \mathbb{R}^3$, the induced metric from the embedding is not Ricci flat.

\comment{
The torus can also be described in terms of an elliptic curve.
Let us briefly sketch the map between these two pictures. 
Define the Weierstrass $\wp$-function
\begin{equation}
\wp(z)=\frac{1}{z}+\sum_{\lambda\in\Lambda\,,\,\lambda\neq 0}\left(\frac{1}{(z+\lambda)^2}-\frac{1}{\lambda^2}\right)
\end{equation}
Recall that $\wp(z)$ and its derivative $\wp^\prime(z)$ fulfill the Weierstrass equation: 
\begin{equation}
\wp^{\prime\,2}(z)=4\wp^3(z)-g_2\wp(z)-g_3\,,
\end{equation}
with 
\begin{equation}\label{eq:ws}
g_2=60\sum_{\substack{
\lambda\in\Lambda \\ \lambda\neq 0}}\frac{1}{\lambda^4}\,,\quad  g_3=140\sum_{\substack{
\lambda\in\Lambda \\ \lambda\neq 0}}\frac{1}{\lambda^6}\,.
\end{equation}
One can see this as a homogeneous cubic polynomial in $\mathbb{P}^2$. Take $[X,Y,Z]$ homogeneous coordinates on $\mathbb{P}^2$, then the torus is cut by the homogeneous cubic polynomial 
\begin{equation}
Y^2 Z=4 X^3-g_2 X Z^2-g_3 Z^3\,,
\end{equation}
which coincides with~\eqref{eq:ws} in the patch $Z=1$ once we identify $X=\wp(z)$ and $Y=\wp^\prime(z)$. Note that the previous Equation gives two values of $Y$ for each value of $X$ except for three points (where $4 X^3-g_2 X-g_3=0$), we therefore obtain two Riemann sheets connected through branch cuts. if we focus in one of the sheets and take $X$ to be the coordinate on it. We would like to see how the Ricci flat torus metric looks in terms of this new coordinate. 
\begin{equation}
dz\, d\bar{z}=\left(\frac{dz}{dX}\right)\left(\frac{d\bar{z}}{d\bar{X}}\right)dX\, d\bar{X}
\end{equation}
but since $X=\wp(z)$, we have $dz/dX=1/\wp^\prime(z)=1/Y$, 
\begin{equation}
dz\, d\bar{z}=\frac{dX\, d\bar{X}}{|Y|^2}
\end{equation}
Another manner to deduce the form of the Ricci flat metric over the Riemann sheet is to deduce directly from the holomorphic top form $\Omega$. In the $\mathbb{P}^2$ patch $Z=1$ the top form over the Torus reduces to 
\begin{equation}
\Omega=\int_{T^2} \frac{d X \wedge d Y}{Y^2-4 X^3+g_2 X +g_3}=\frac{dX}{2 Y}
\end{equation}
therefore the Ricci flat Kaehler form is 
\begin{equation}
\omega\sim \Omega\wedge\bar{\Omega}= \frac{d X \wedge d \bar{X}}{4 |Y|^2}
\end{equation}

\subsubsection{The torus $T^2$ as a cubic in $\mathbb{P}^2$}
}

As in~\eref{eq:cubic}, consider the torus $T^2$ defined by a ternary cubic equation in $\mathbb{P}^2$:
\begin{equation}\label{eq:t2}
z_1^3+z_2^3+z_3^3=0 ~,
\end{equation}
with homogeneous coordinates $[z_1:z_2:z_3]$.

\begin{figure}[h!]
\centering
\includegraphics[scale=.35]{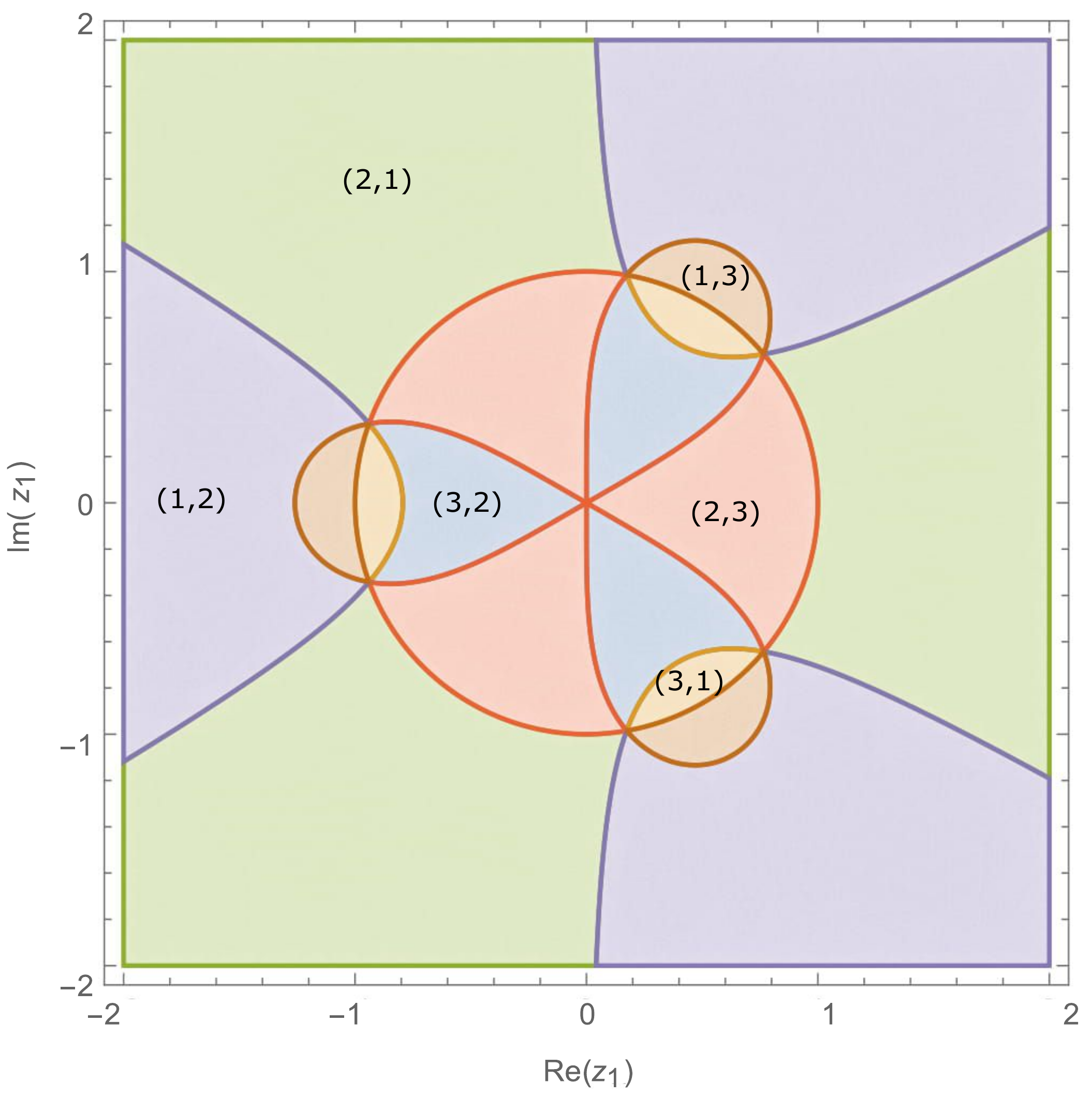}~~\includegraphics[scale=.35]{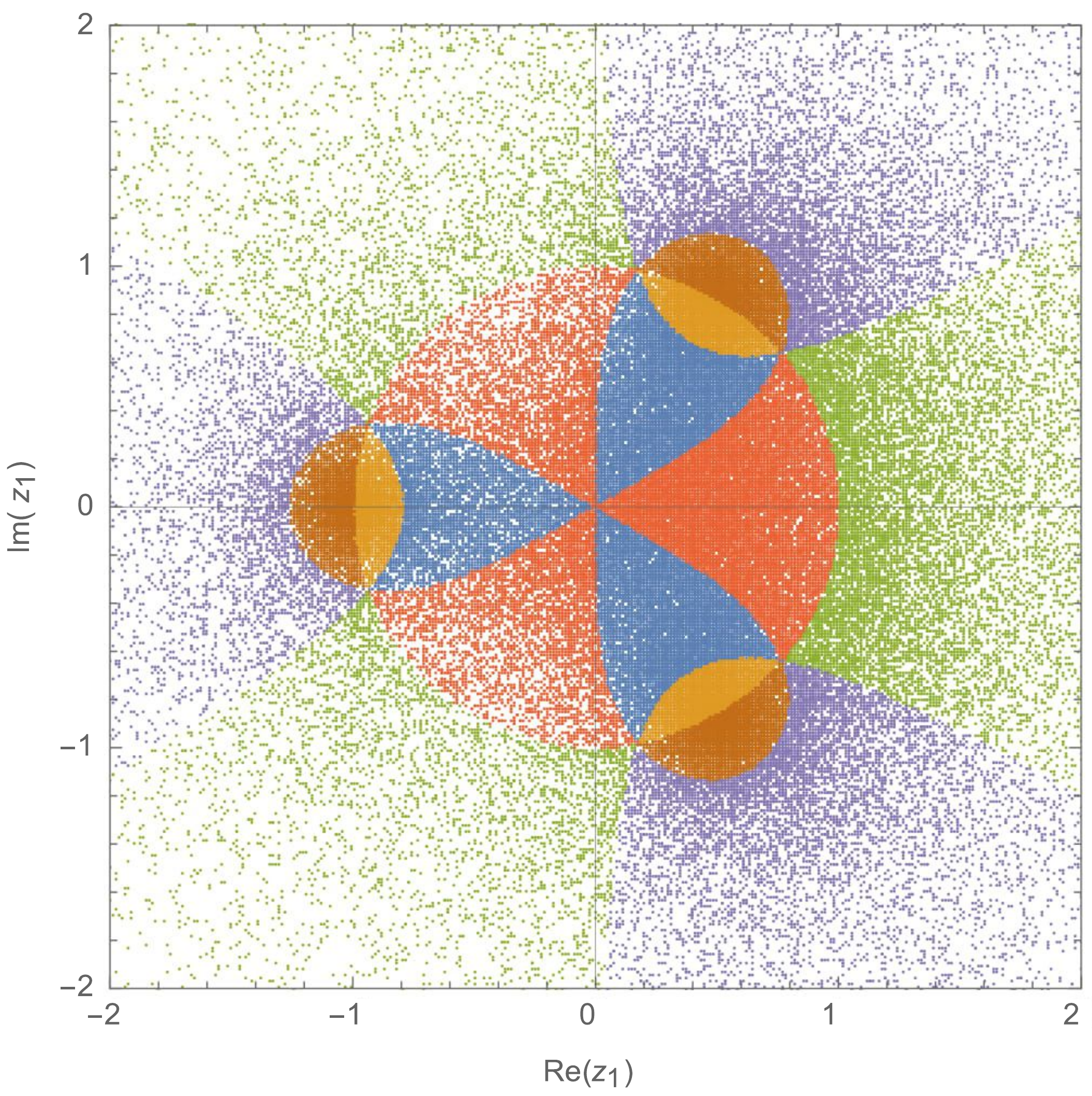}\\[20pt]
\caption{\small\textit{The torus can be viewed as three Riemann sheets glued together through branch cuts stretching between the cubic roots of $-1$ and the point at infinity. Three copies of the complex plane span the chart (3,2). The different regions in color correspond to the intersection with other patches. In the plot on the right, we present the distribution of sample points following the method indicated in Section~\ref{sec:points}.}}
  \label{fig:T2patches}
  \vspace{10pt}
\end{figure}

We can consider the Fubini--Study metric in $\mathbb{P}^2$ restricted to the torus.
In particular, let us consider the patch where $z_3=1$ and take $z_2$ to be a function of the coordinate $z_1$. In terms of the patching scheme discussed in Section~\ref{sec:points}, this is the $(3,2)$ patch of the torus. There are six patches in total. In Figure~\ref{fig:T2patches}, we have depicted the $(3,2)$ patch corresponding to the complex plane spanned by the complex coordinate $z_1$. We have included the overlapping regions with the other five patches as well as the distribution of points generated using the hypersphere method. As expected from the symmetry, one gets roughly the same number of points in every preferred patch. 
Note that for each value of $z_1$, one generically obtains three different values for $z_2$, which lie in the torus defined by~\eqref{eq:t2}.
However, one can concentrate on one of the Riemann sheets, particularly since the values of the metric at each of the three roots coincide.

\begin{figure}[h!]
\centering
\includegraphics[scale=.6]{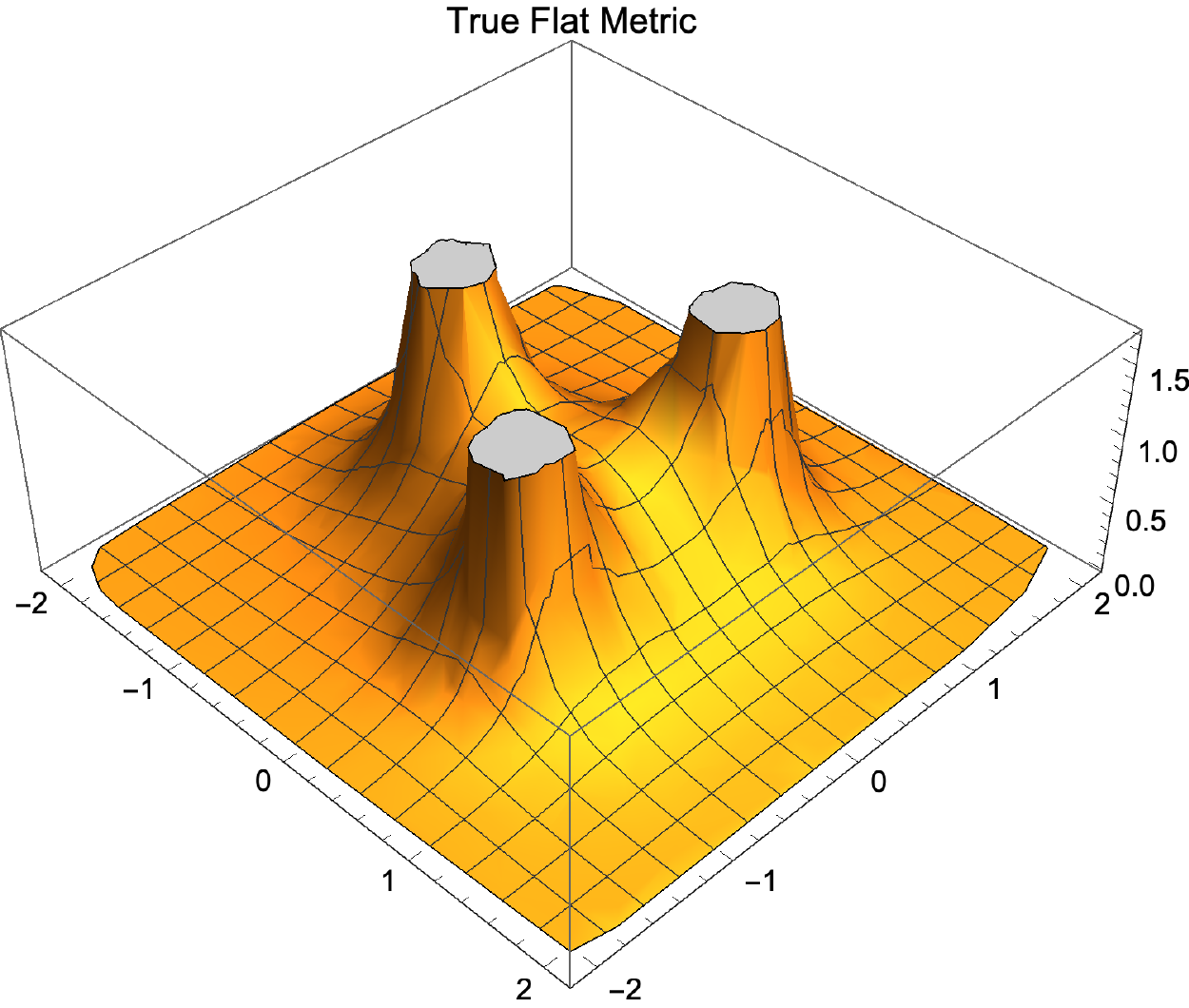}~~\includegraphics[scale=.3]{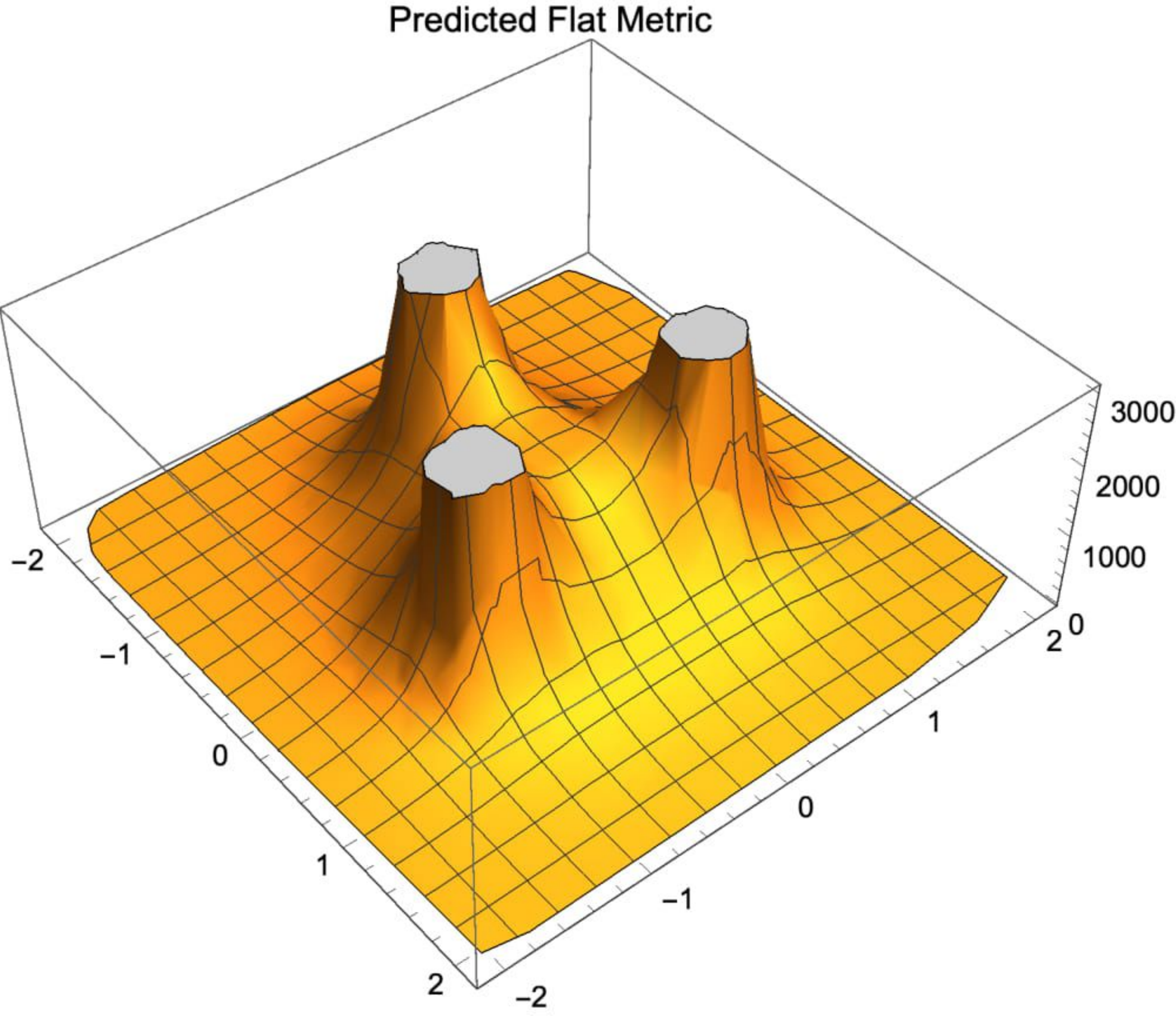}\\[20pt]
\caption{\small\textit{True and predicted Ricci flat metrics on the complex torus \eqref{eq:t2}. The axes are given by the real and imaginary parts of the independent coordinate $z_1$. The three peaks appear at points $z_1^3=-1$ for which the coordinate $z_2=0$. Note that these apparent singularities for the metric are not there if one takes $z_2$ as the dependent coordinate for those points. The predicted metric differs from the true metric by an overall scaling thus yielding a vanishing Ricci curvature.}}
  \label{fig:T2fig}
  \vspace{10pt}
\end{figure}

The Fubini--Study metric in the patch under consideration reads
\begin{equation}\label{eq:fst2} 
ds^2=\left(\frac{(1 +|z_1|^4+ |z_2|^4)}{\pi  |z_2|^4 (1 +  |z_1|^2+ |z_2|^2)^2}\right)dz_1 d\bar{z}_1 ~,
\end{equation}
with $z_2=(-1-z_1^3)^{1/3}$.
The Ricci flat metric can be obtained from the following relation 
\begin{equation}\label{eq:flatt2}
J \sim \Omega \wedge \bar{\Omega} = \frac{d z_1 \wedge d \bar{z}_1}{9|z_2|^4} ~.
\end{equation}
The top holomorphic form $\Omega$ on a torus is a $(1,0)$ form; $J$ is the K\"ahler form.
For the metric, we consider an ansatz of the form 
\begin{equation}
J^\prime=\frac{(\alpha_1+\alpha_2 |z_1|^4+\alpha_3 |z_2|^4)dz_1 \wedge d\bar{z}_1}{\pi  |z_2|^4 (\beta_1 +  \beta_2 |z_1|^2+ \beta_3|z_2|^2)^2}\,. 
\end{equation}
In order to approach the flat metric starting from this ansatz, we devise an algorithm that runs over the space of real parameters $\{\alpha_i,\beta_i\}$ in order to minimize the norm squared of the Ricci tensor.
Indeed, we find that the optimal parameters are of the order of one part in $10^7$ with the exception of $\alpha_2=654.92$ and $\beta_2=378.08$.
Therefore the metric obtained is given by~\eqref{eq:flatt2} up to an overall coefficient.
This is illustrated in Figure~\ref{fig:T2fig}.


\subsection{The K3 surface}\label{sec:k3}
\begin{figure}[h!]
\centering
\includegraphics[scale=.35]{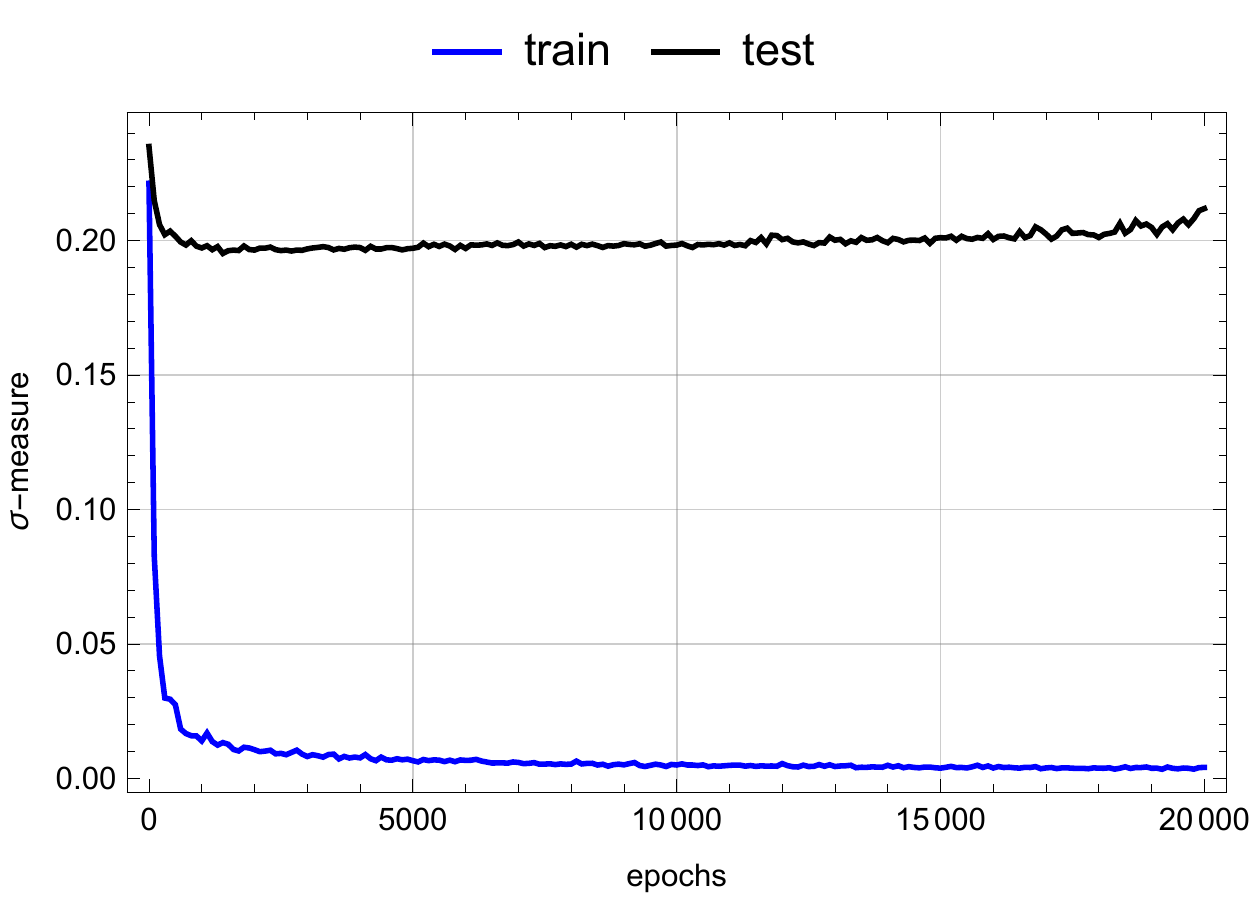}~~\includegraphics[scale=.35]{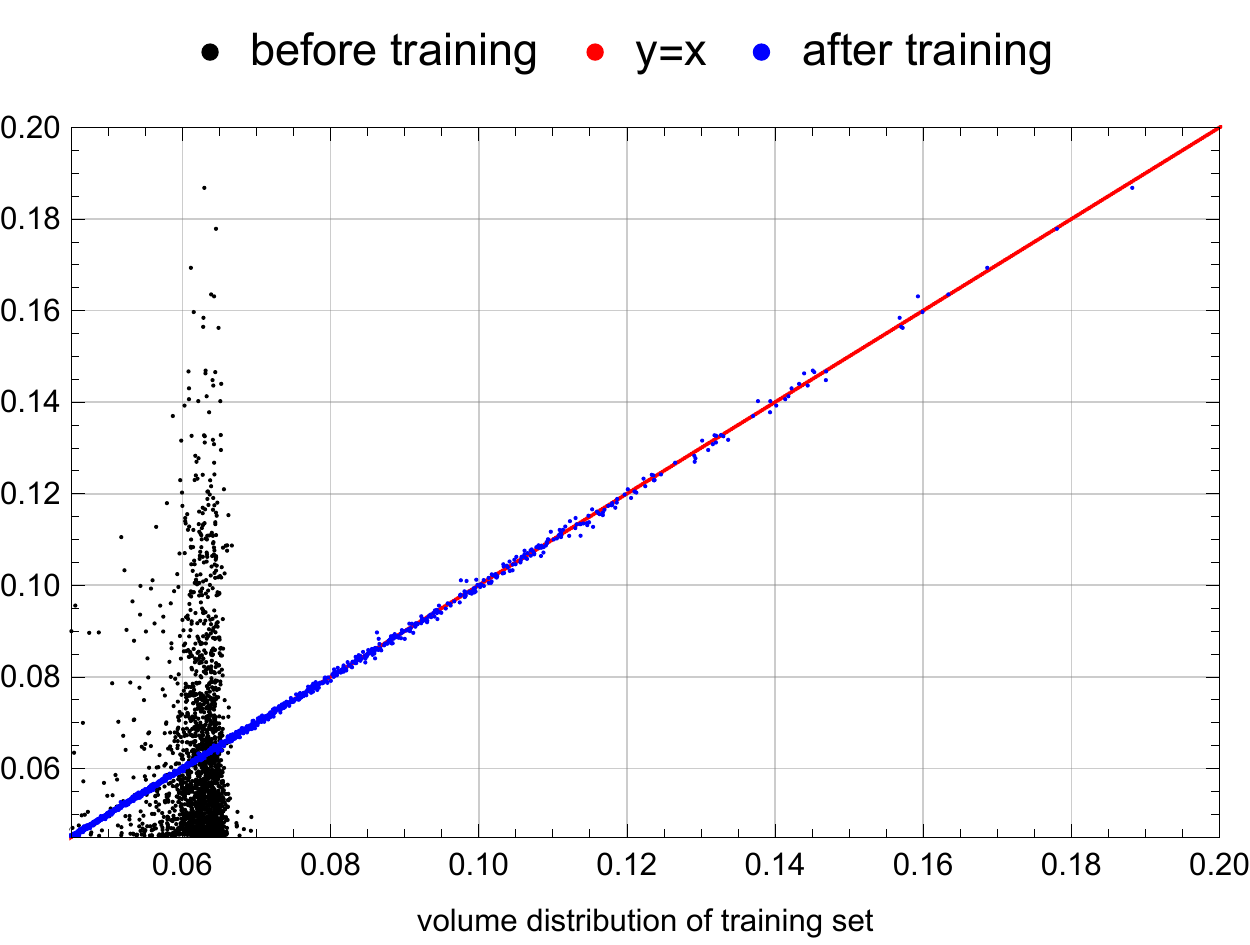}~~\includegraphics[scale=.35]{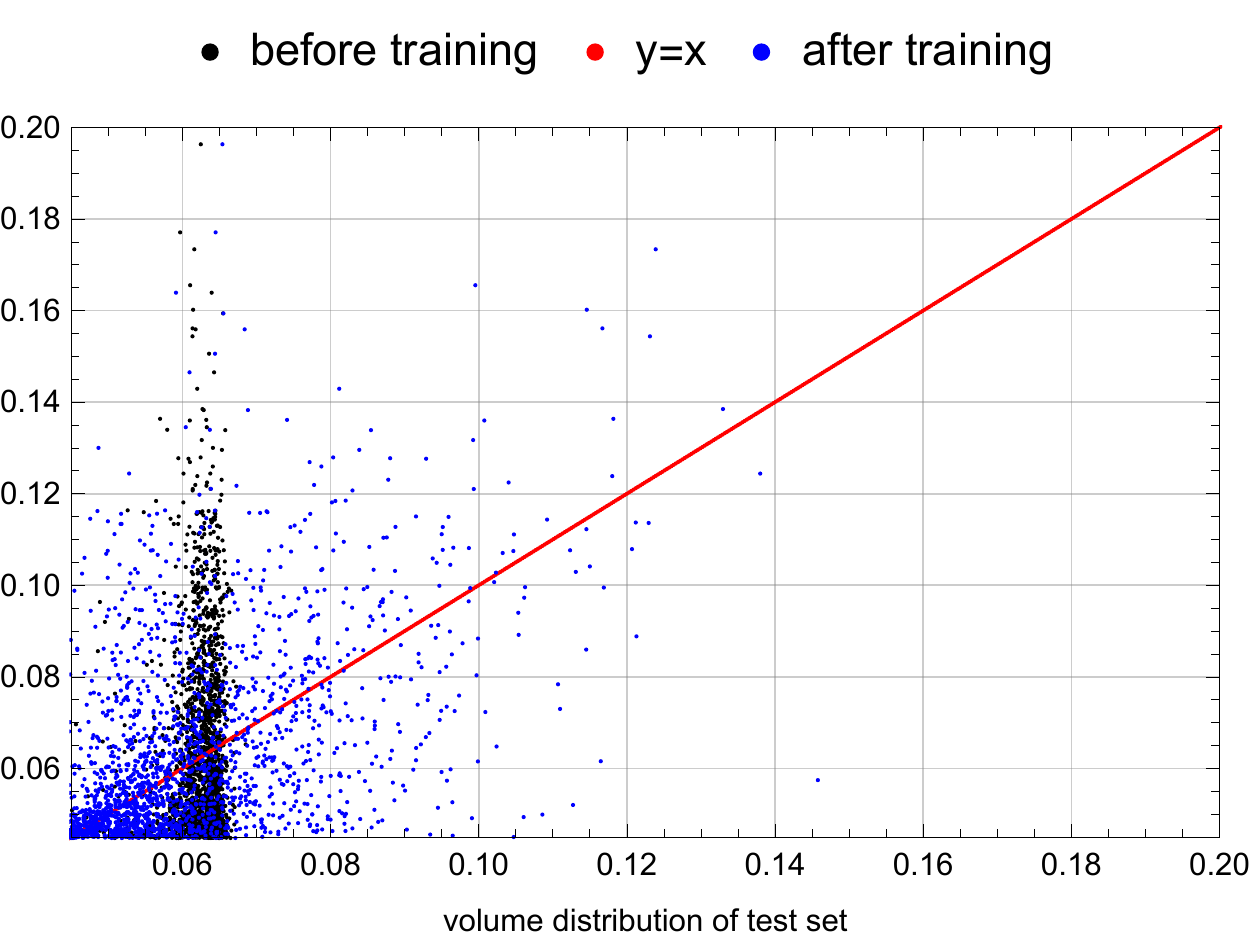}\\[20pt]
 \includegraphics[scale=.35]{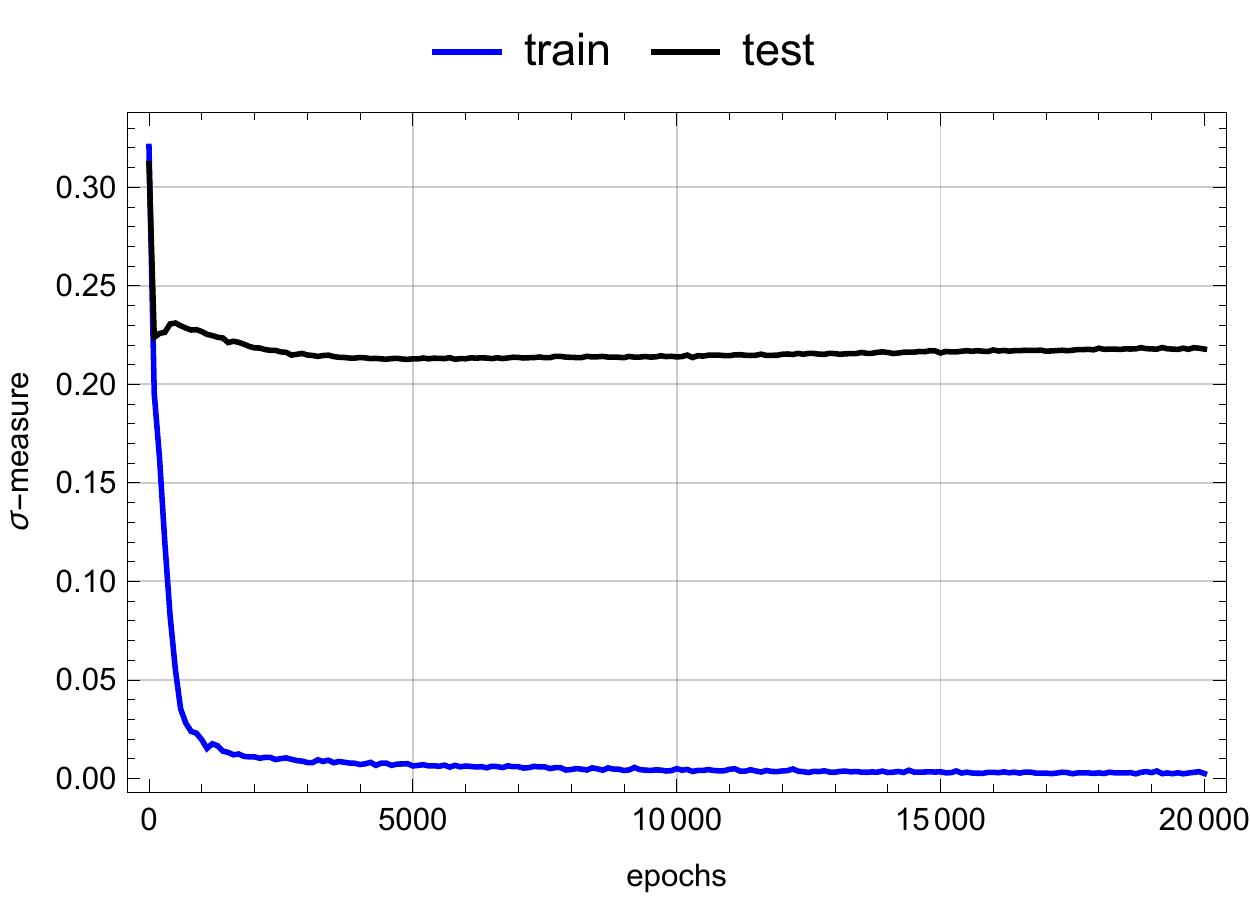}~~\includegraphics[scale=.35]{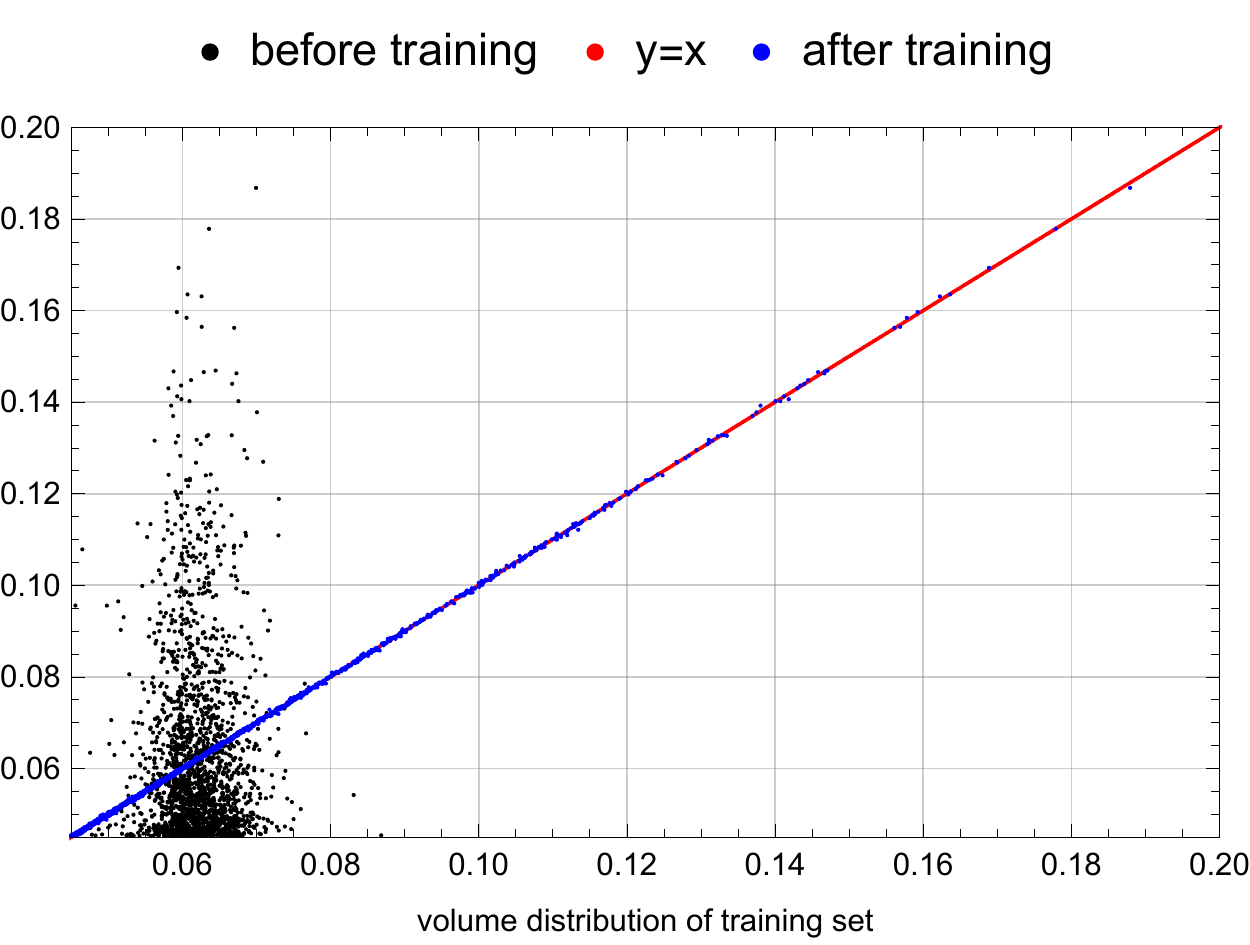}~~\includegraphics[scale=.35]{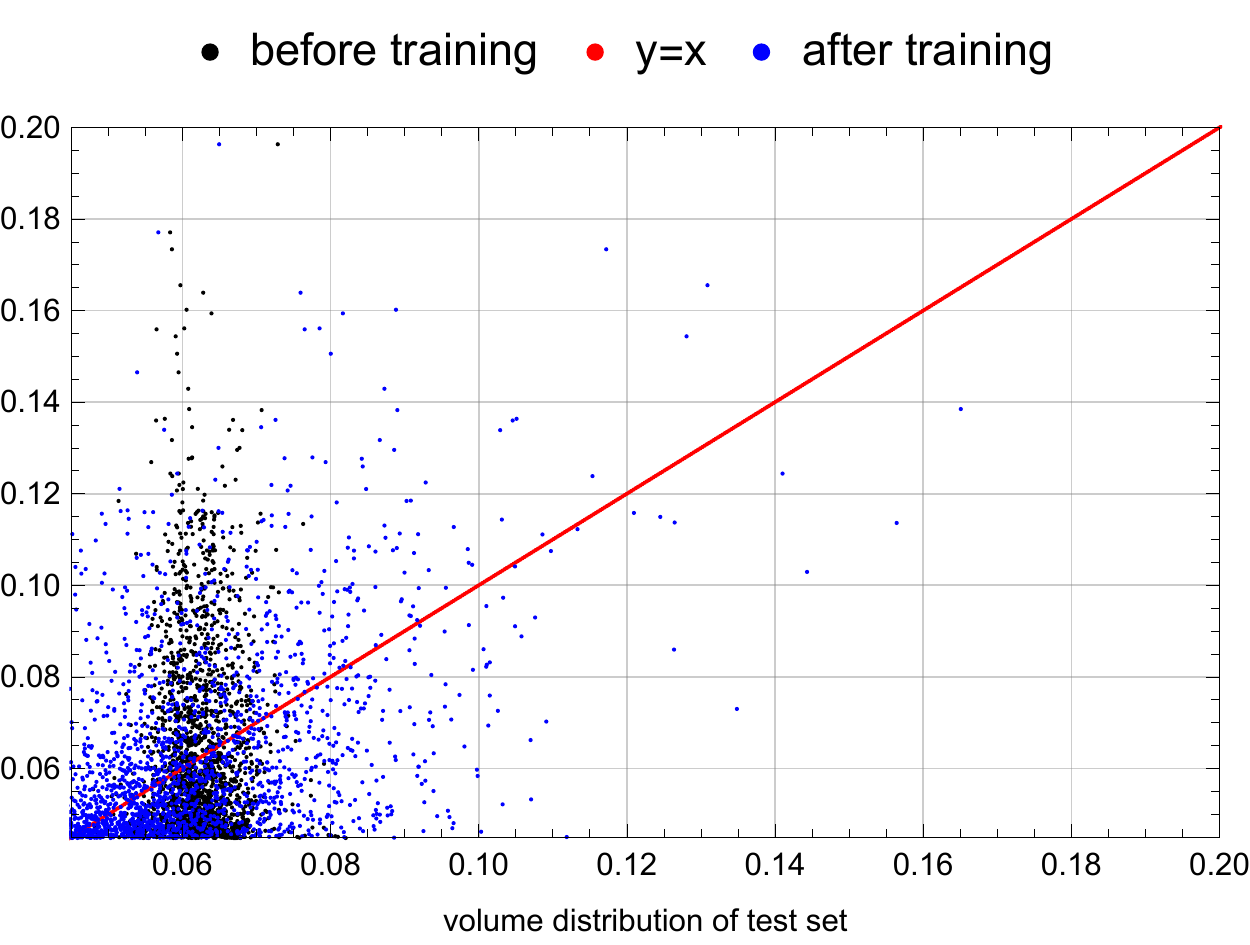}\\[20pt]
  \includegraphics[scale=.35]{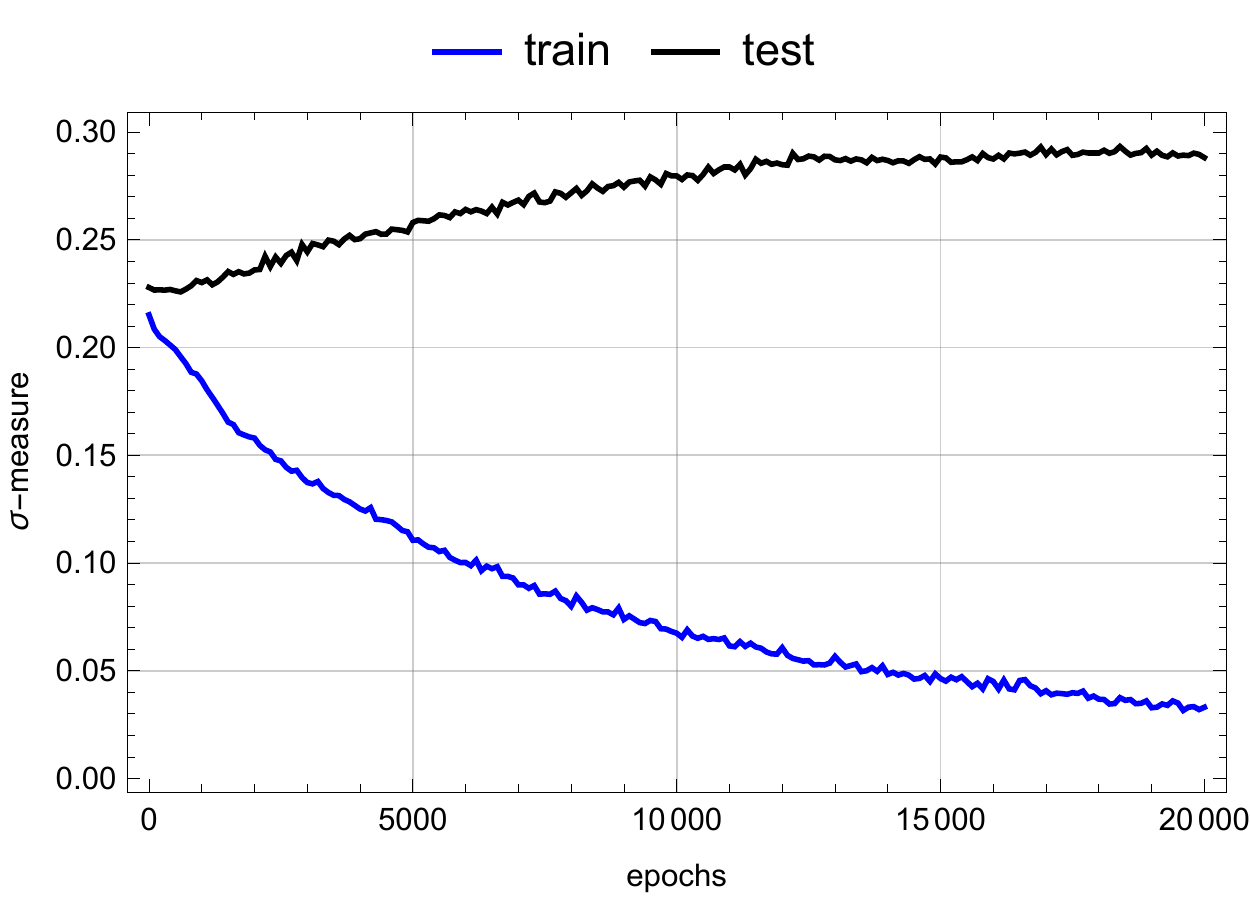}~~\includegraphics[scale=.35]{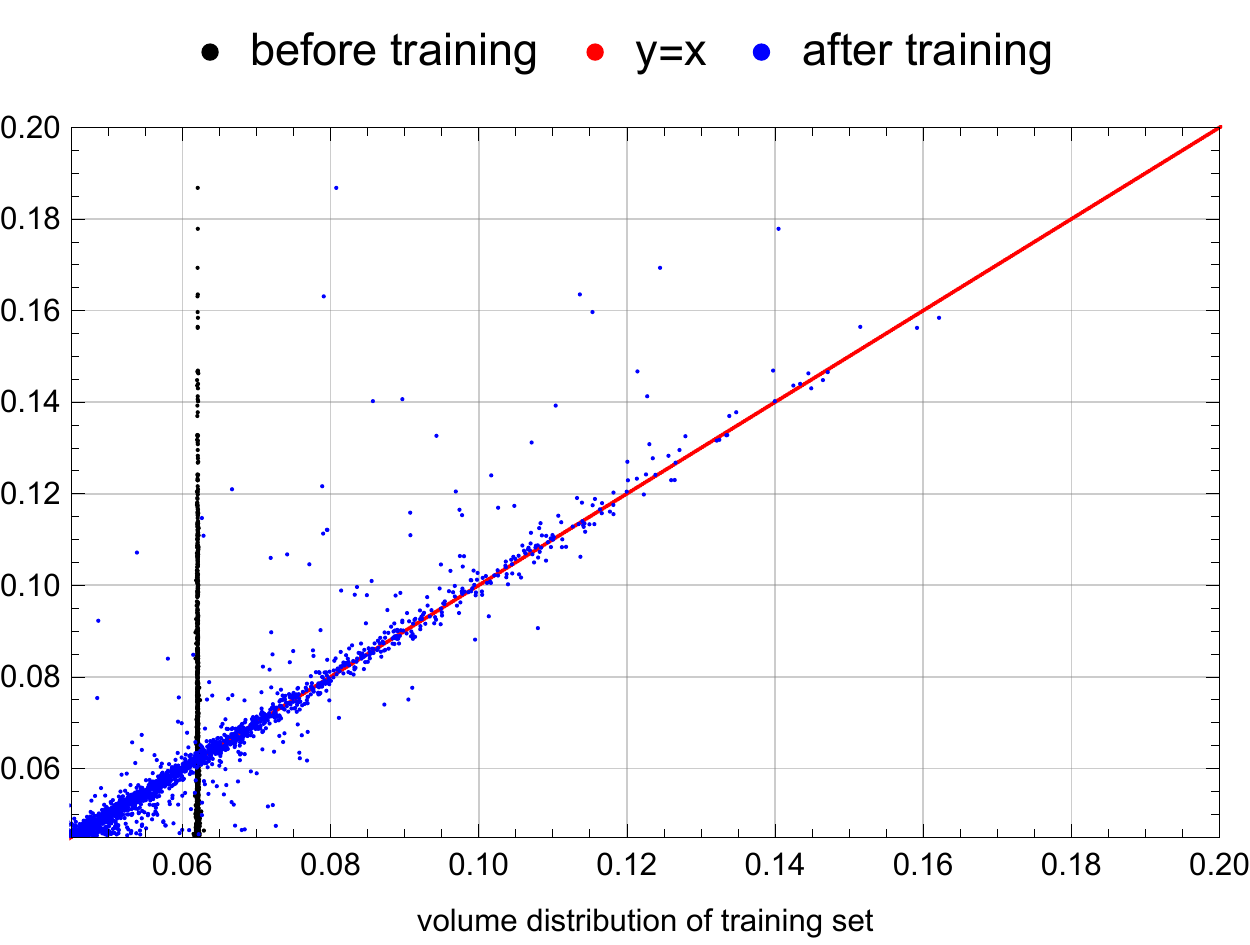}~~\includegraphics[scale=.35]{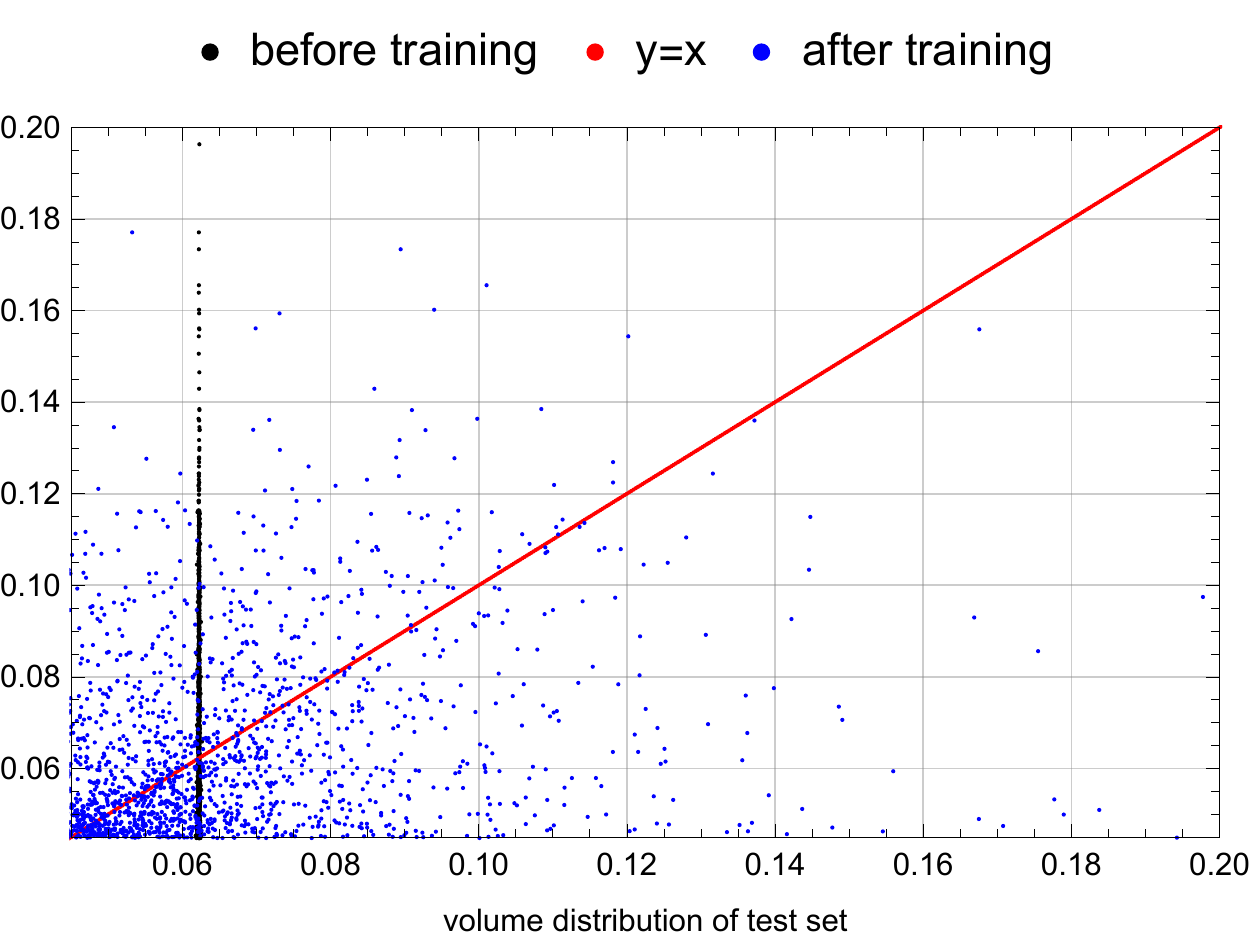}\\[20pt]
\caption{\small\textit{Evolution of $\sigma$-measures and normalized volume distributions 
for the quartic K3 surface~\eqref{eq:quartic}.
Figures in the first column show the evolution of $\sigma$ for both training and test sets, for three activation functions, namely, ReLU, $\tanh$, and logistic sigmoid, which correspond to the three rows, while figures in the second column show the volume distribution for the training points, before and after training.
Figures in the third column depict the volume distributions for the test set before and after training.}}
  \label{fig:sigmasK3}
  \vspace{-10pt}
\end{figure}

For the K3 surface, we consider the hypersurface equation~\eqref{eq:quartic} in $\mathbb{P}^3$. We have in total $12$ $(m,l)$ patches for the K3 hypersurface. In this case, we have trained using the $\sigma$-measure only.
We have considered different activation functions, namely ReLU, $\tanh$, and logistic sigmoid. Both the $\tanh$ and the logistic sigmoid activation functions are smooth and for these we expect a smooth behavior on the Ricci tensor derived from the corresponding neural network approximation functions. 
For all the activation functions considered, we have divided the set of points into a training and a test set to evaluate the ability of the network to generalize to unseen data.
The results are summarized in Figure~\ref{fig:sigmasK3}. There, we show the normalized volume distribution $d{\rm Vol}_J/{\rm Vol}_J$ obtained from the neural network. This is compared to the expectations for $d{\rm Vol}_\Omega/{\rm Vol}_\Omega$. Hence, the closer the distribution of points is to the line $x=y$, the smaller the $\sigma$-measure obtained. 
We also include the evolution of the sigma measure on training and test sets as the training evolves.
In general, we observe that as the training evolves, we get a very good agreement with the expectation for the normalized volume differentials in all three cases.
However, only for the ReLU and the $\tanh$ activation functions do we obtain a drop in $\sigma$ for the test set as training evolves.
However, none of the choices under consideration is able to generalize in such a way that $\sigma$ after training is of the same order for both sets. We address this matter in the upcoming section, noticing that for larger training and test sets the networks are indeed capable of generalization.
In the case of the K3, the best $\sigma$ values are obtained for the $\tanh$ activation function, for which we obtain a final $\sigma$ of $0.025$ for training and $0.21$ for the test set. 

\begin{figure}[h!]
\centering
\includegraphics[scale=.35]{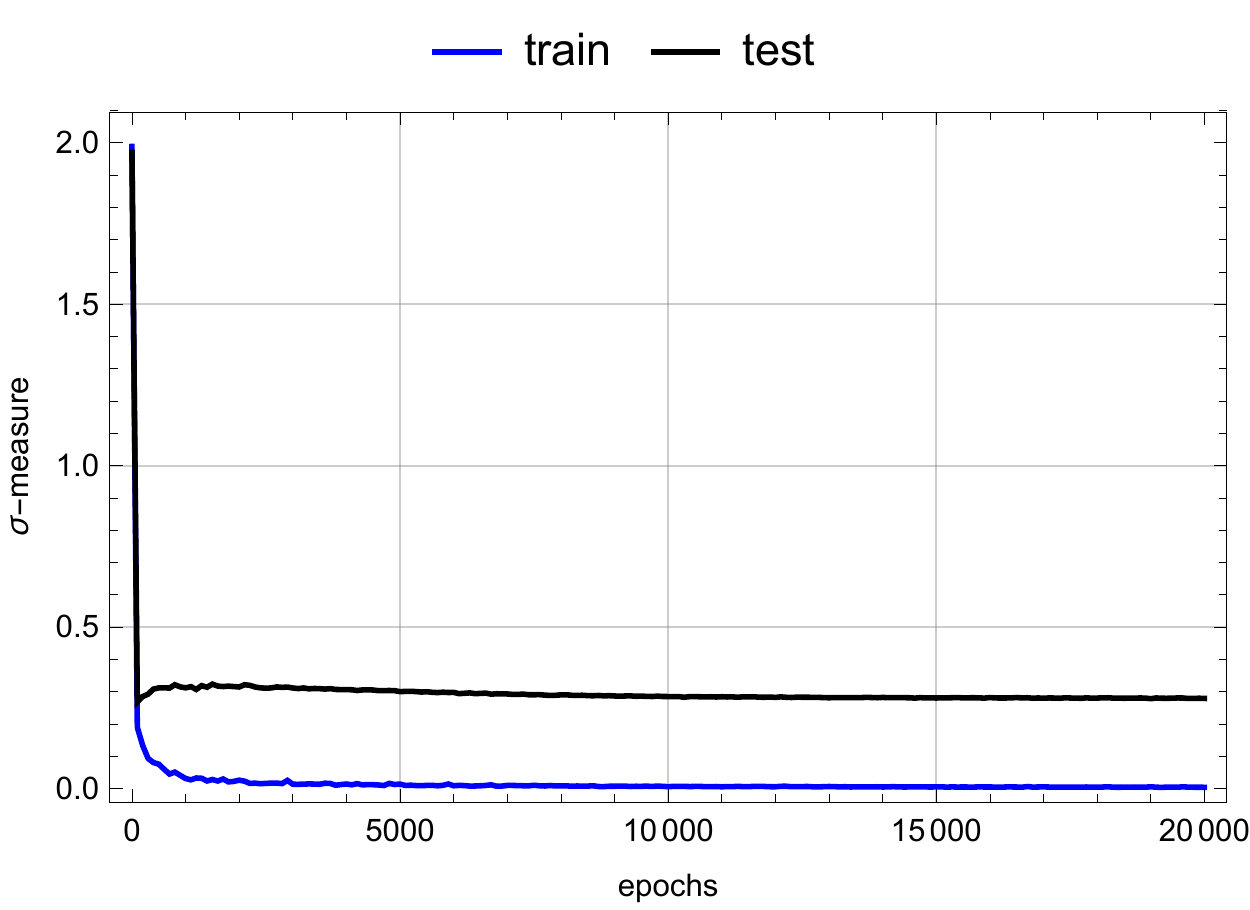}~~\includegraphics[scale=.35]{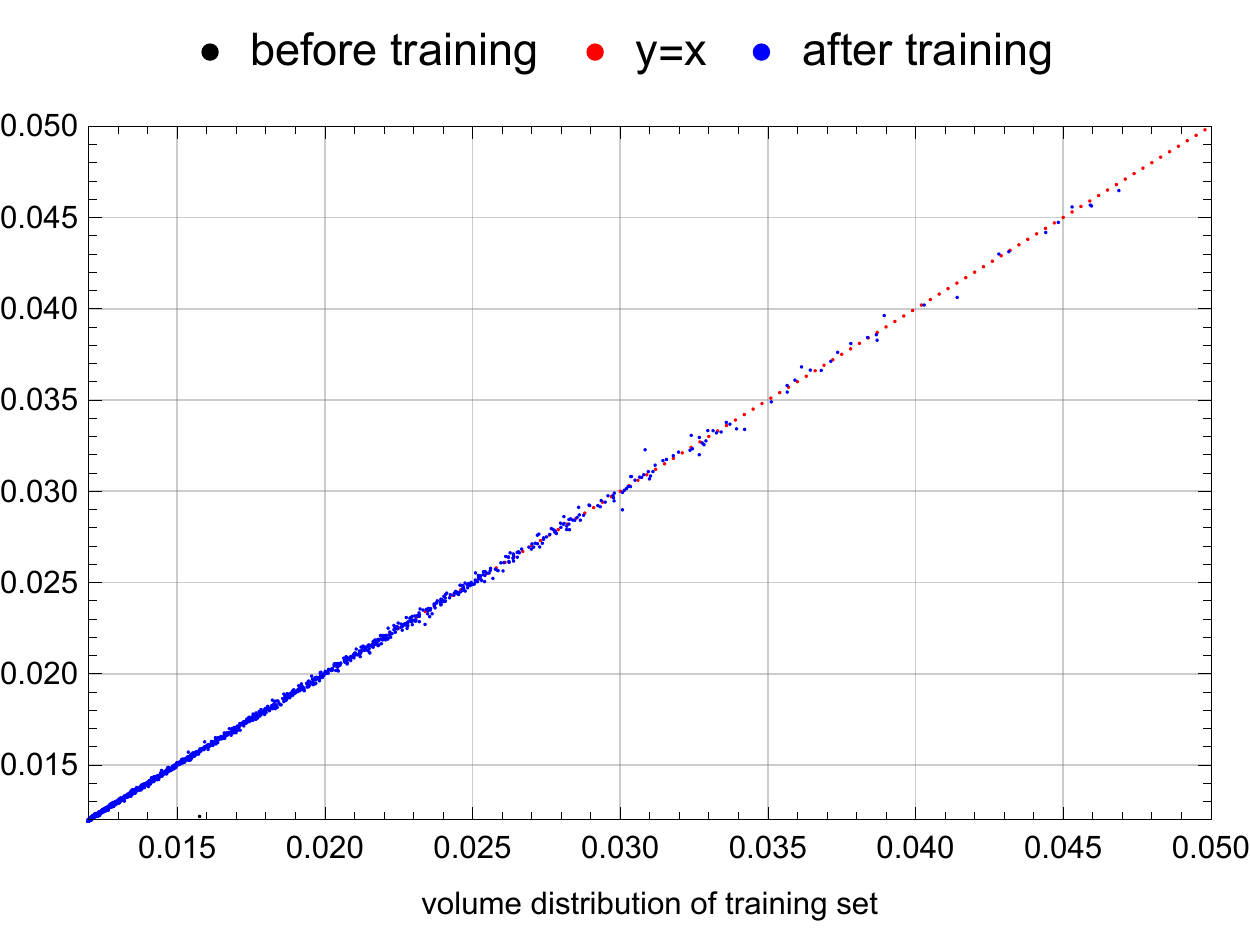}~~\includegraphics[scale=.35]{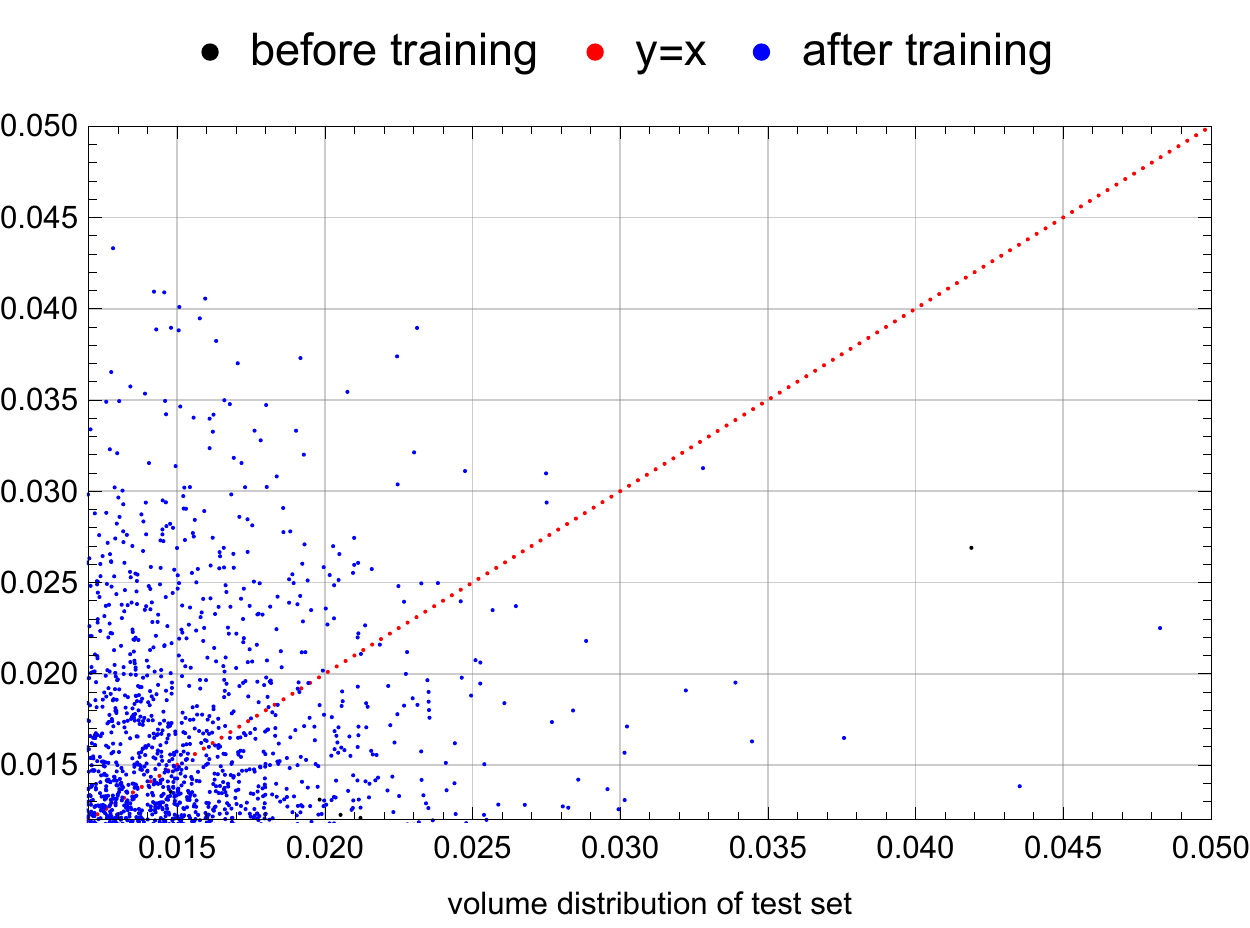}\\[20pt]
 \includegraphics[scale=.35]{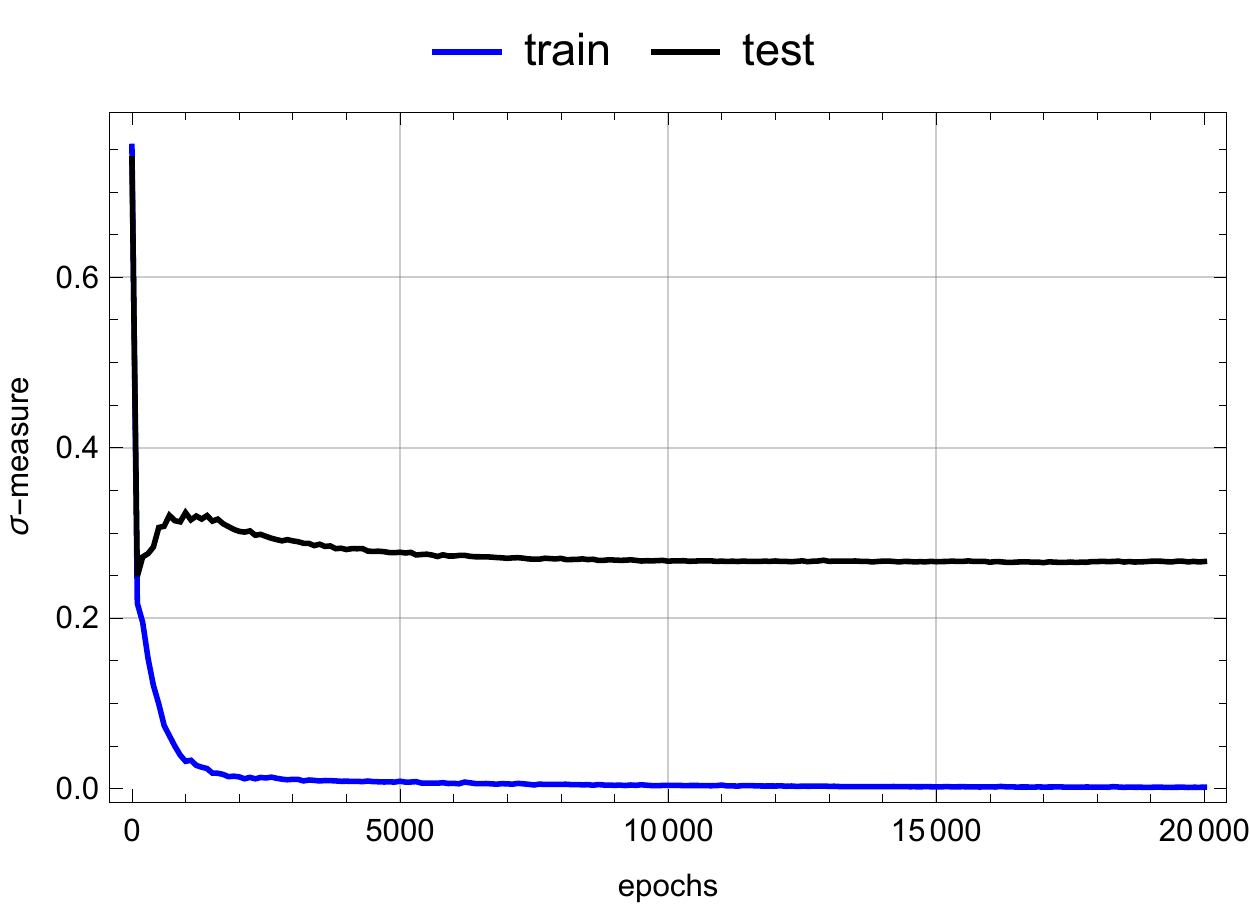}~~\includegraphics[scale=.35]{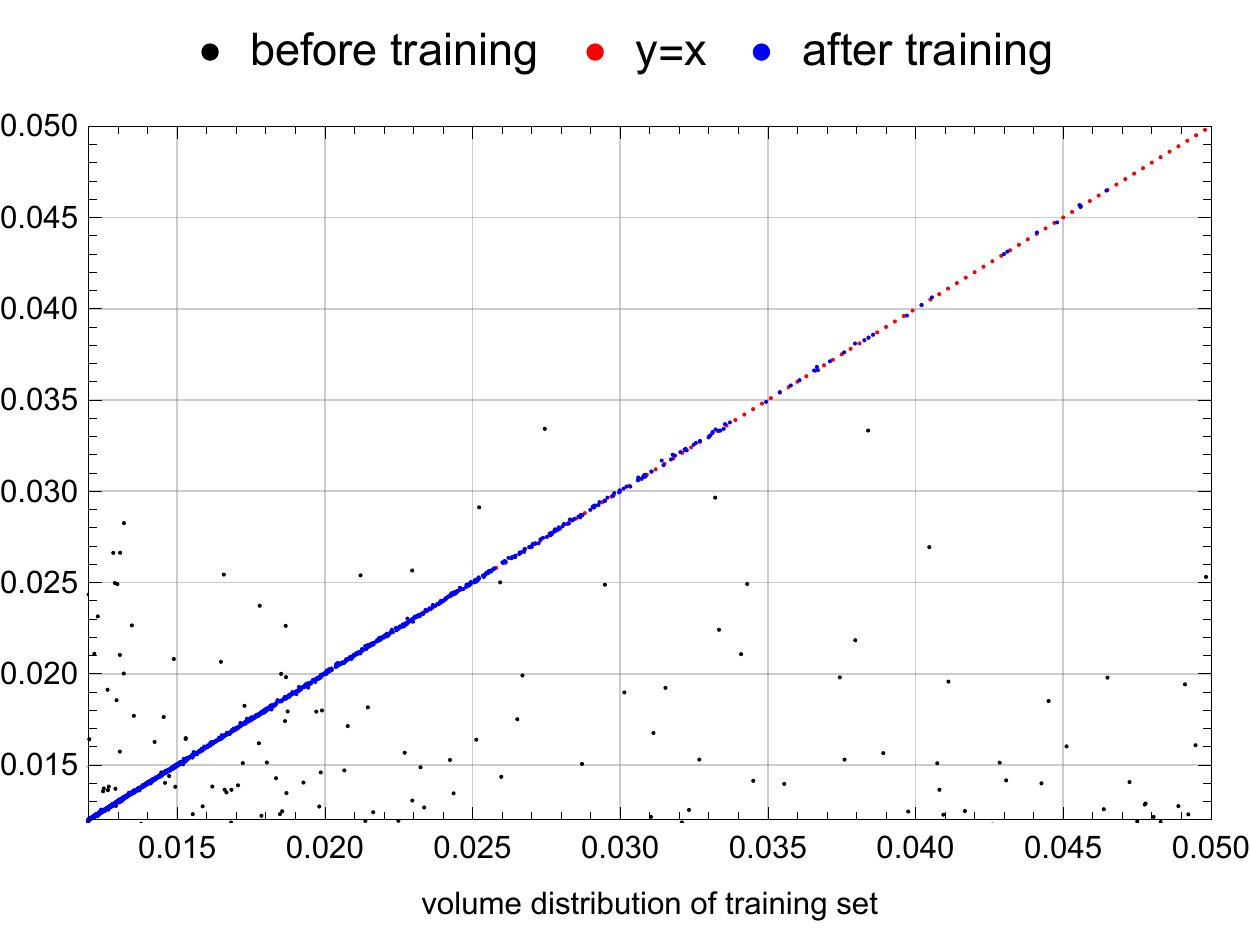}~~\includegraphics[scale=.35]{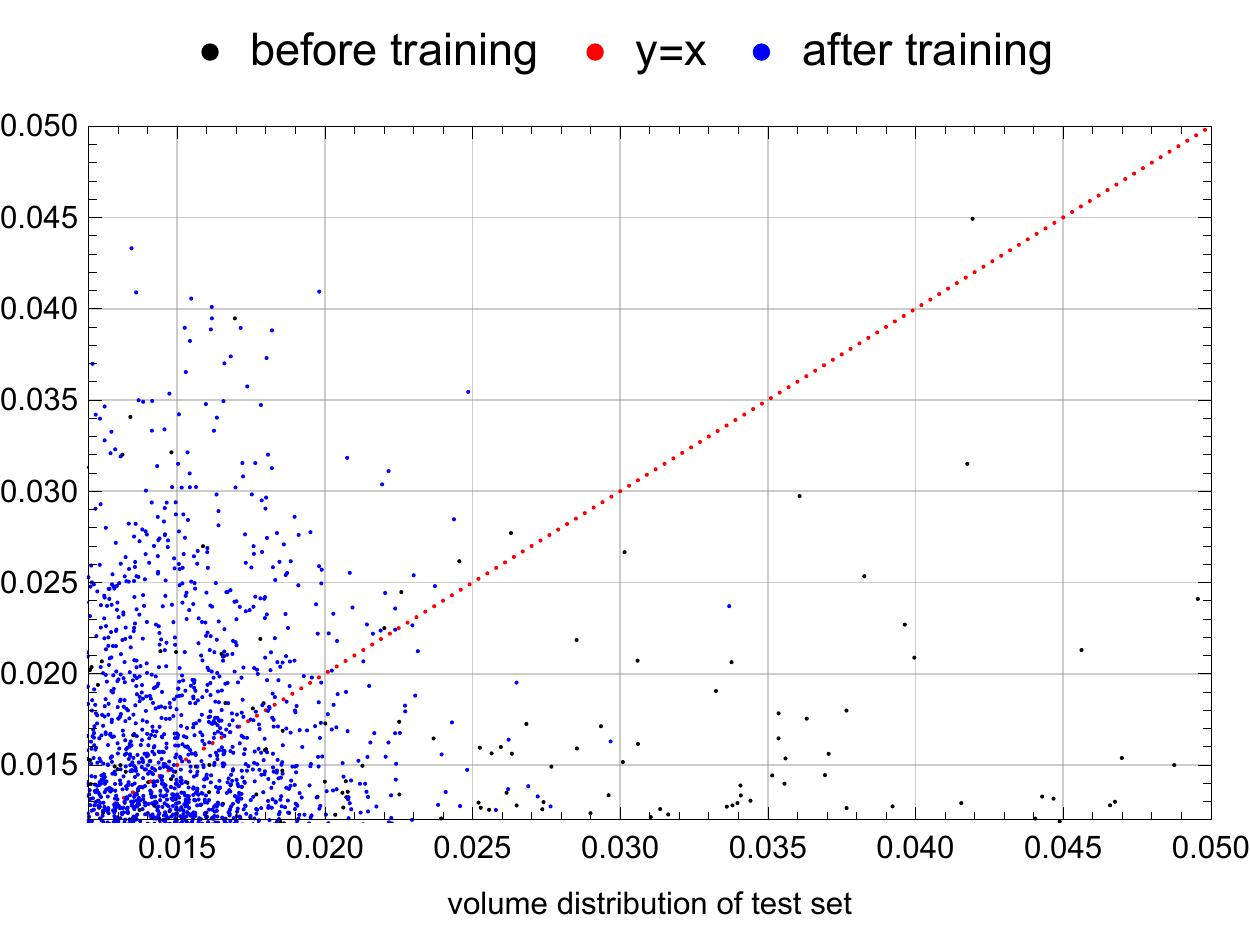}\\[20pt]
  \includegraphics[scale=.35]{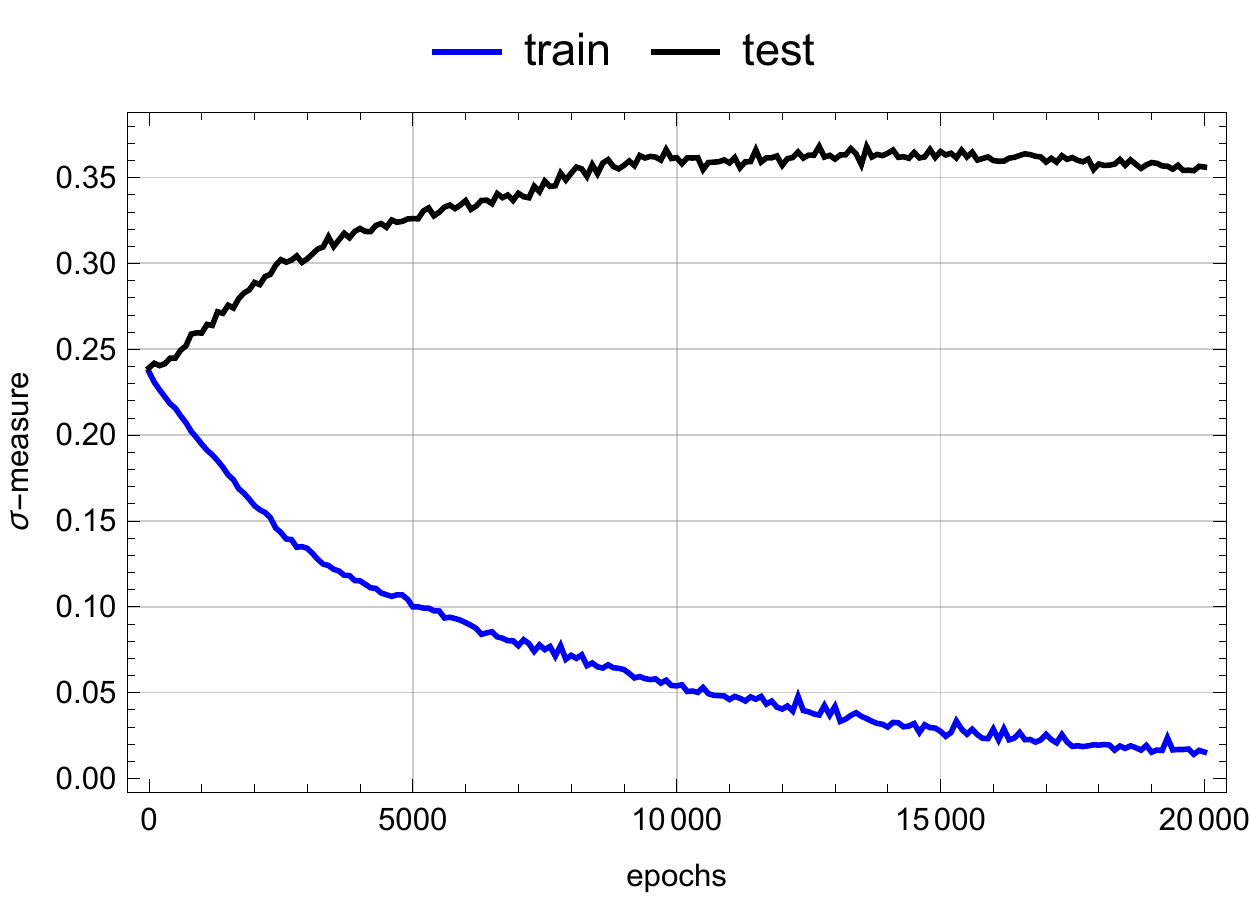}~~\includegraphics[scale=.35]{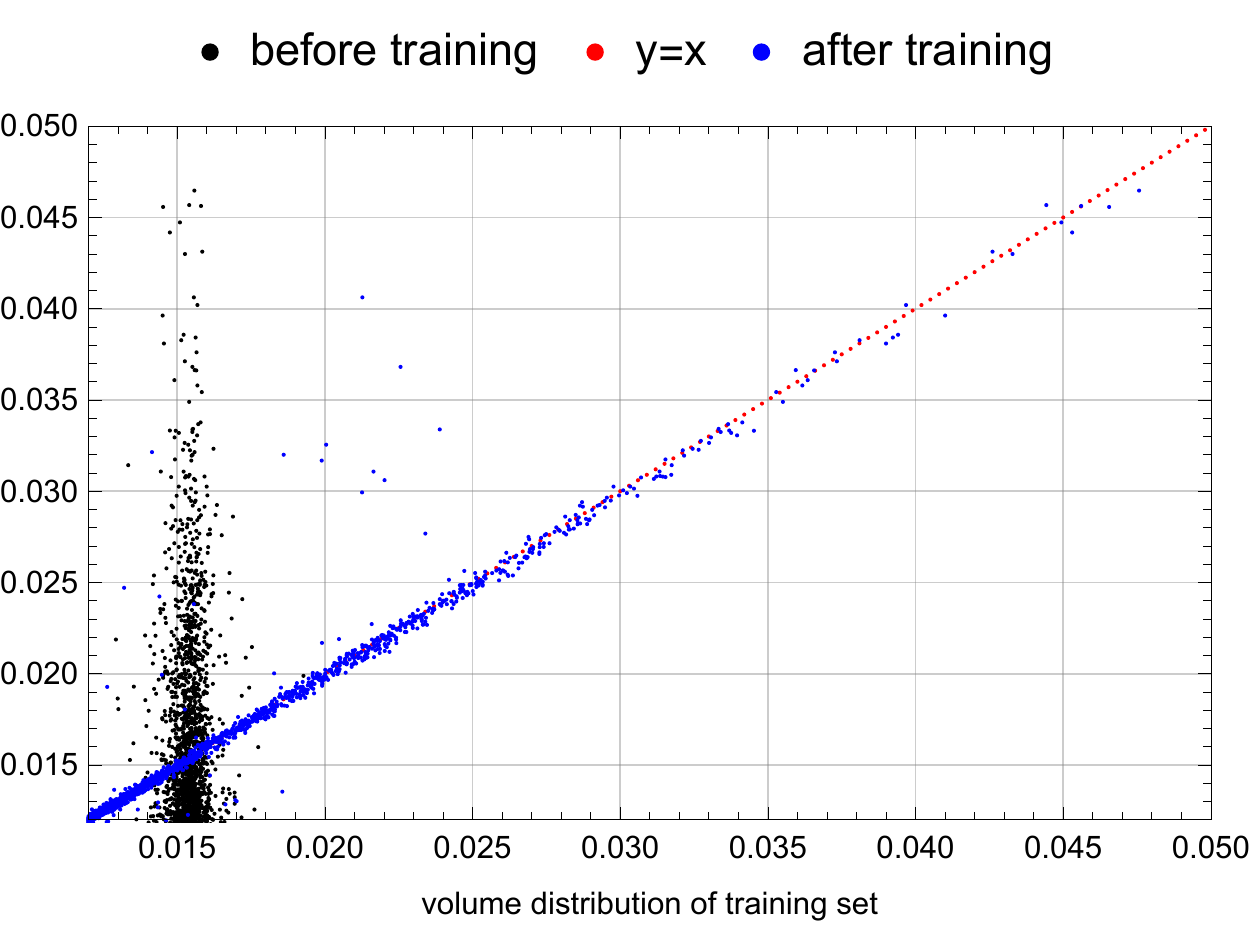}~~\includegraphics[scale=.35]{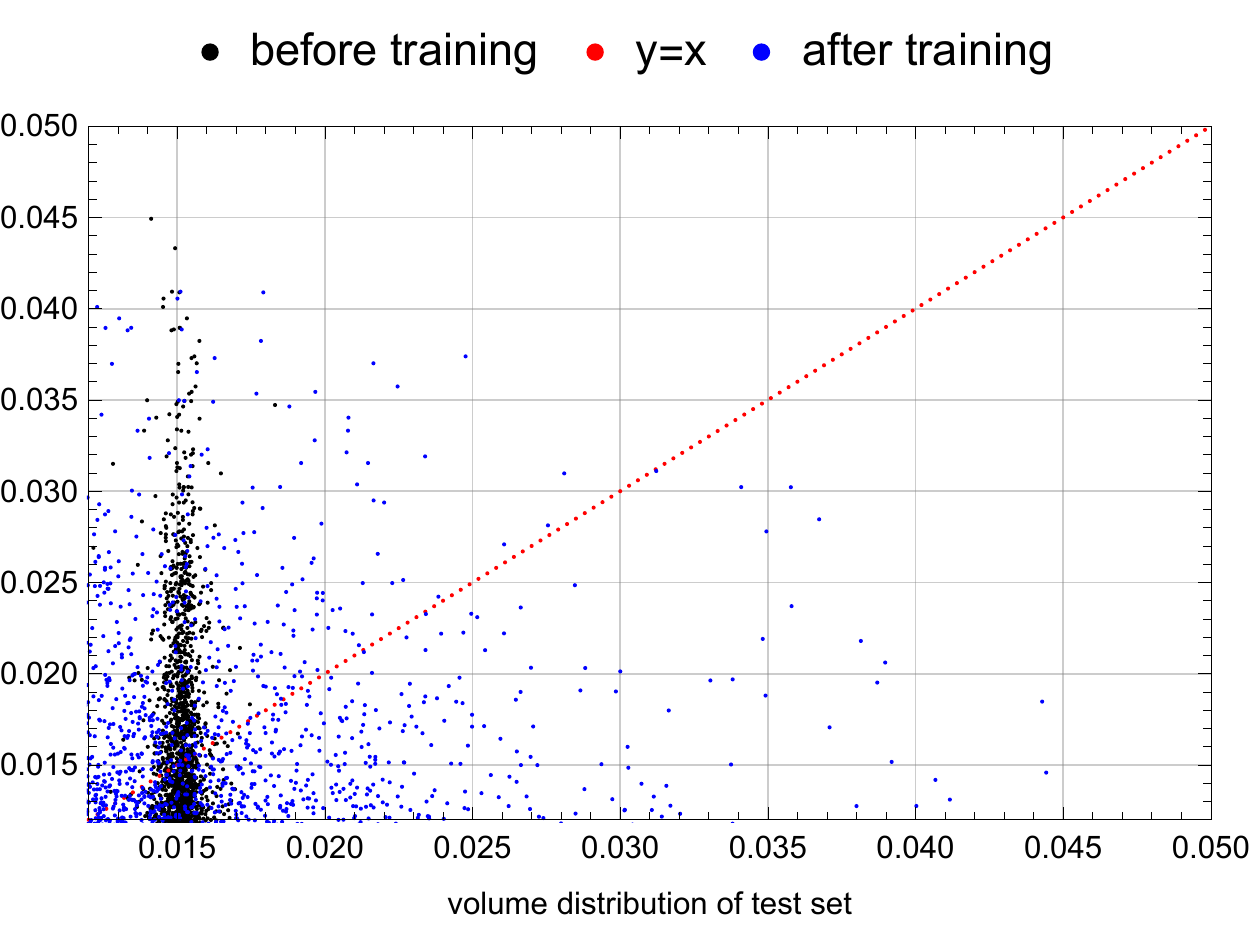}\\[20pt]
\caption{\small\textit{Evolution of $\sigma$-measures and normalized volume distributions 
for the Fermat quintic~\eqref{eq:quintic}.
Figures in the first column show the evolution of $\sigma$ for both training and test sets, for three activation functions, namely, ReLU, $\tanh$, and logistic sigmoid, which correspond to the three rows, while figures in the second column show the volume distribution for the training points, before and after training.
Figures in the third column show the same for the test points.}}
  \label{fig:sigmasQ}
  \vspace{-10pt}
\end{figure}
\newpage
\subsection{The Fermat quintic}\label{sec:quintic}
\begin{figure}[h!]
\centering
\includegraphics[scale=.413]{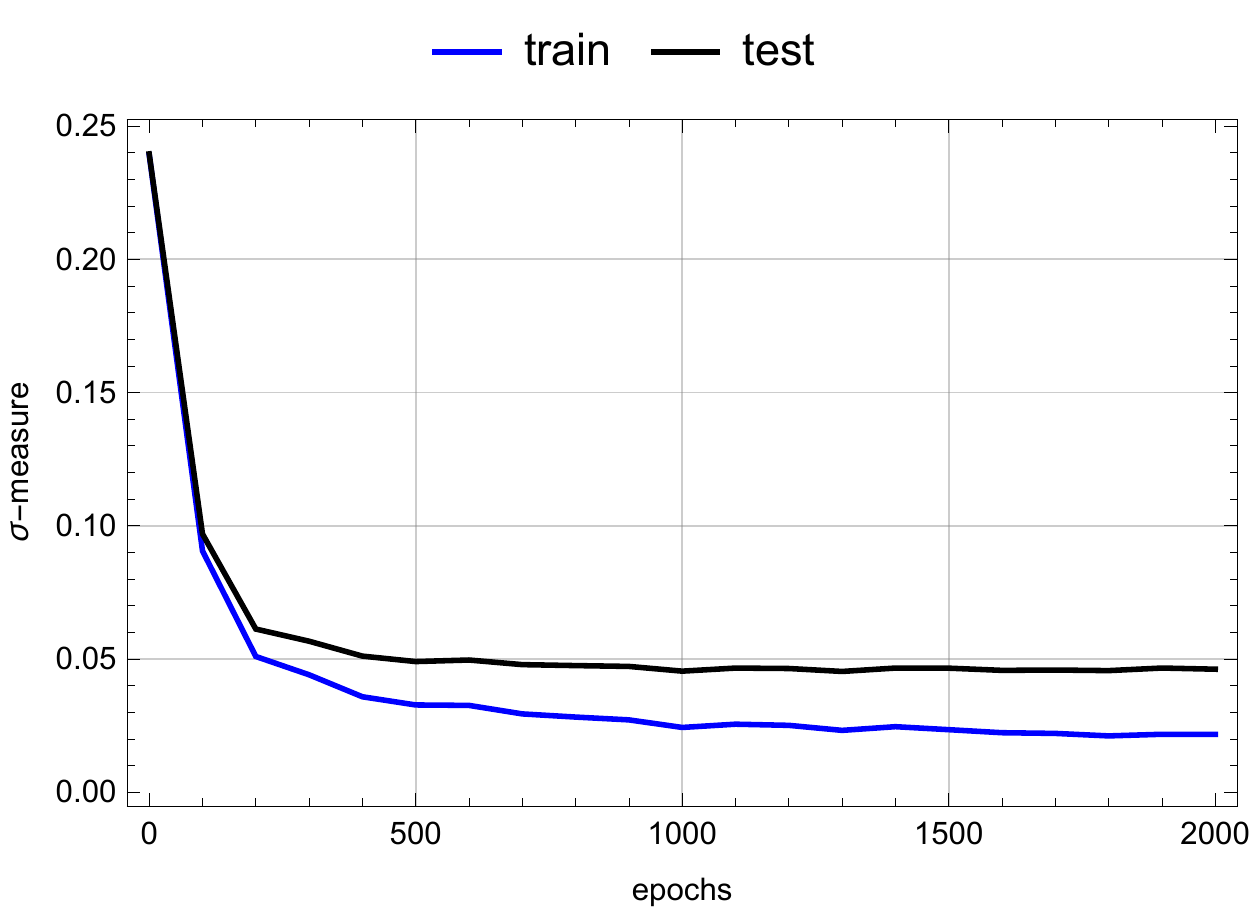}~~\includegraphics[scale=.4]{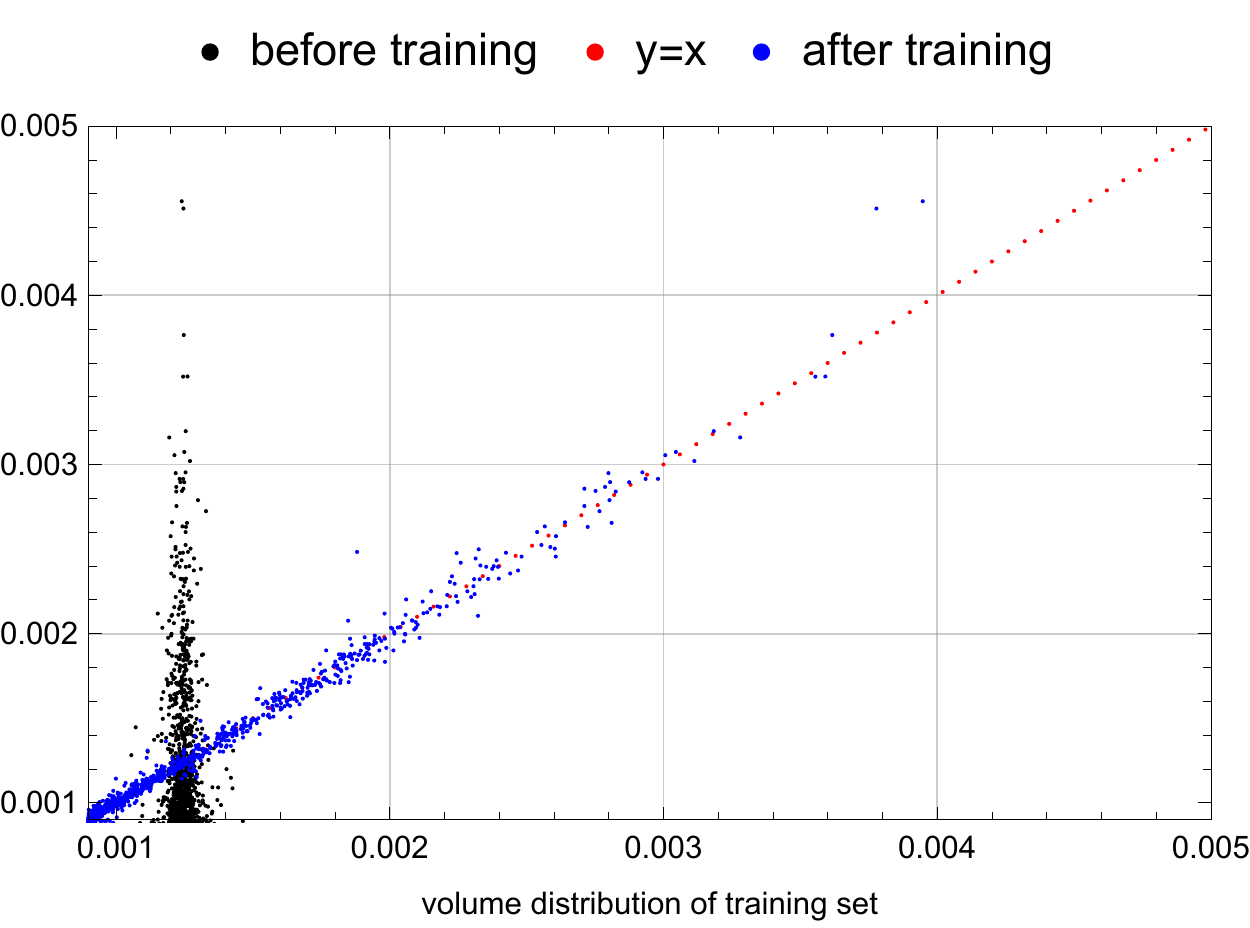}~~\includegraphics[scale=.4]{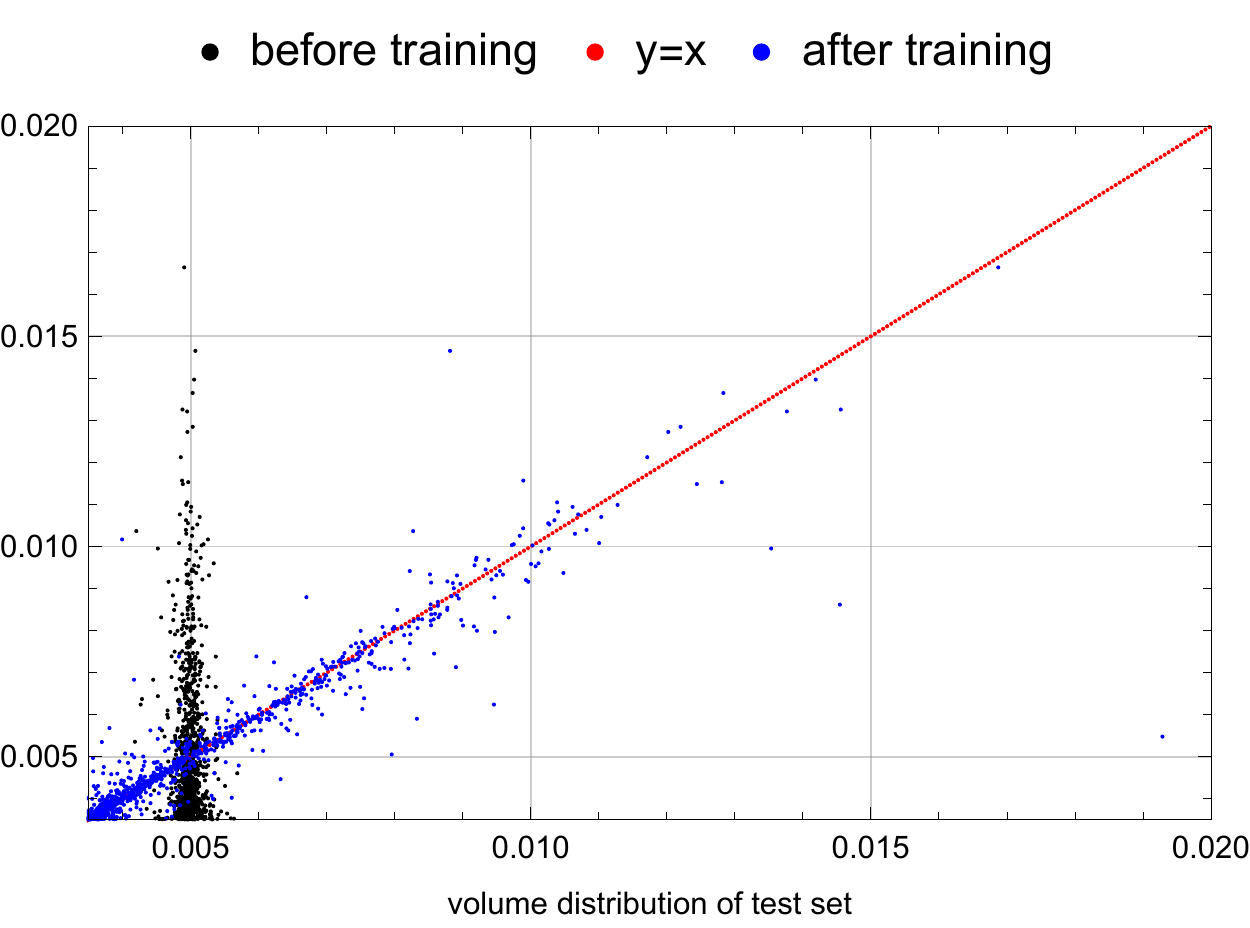}\\[20pt]
\includegraphics[scale=.413]{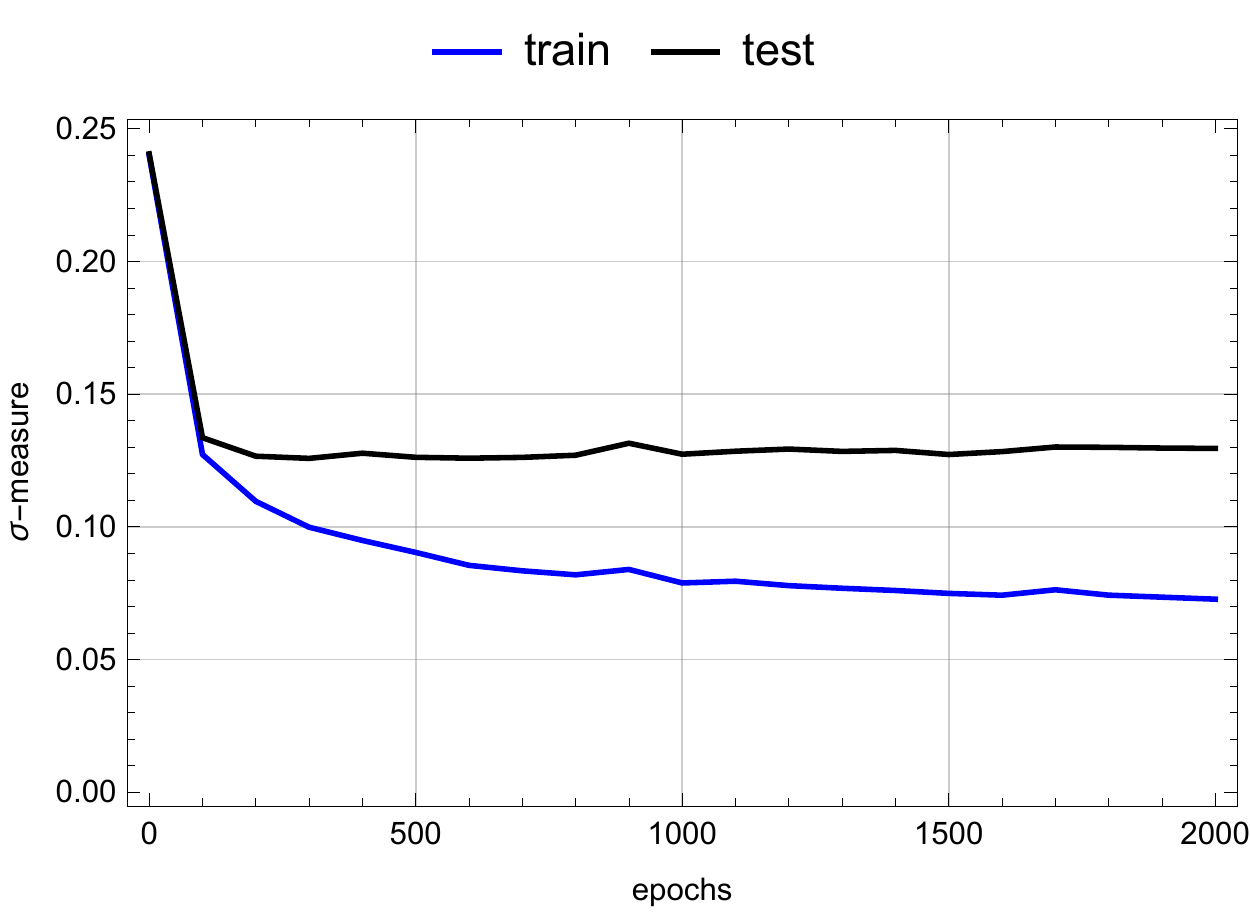}~~\includegraphics[scale=.4]{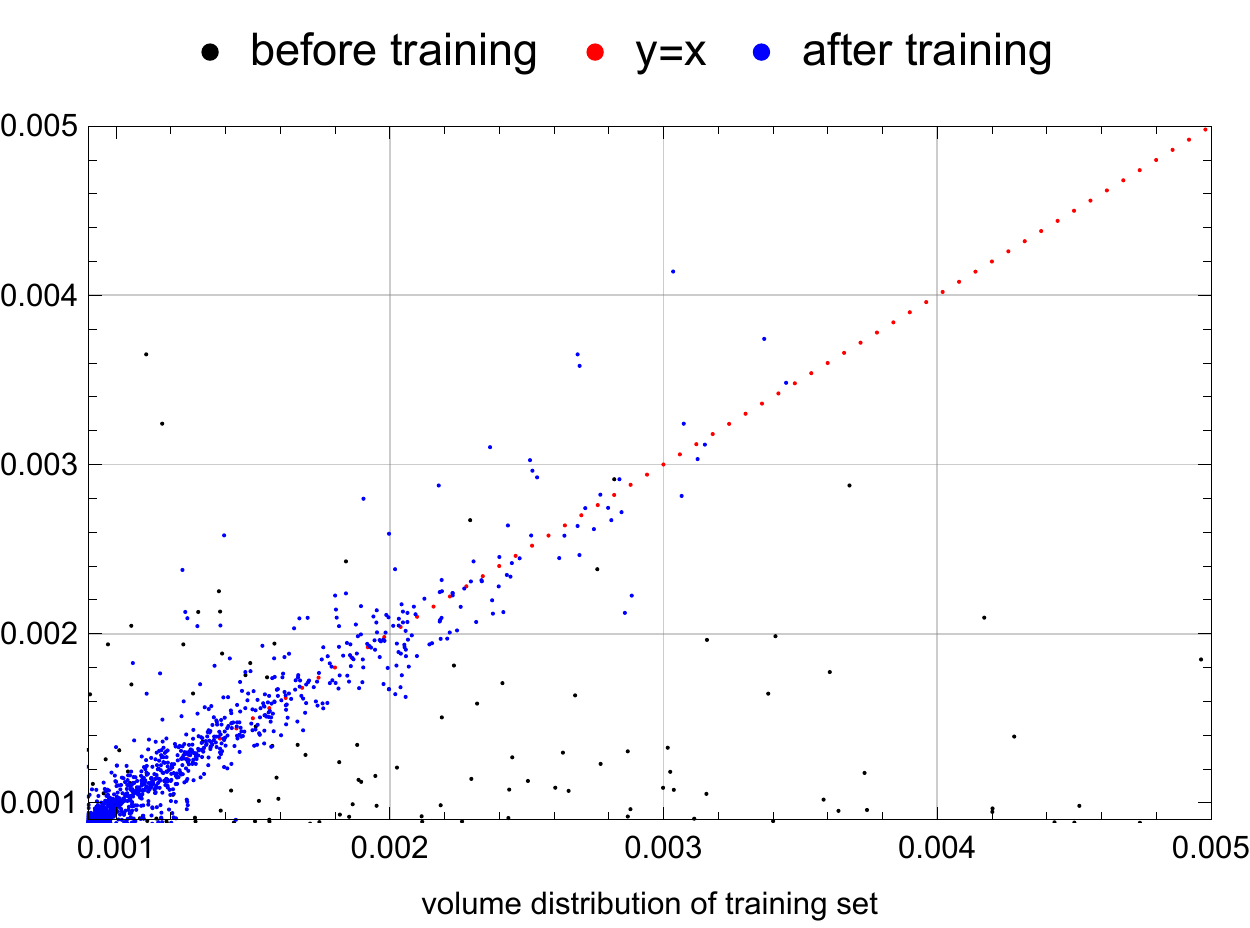}~~\includegraphics[scale=.4]{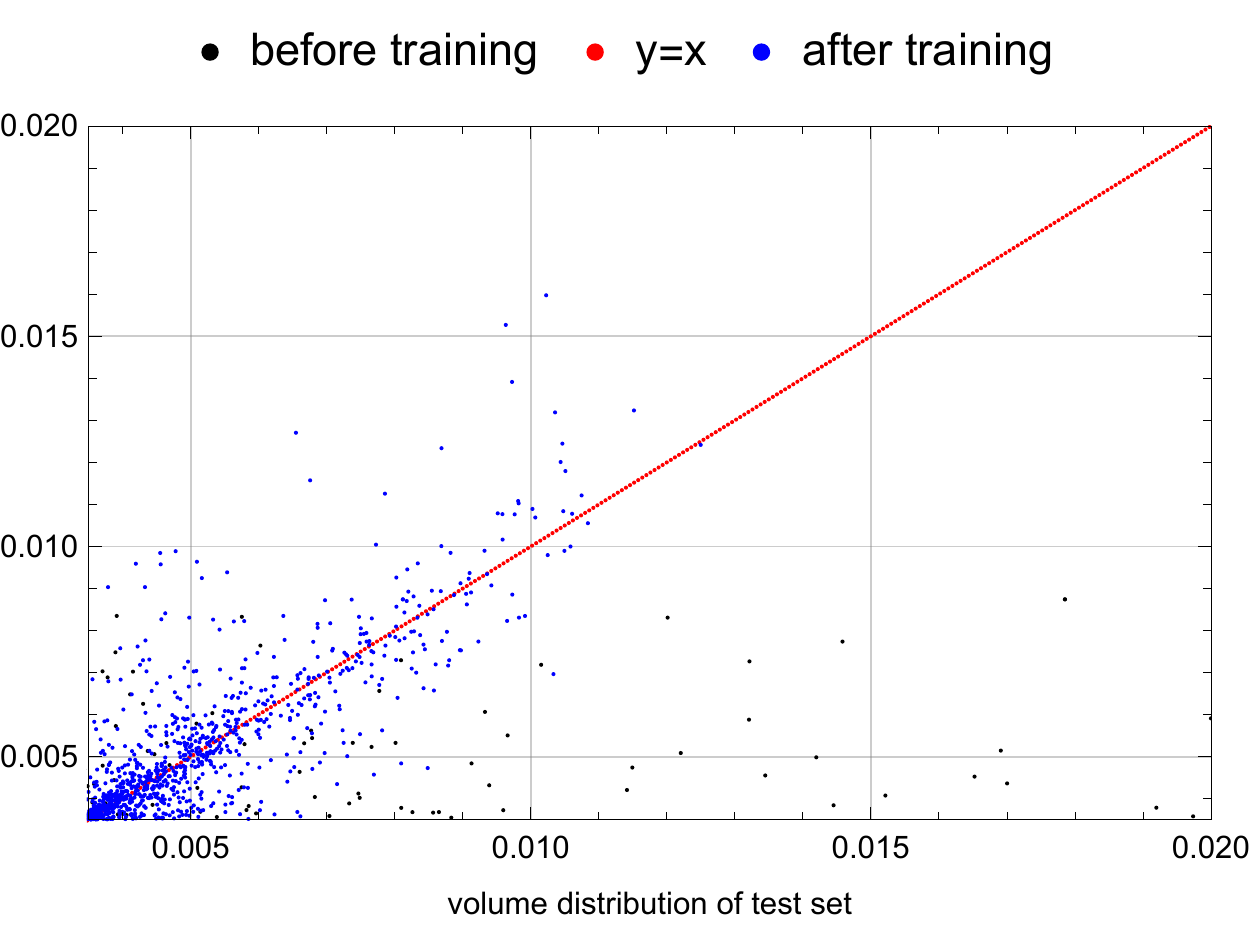}\\[20pt]
\caption{\small\textit{Evolution of $\sigma$-measures and normalized volume distributions 
for the Fermat quintic  hypersurface~\eqref{eq:quintic}. In this case we have used $500,000$ data points for training and $125,000$ for testing. In the first row we present the results for a logistic sigmoid and in the second row the results for a $\tanh$ activation function. 
The plots to the left shows the evolution of the $\sigma$-measure as training advances. The central and right figures show the volume distribution for the training points and test points respectively, before and after training.}}
  \label{fig:manypts}
  \vspace{-10pt}
\end{figure}

The Fermat quintic is described by~\eqref{eq:quintic}. In this case we have $20$ different $(m,l)$ patches. 
For test and training sets of $2000$ points, we train employing the $\sigma$-measure only. 
We observe that $\sigma$ only drops in generalization experiments for the ReLU and $\tanh$ activation functions. The results are presented in Figure~\ref{fig:sigmasQ}. 
For the ReLU activation function, we obtain a $\sigma$ of $0.004$ and $0.27$ for training and test sets respectively.
Similarly, for the $\tanh$ activation function, we find $\sigma$ values of $0.0016$ and $0.26$. As in the K3 case, we observe that the networks are good at approximating the outputs for the training data, but we do not see a similar performance on the test set. We can show that this is an effect due to the small number of points that we are employing for training. In Figure~\ref{fig:manypts}, we present the results for a training dataset of $500,000$ points and a test set of $125,000$ points. Here, we have considered the $\sigma$-measure only and employed the logistic sigmoid and $\tanh$ activation functions. It is worth remarking the significant improvement of the neural network approximation function predictions on the test data points, as the normalized volume distributions follow the line $y=x$ more clearly that in the previous cases. This can also be seen in the evolution of the $\sigma$-measure on both datasets as training evolves. After the training process, we obtain for the logistic sigmoid activation function a value of $\sigma=0.022$ on the training set and $\sigma=0.046$ on the test set. For the $\tanh$ activation function we find $\sigma=0.073$ on the training set and $\sigma=0.13$ on the test dataset.

\begin{figure}[h!]
\centering
\includegraphics[scale=.19]{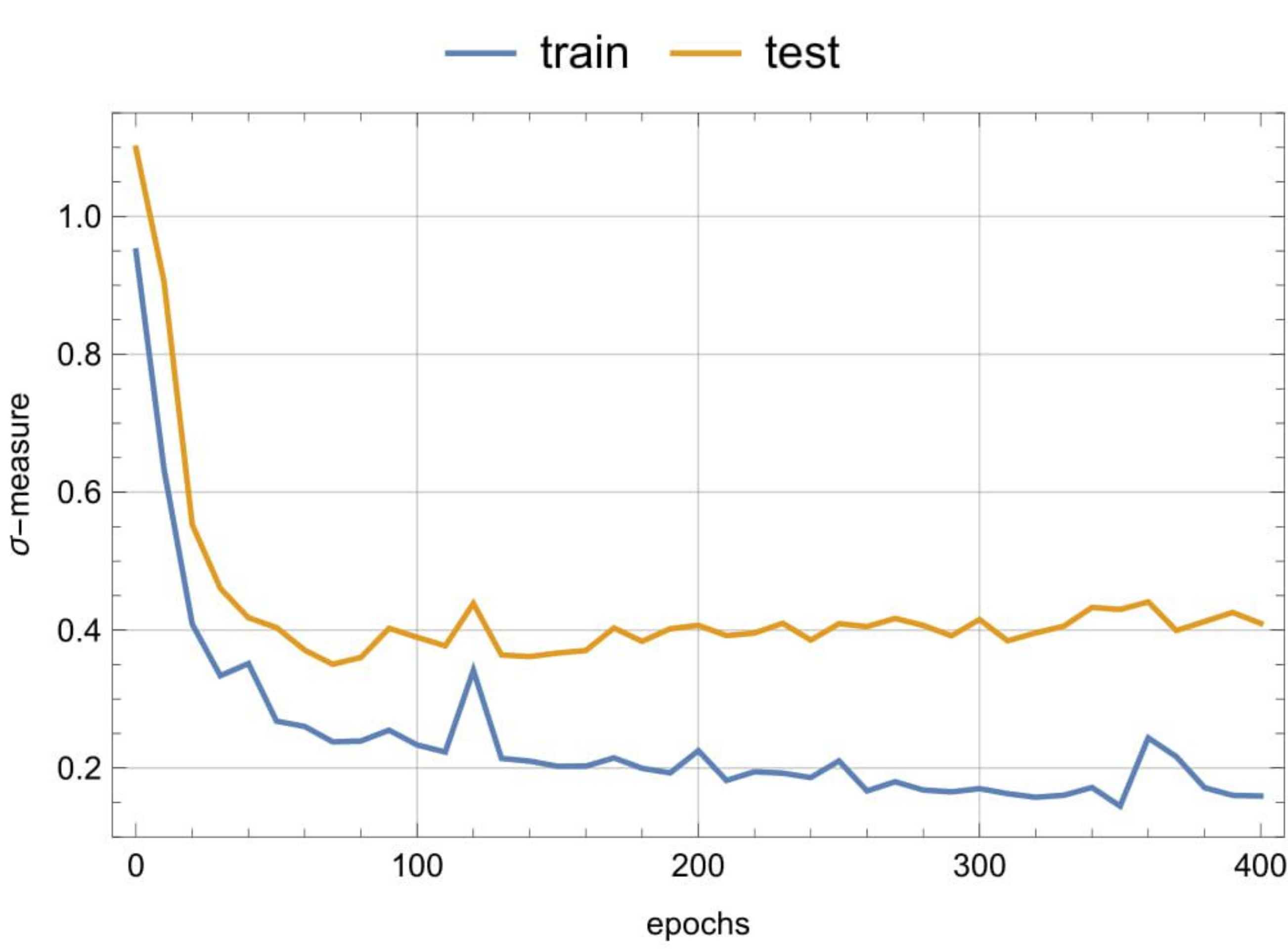}~~\includegraphics[scale=.19]{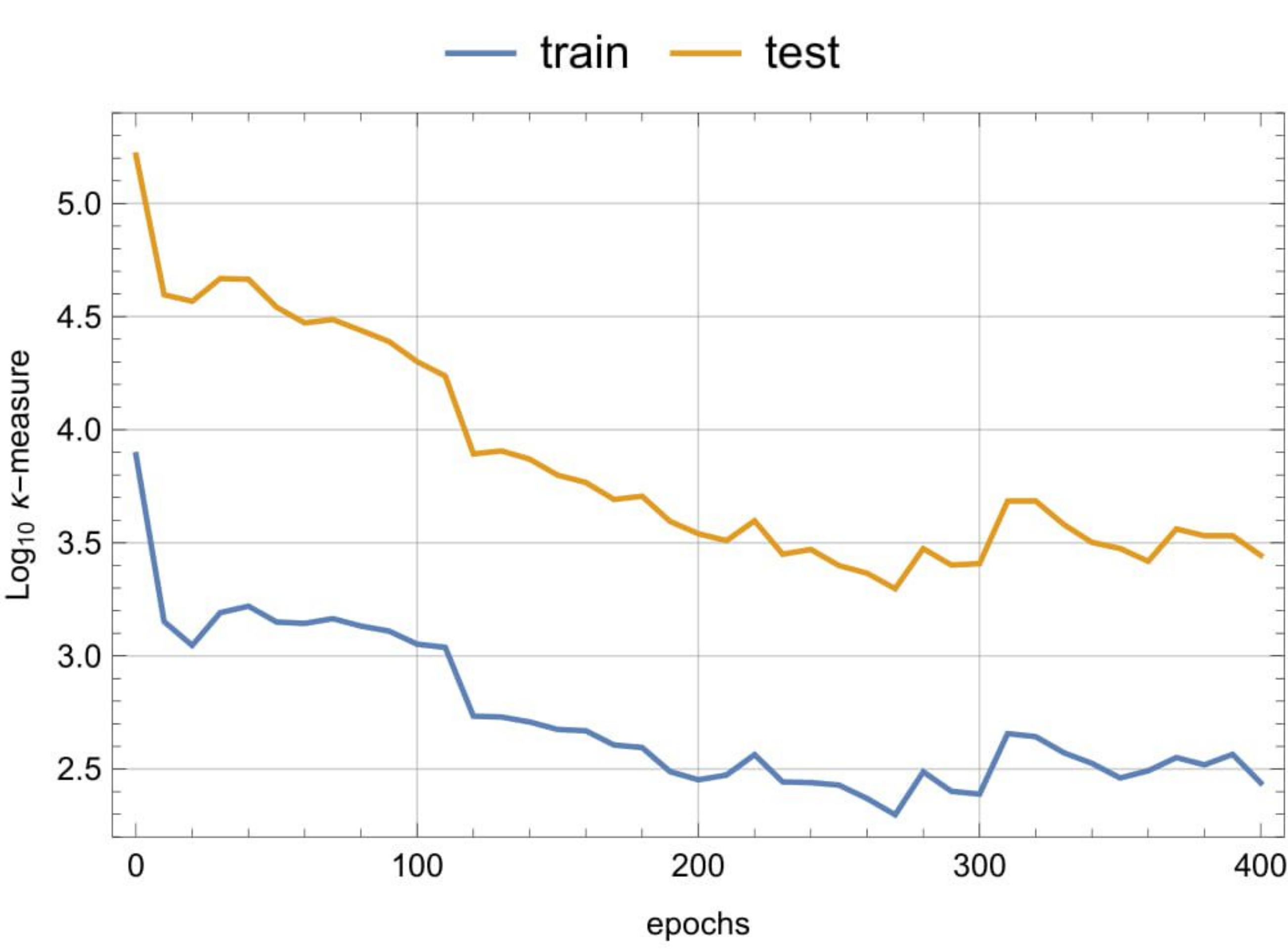}~~\includegraphics[scale=.19]{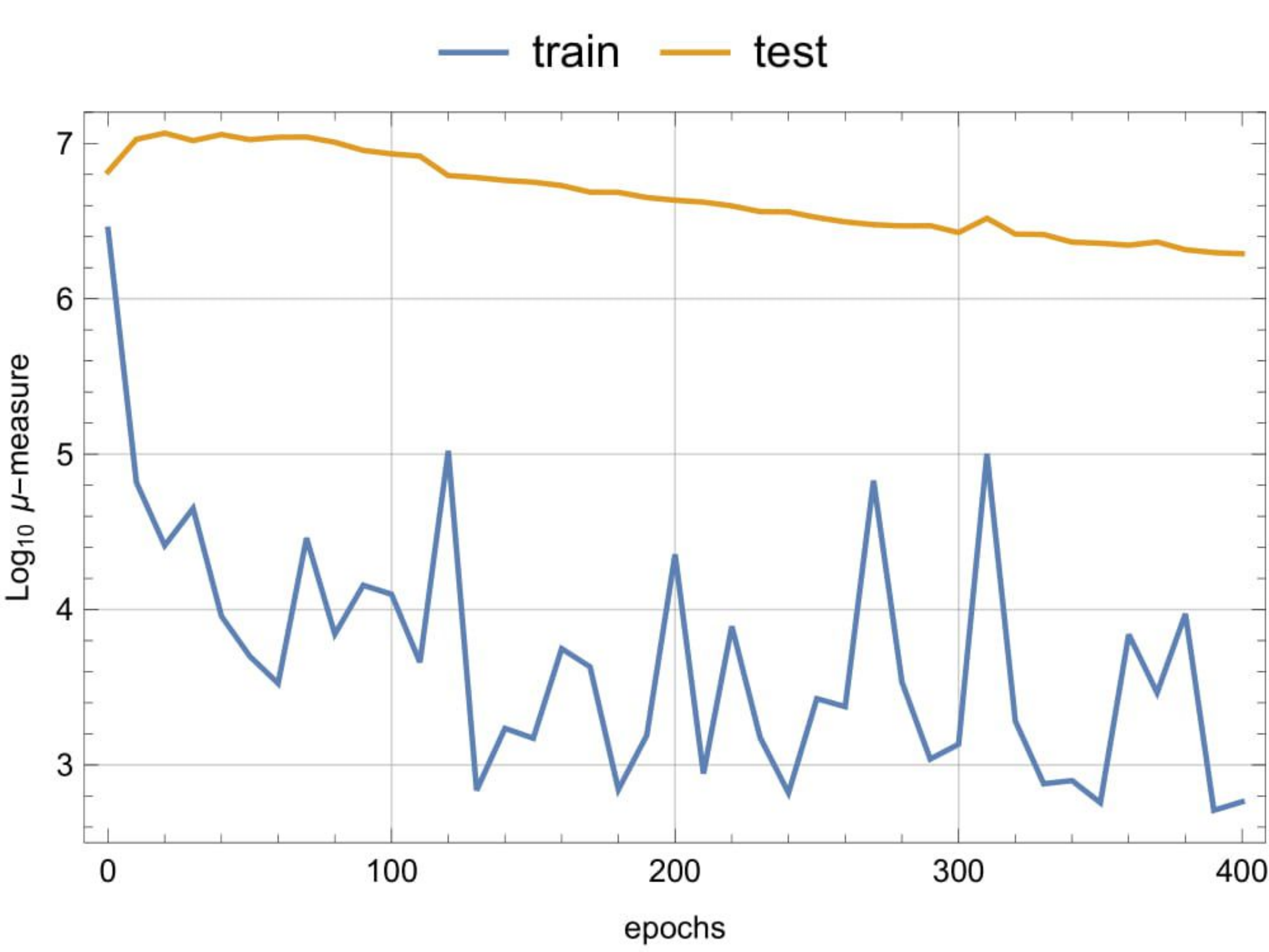}\\[10pt]
\caption{\small\textit{Evolution of the $\sigma$, $\kappa$ and $\mu$-measures as defined in Section 6.2, for the Fermat quintic ~\eqref{eq:quintic}. These were computed on points used for training as well as on a mutually exclusive test set, illustrating that the metric we obtain is indeed an approximation of the K\"ahler, Ricci flat metric on the Fermat quintic.}}
  \label{fig:alllossesQ}
  \vspace{10pt}
\end{figure}

Additionally, we explored the behavior of the $\mu$- and $\kappa$-measures for the quintic. The training process with non-zero weights $\alpha_\sigma$, $\alpha_\kappa$ and $\alpha_\mu$ in the loss function~\eqref{eq:loss} is more involved as the matrix products in \texttt{PyTorch} have to be converted to real numbers. This reflects in longer computation times. For this reason, instead of training over $20,000$ epochs as in the training with $\sigma$-measure only, we have chosen to train only during $400$ epochs. The hyperparameters $\alpha_\sigma$, $\alpha_\kappa$, and $\alpha_\mu$ are chosen from normalization of the loss values obtained in the first training epoch. For our case, we have chosen $\alpha_\sigma=1$, $\alpha_\kappa=1.25\times 10^{-4}$, and $\alpha_\mu=3.16\times 10^{-7}$. In Figure~\ref{fig:alllossesQ}, we have included the results for the various loss functions using the same test and training points as in the previous experiments. We observe that as the network training evolves, we see a decrease of two orders of magnitude in the $\kappa$-measure and three orders of magnitude in the $\mu$-measure. Note however that the performance on the test set is significantly worse that on the training set. 

\newpage
\subsection{The Dwork family}\label{sec:dwork}
\begin{figure}[h!]
\centering
\includegraphics[scale=.33]{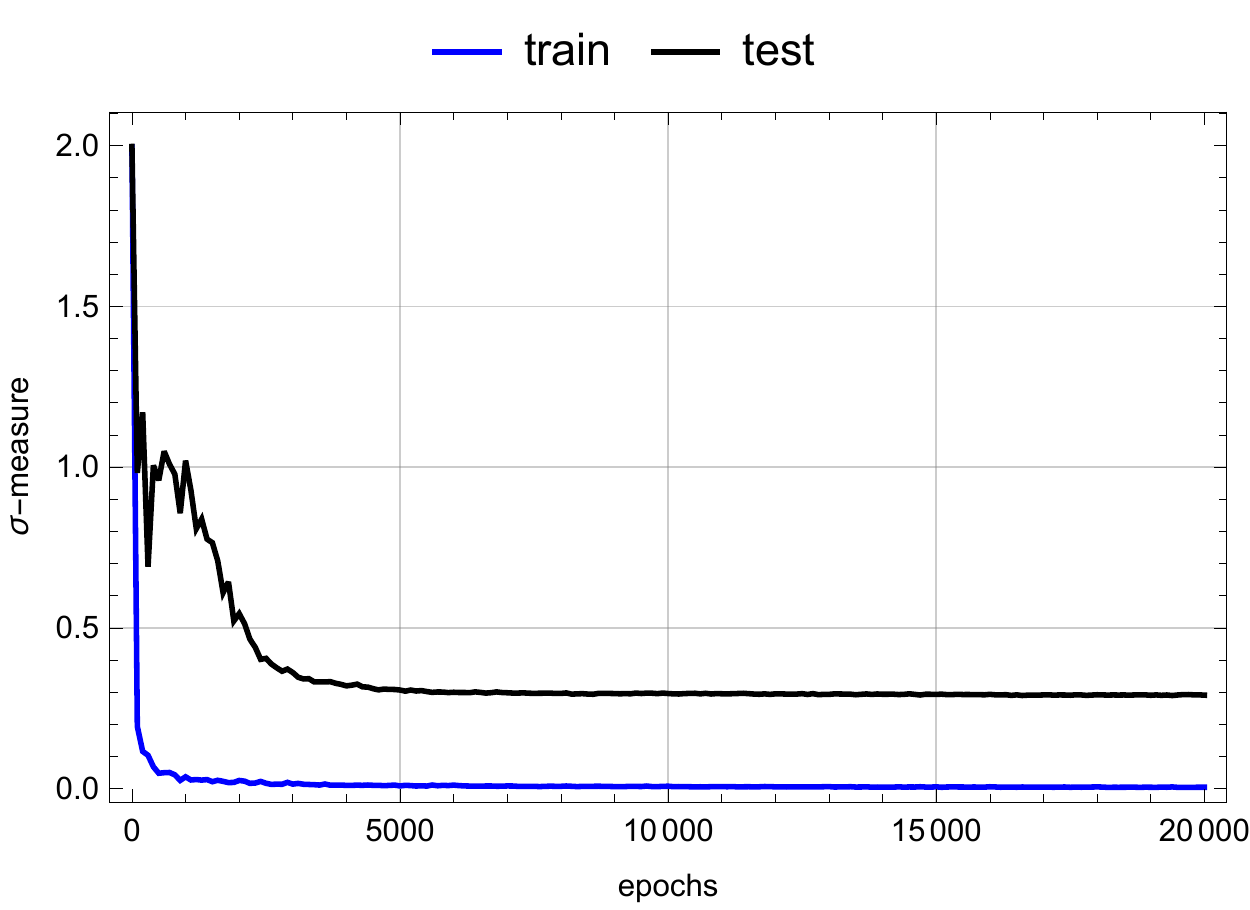}~~\includegraphics[scale=.33]{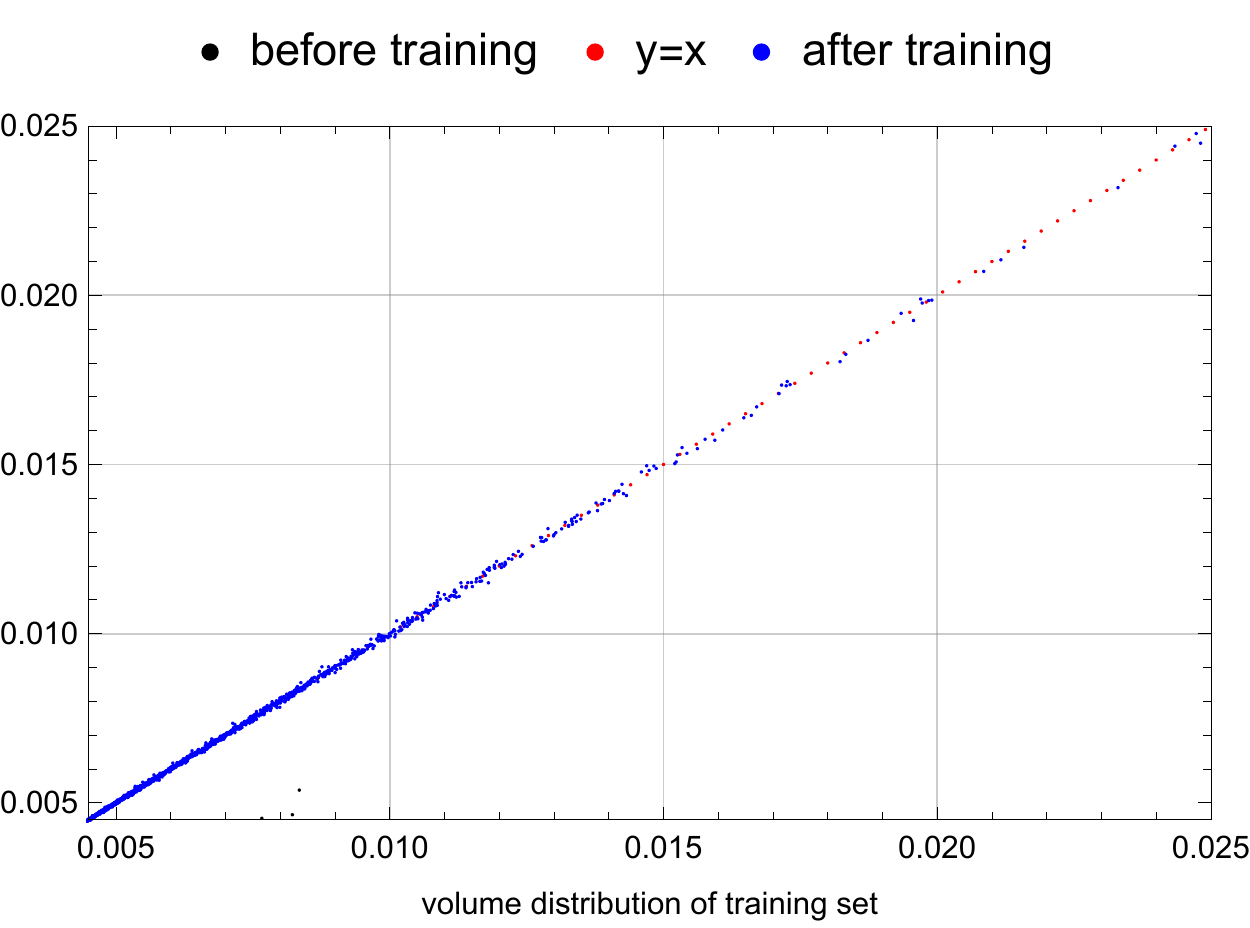}~~\includegraphics[scale=.33]{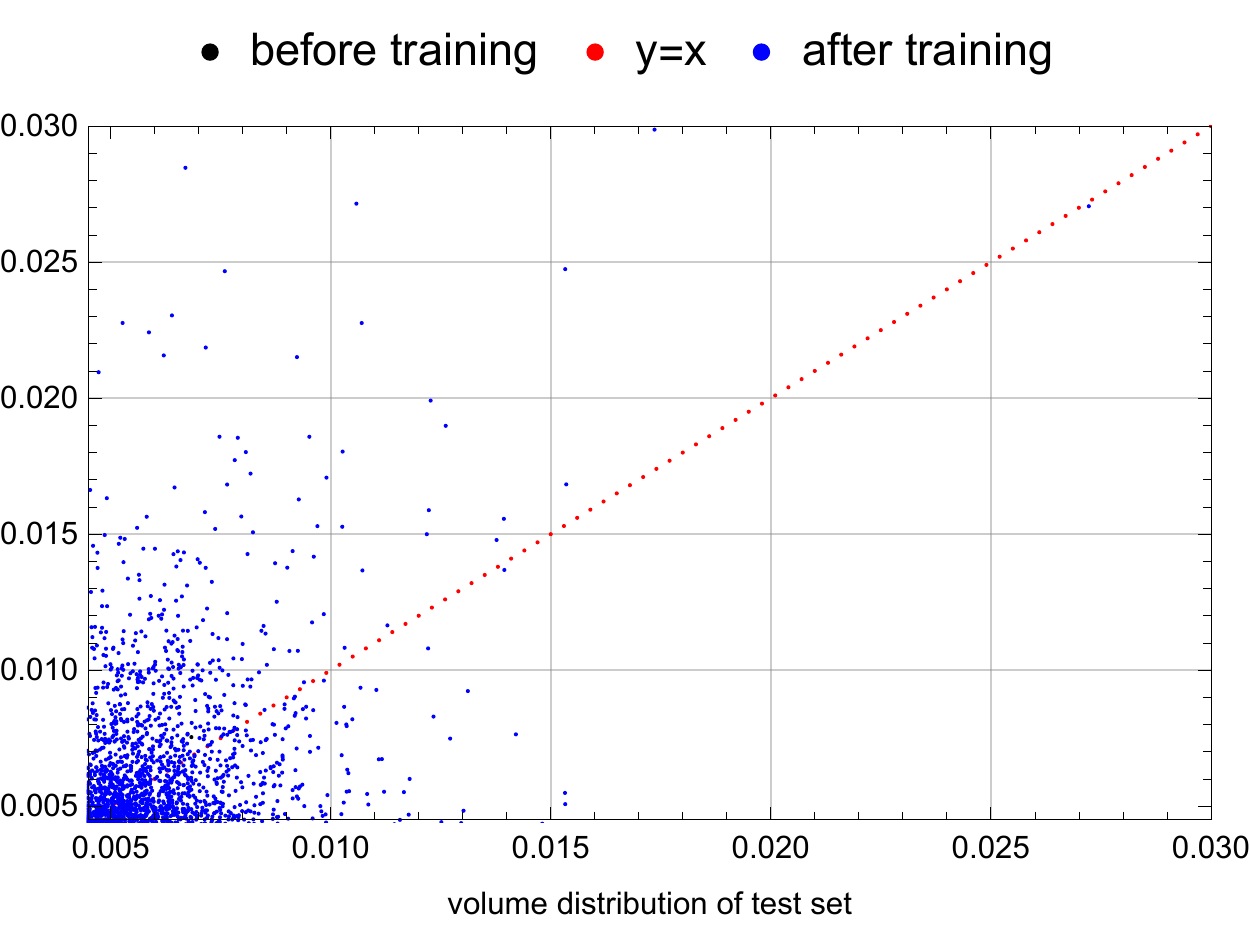}\\[20pt]
 \includegraphics[scale=.33]{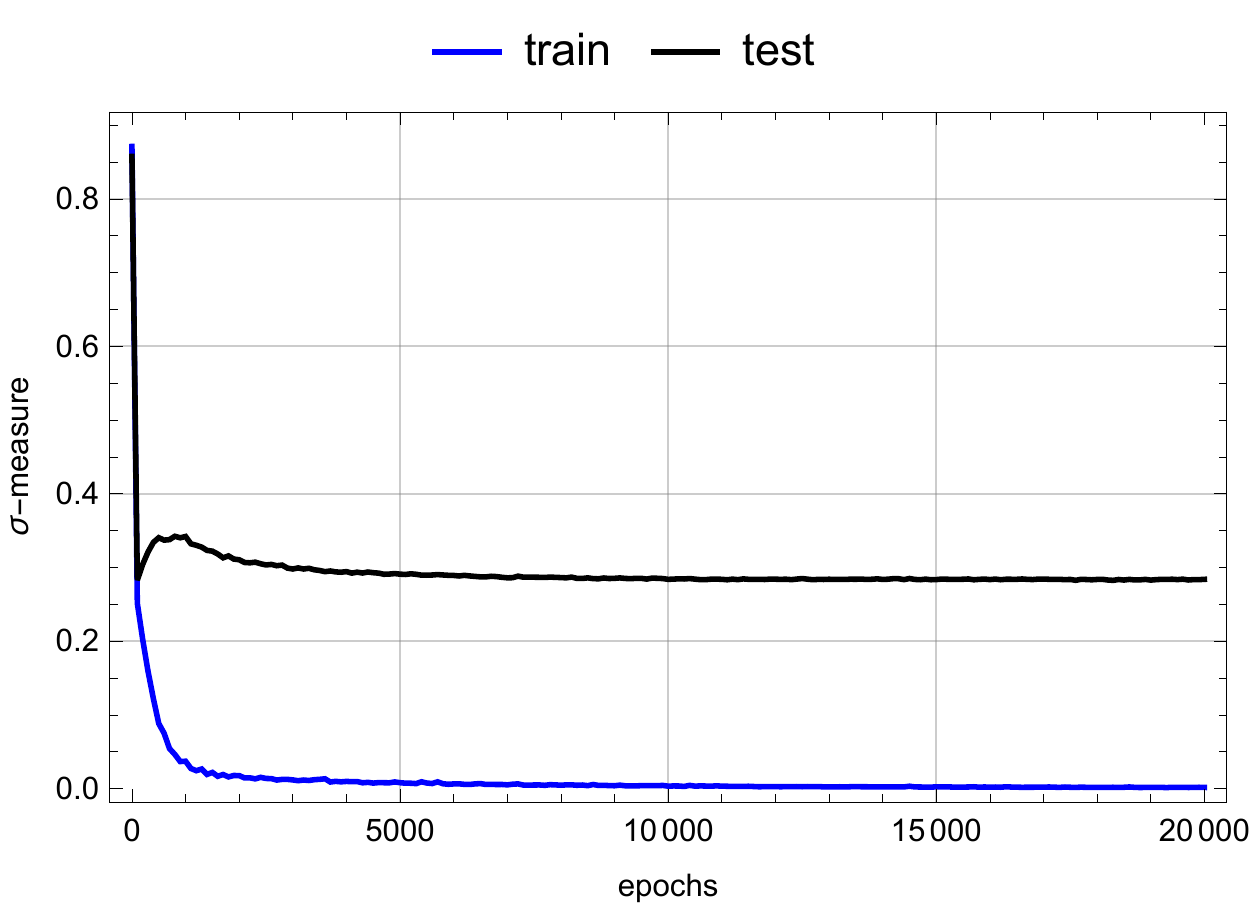}~~\includegraphics[scale=.33]{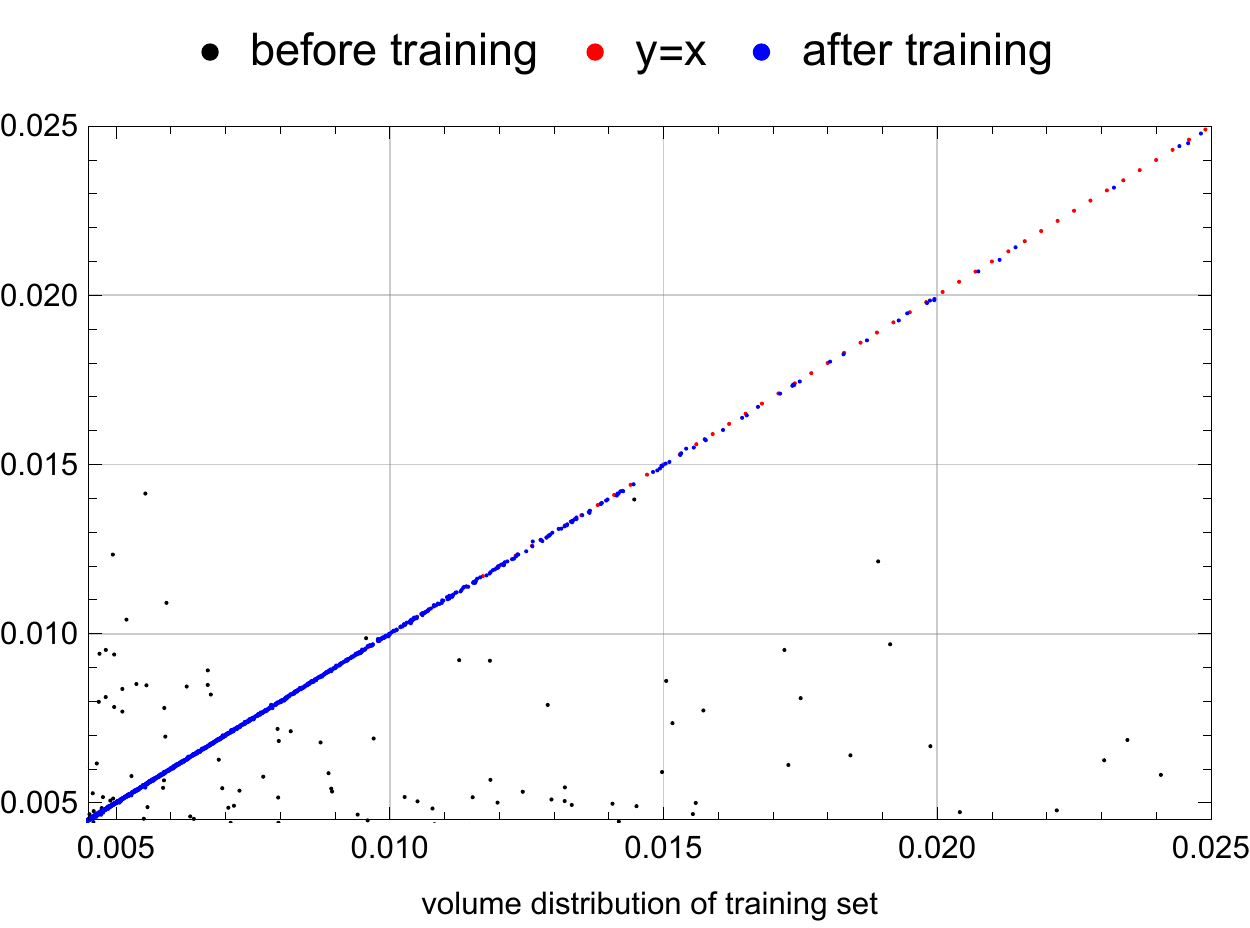}~~\includegraphics[scale=.33]{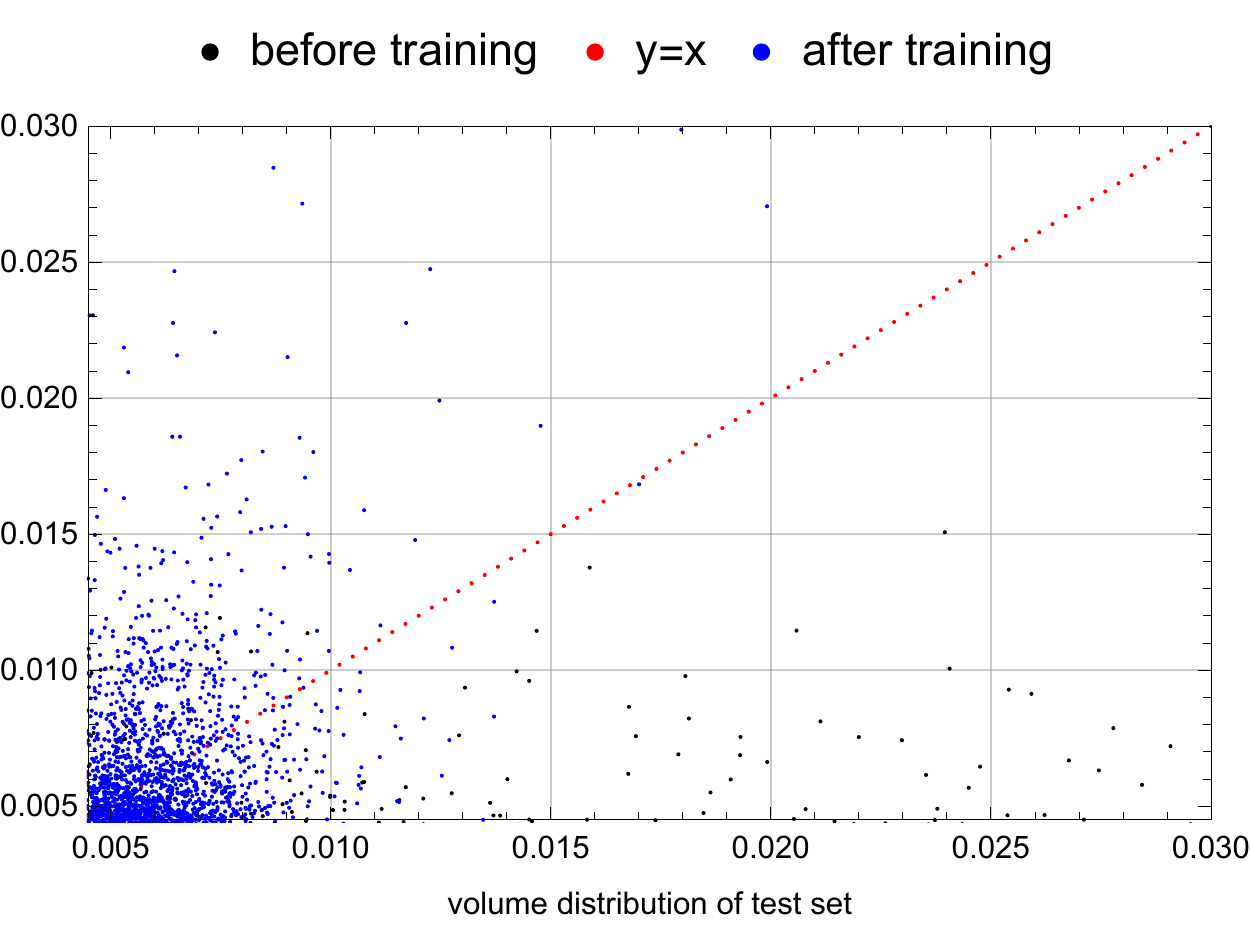}\\[20pt]
  \includegraphics[scale=.33]{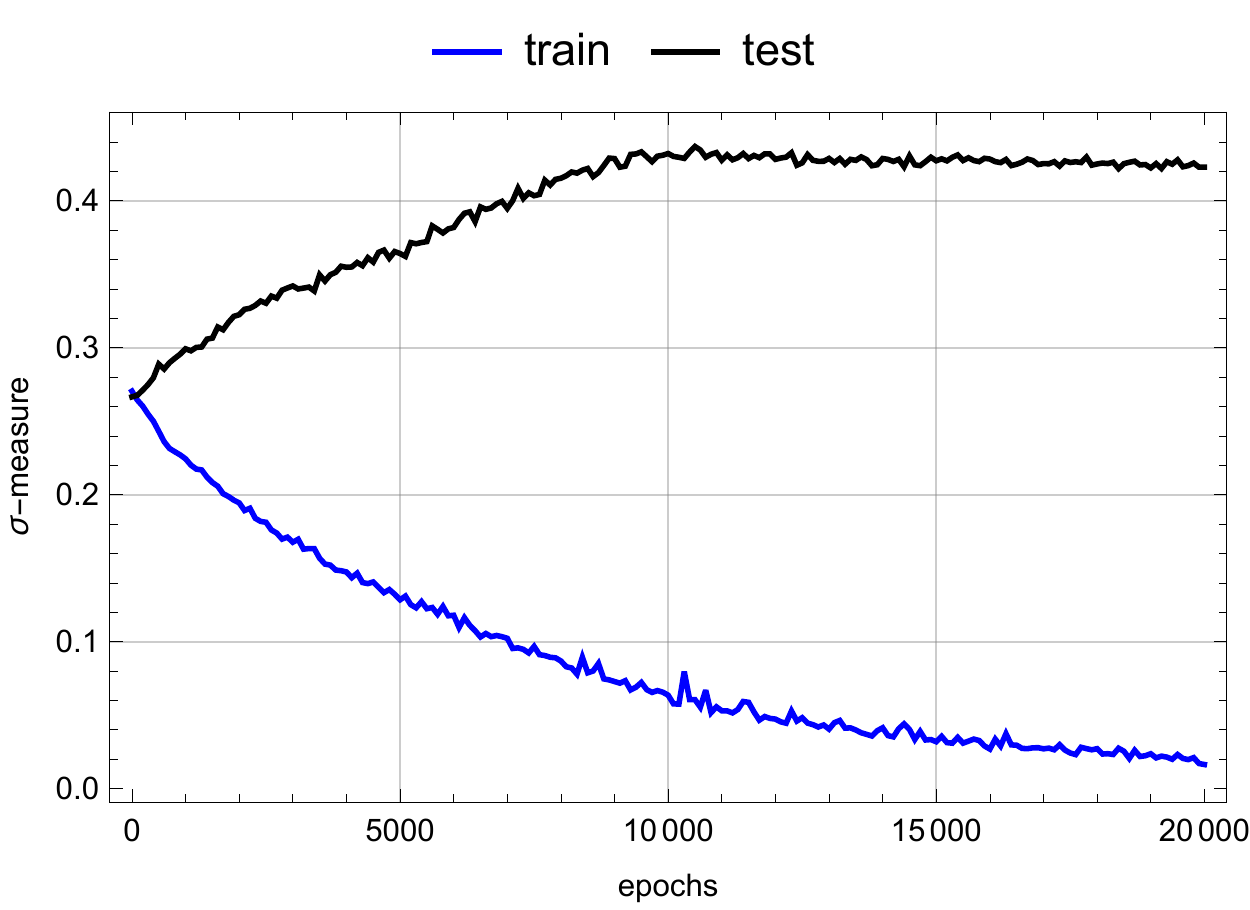}~~\includegraphics[scale=.33]{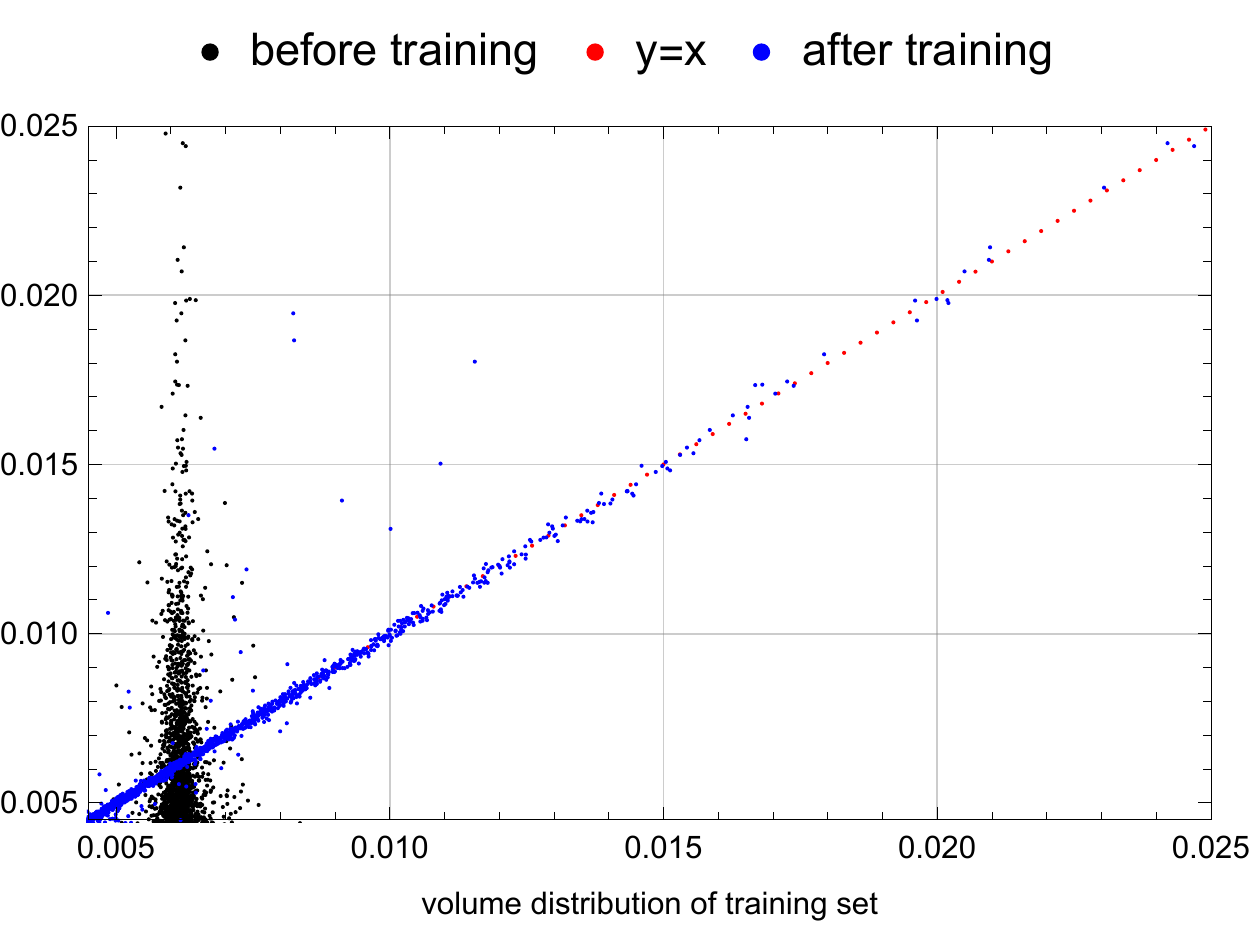}~~\includegraphics[scale=.33]{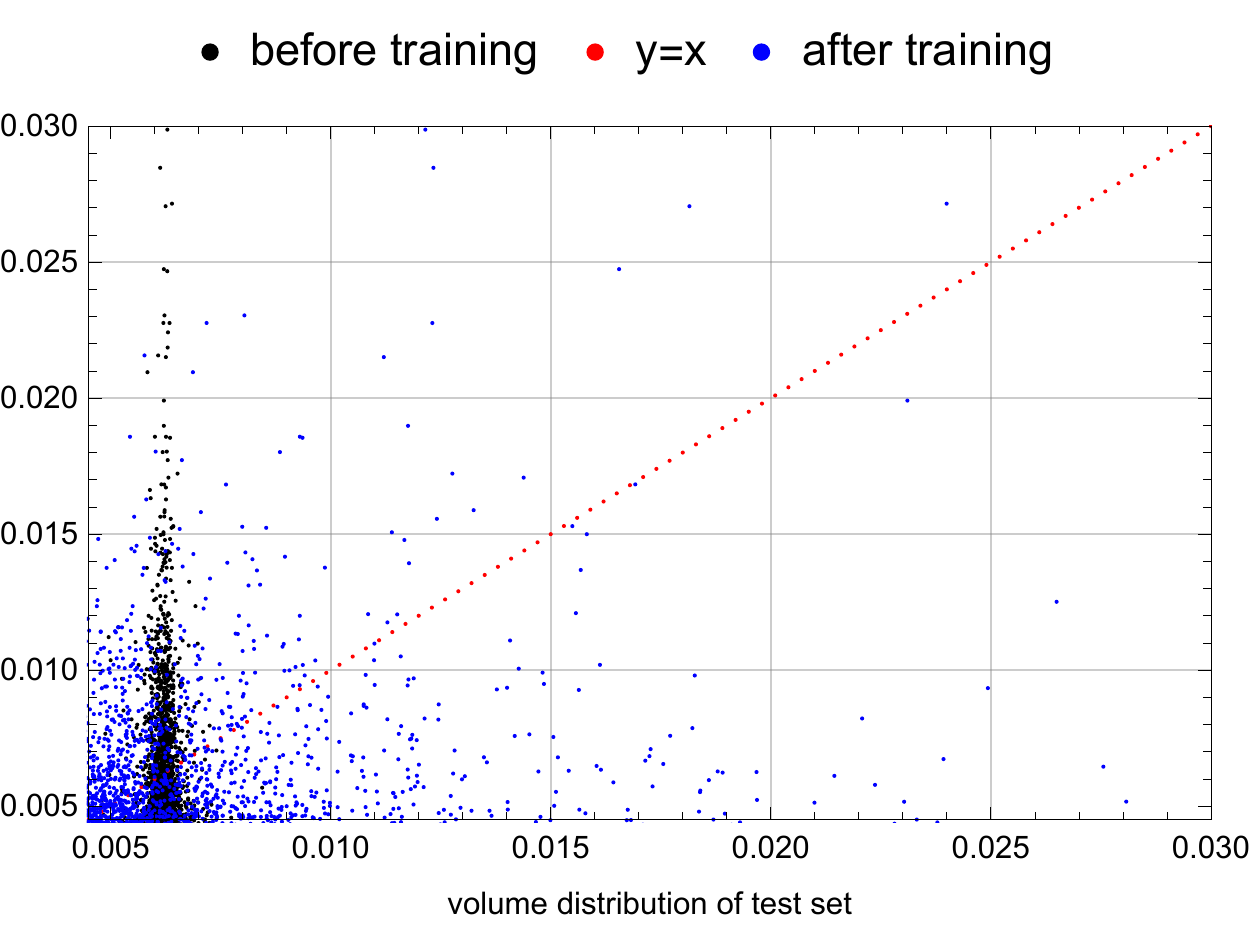}\\[20pt]
\caption{\small\textit{Evolution of $\sigma$-measures and normalized volume distributions 
for a member of the Dwork family of quintics~\eqref{eq:dwork} with $\psi=-\frac{1}{5}$. Figures in the first column show the evolution of $\sigma$ for both training and test sets, for three activation functions, namely, ReLU, $\tanh$, and logistic sigmoid, which correspond to the three rows, while figures in the second column show the volume distribution for the training points, before and after training.
Figures in the third column show the same for the test points.}}
  \label{fig:sigmasQPsi}
  \vspace{-10pt}
\end{figure}

The neural network approximation can be extended to any member of the Dwork family given by~\eref{eq:dwork}.
For this experiment, we take $\psi=-1/5$. Our results are summarized in Figure~\ref{fig:sigmasQPsi}.
In this case we obtain the best $\sigma$ for the $\tanh$ activation function.
After training, the $\sigma$-measure drops to $0.0015$ on the training set and to $0.28$ on the test set.
We also see that the ReLU activation function gives a $\sigma$ of $0.0042$ on the training set and $0.29$ on the test set.

\newpage
\subsection{The Tian--Yau manifold}\label{sec:tianyau}
For the Tian--Yau manifold~\eref{eq:ty}, we train and test with points in a single patch. There are in total $192$ patches. Note that in contrast to the previous hypersurface examples the patches are not all equivalent. This is due to the fact that all permutations among homogeneous coordinates have to act simultaneously on both $\mathbb{P}^3$ ambient spaces in order to leave Equation \eqref{eq:q3} invariant. This leaves us with four inequivalent families of patches. The family we considered for our computations is based on the patch $(4,6;2,7,1)$. Here again we only train using the $\sigma$-measure as the Loss function. 
The network approximation produces $\sigma$-measure values on the training set of $0.007$, 
$0.005$, and  $0.032$ for the ReLU, $\tanh$, and logistic sigmoid, respectively.
In this case we see that $\sigma$ on the test set only drops for the ReLU activation function.
It seems that for this example the network produces a broader dispersion of outputs for the test data than in the previous examples. 
\begin{figure}[h!]
\centering
\includegraphics[scale=.413]{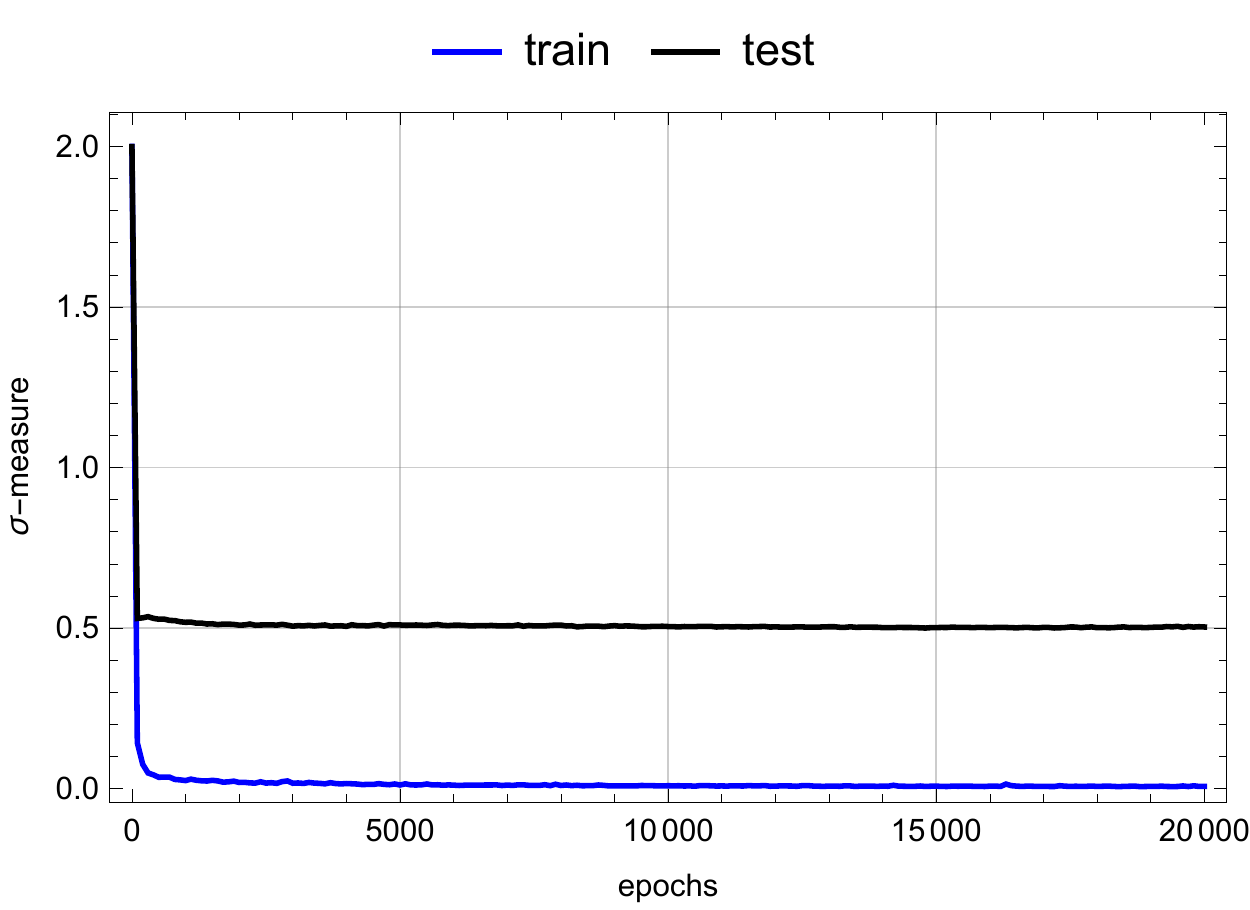}~~\includegraphics[scale=.4]{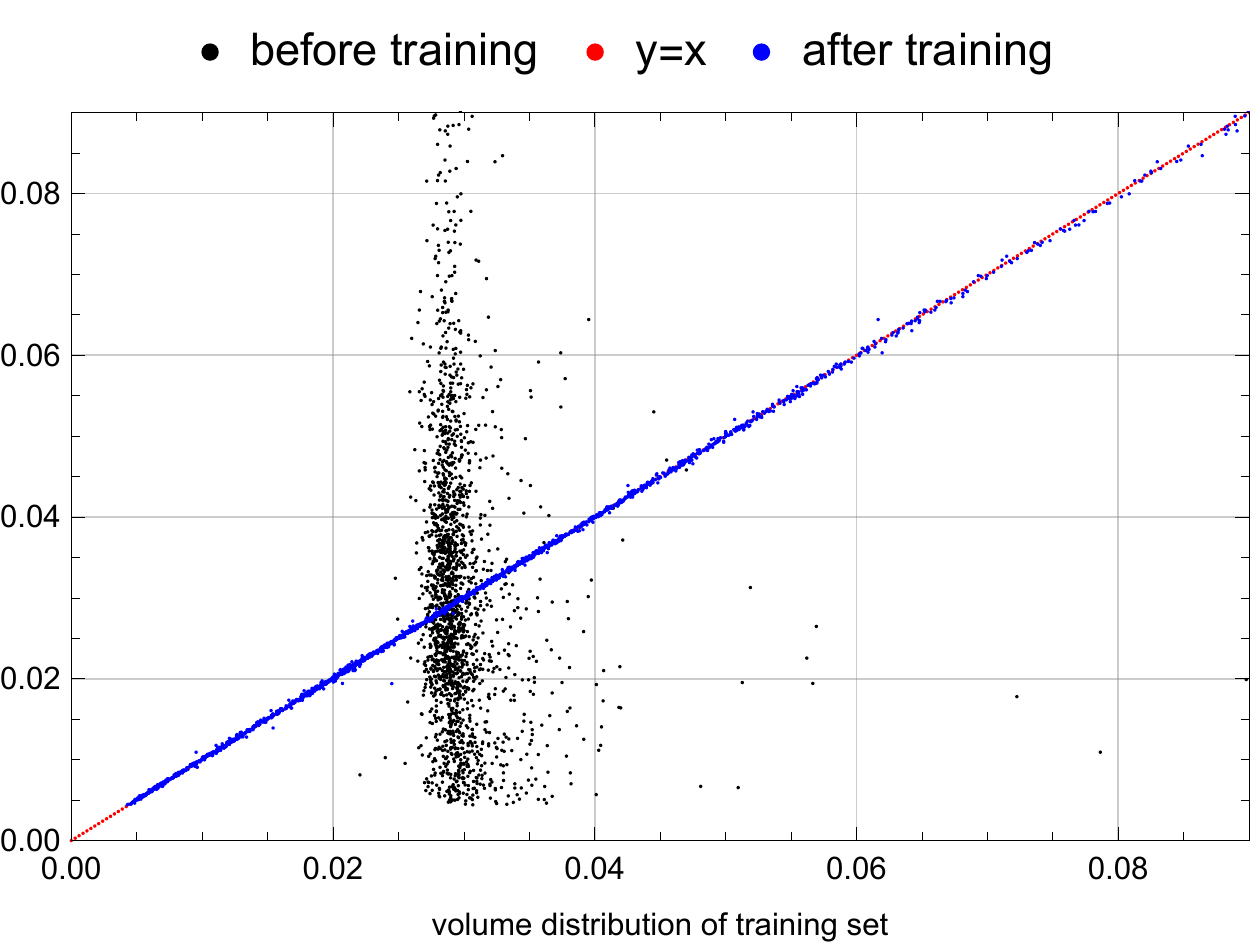}~~\includegraphics[scale=.4]{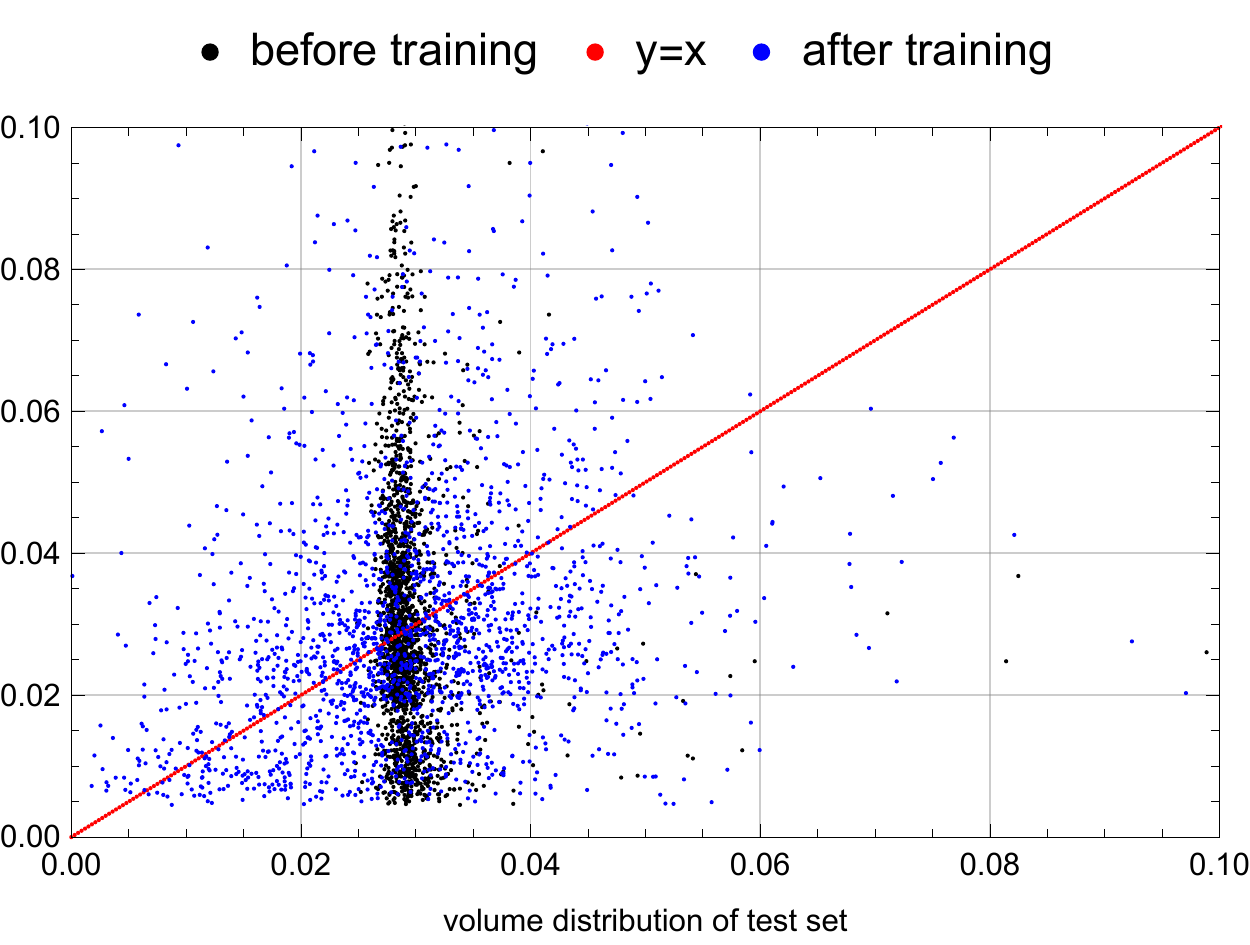}\\[20pt]
 \includegraphics[scale=.413]{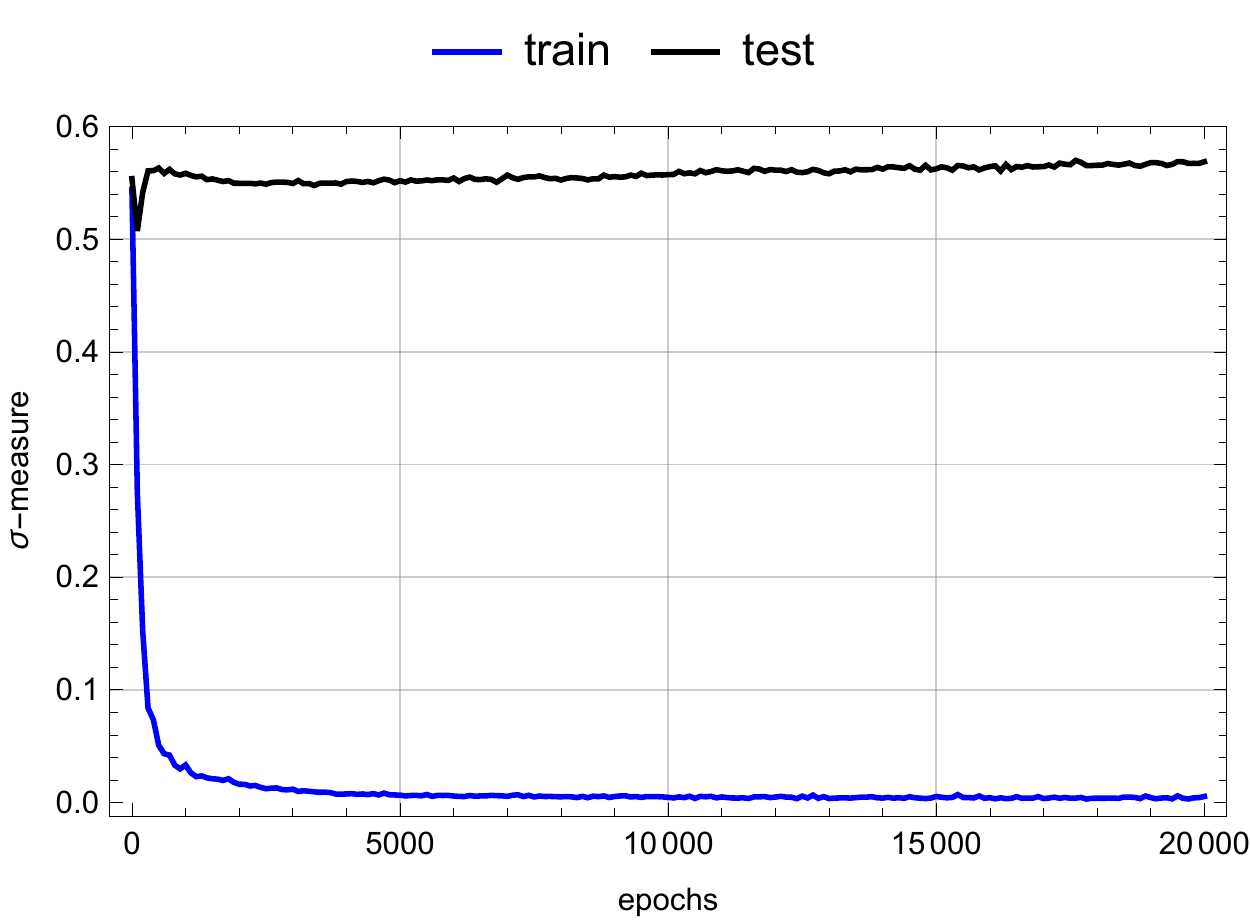}~~\includegraphics[scale=.4]{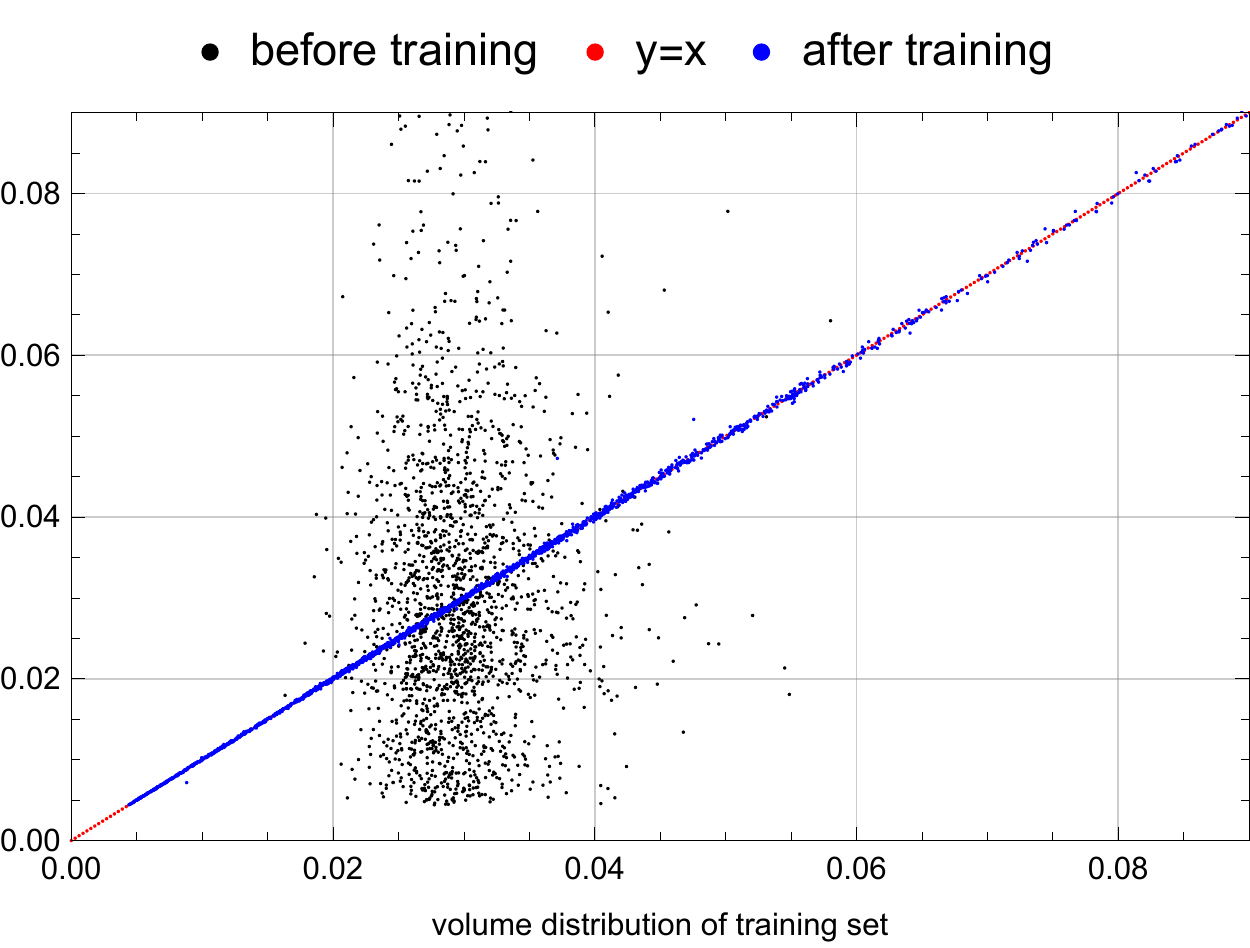}~~\includegraphics[scale=.4]{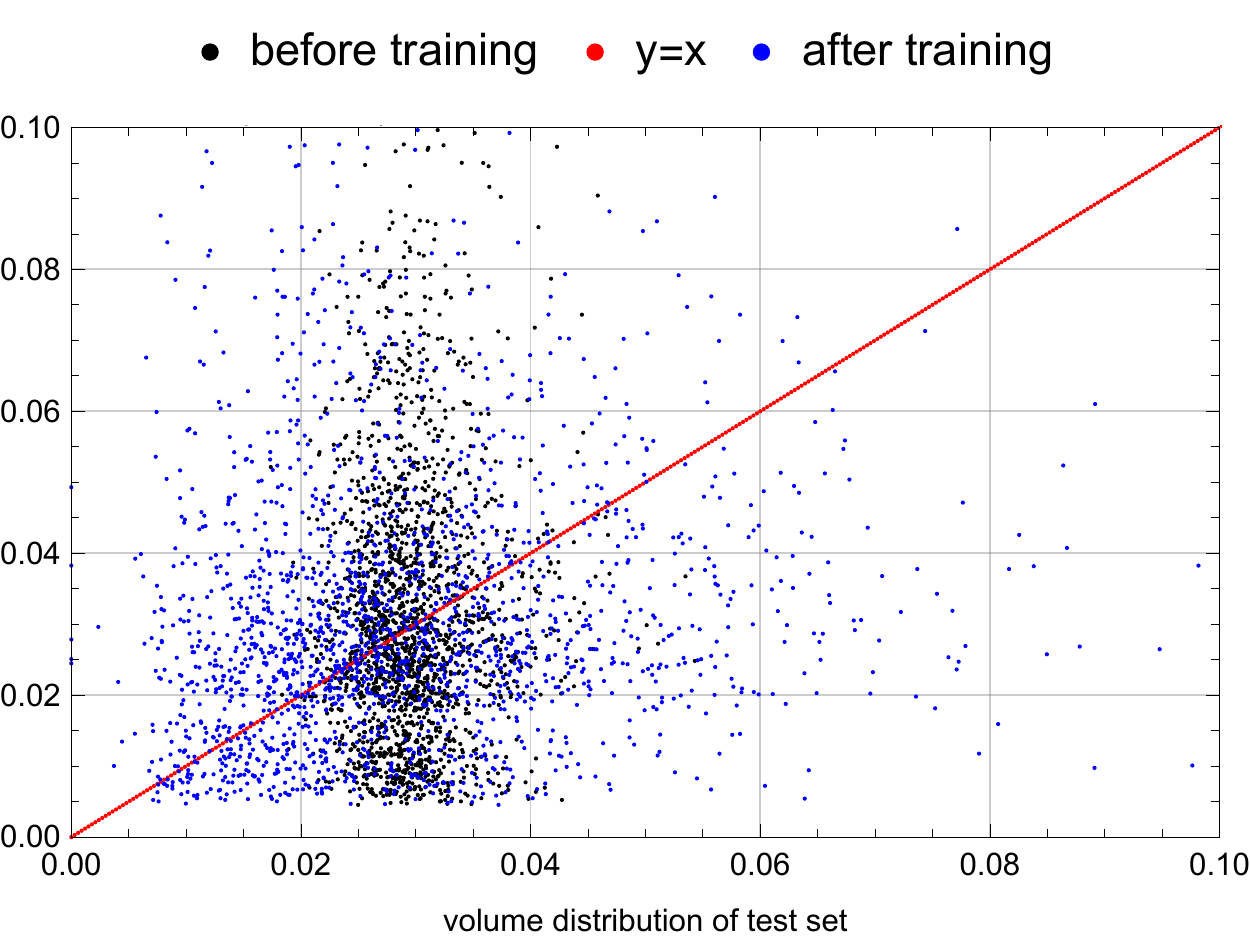}\\[20pt]
  \includegraphics[scale=.413]{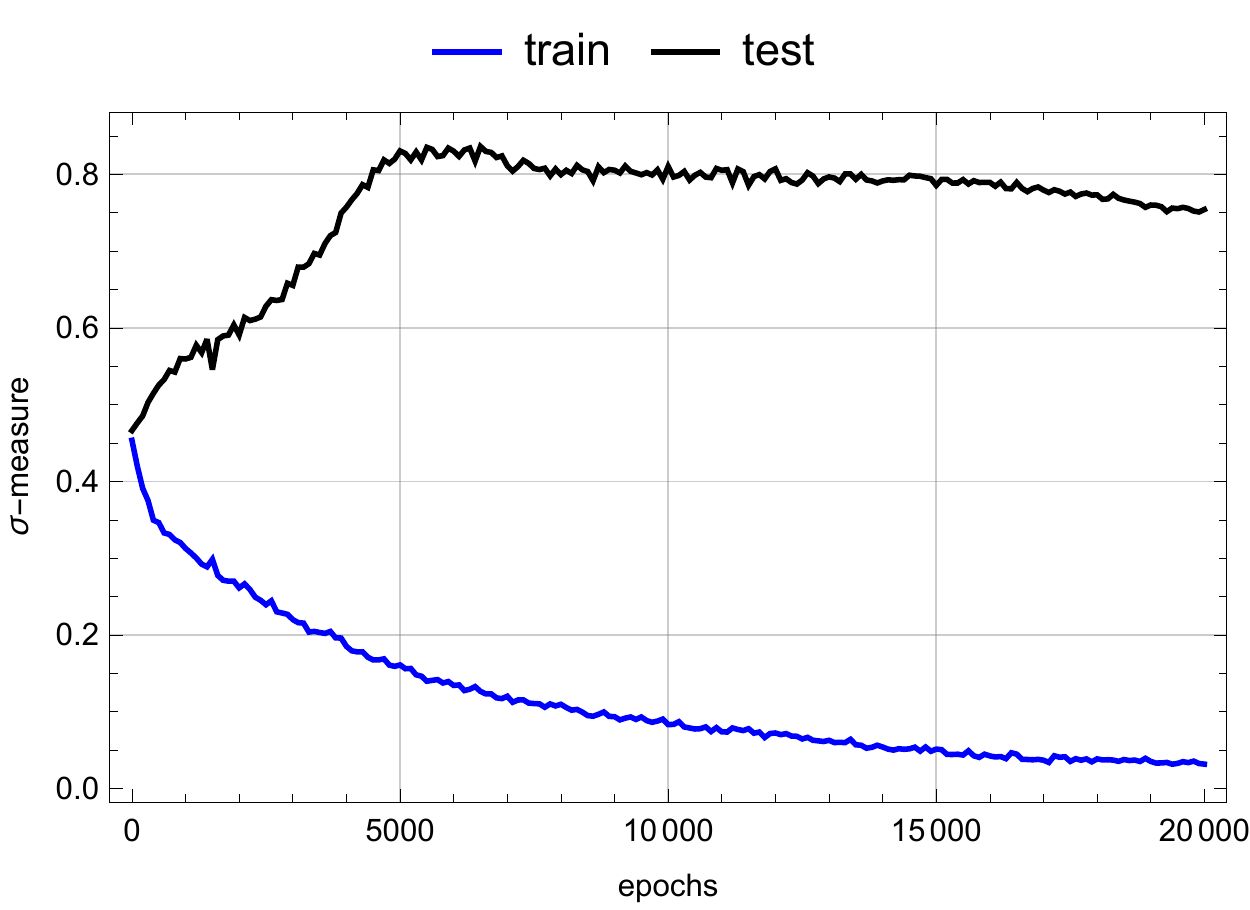}~~\includegraphics[scale=.4]{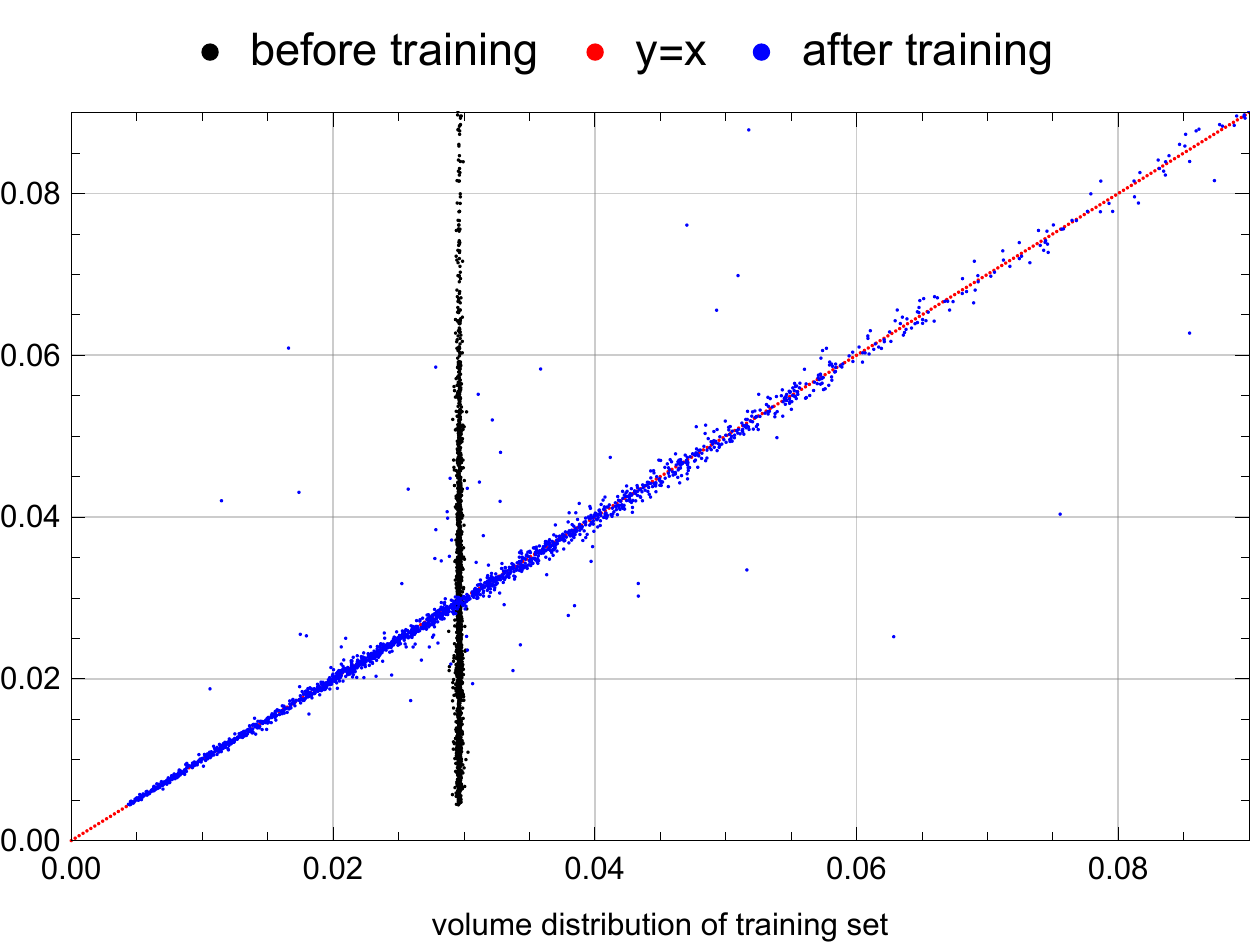}~~\includegraphics[scale=.4]{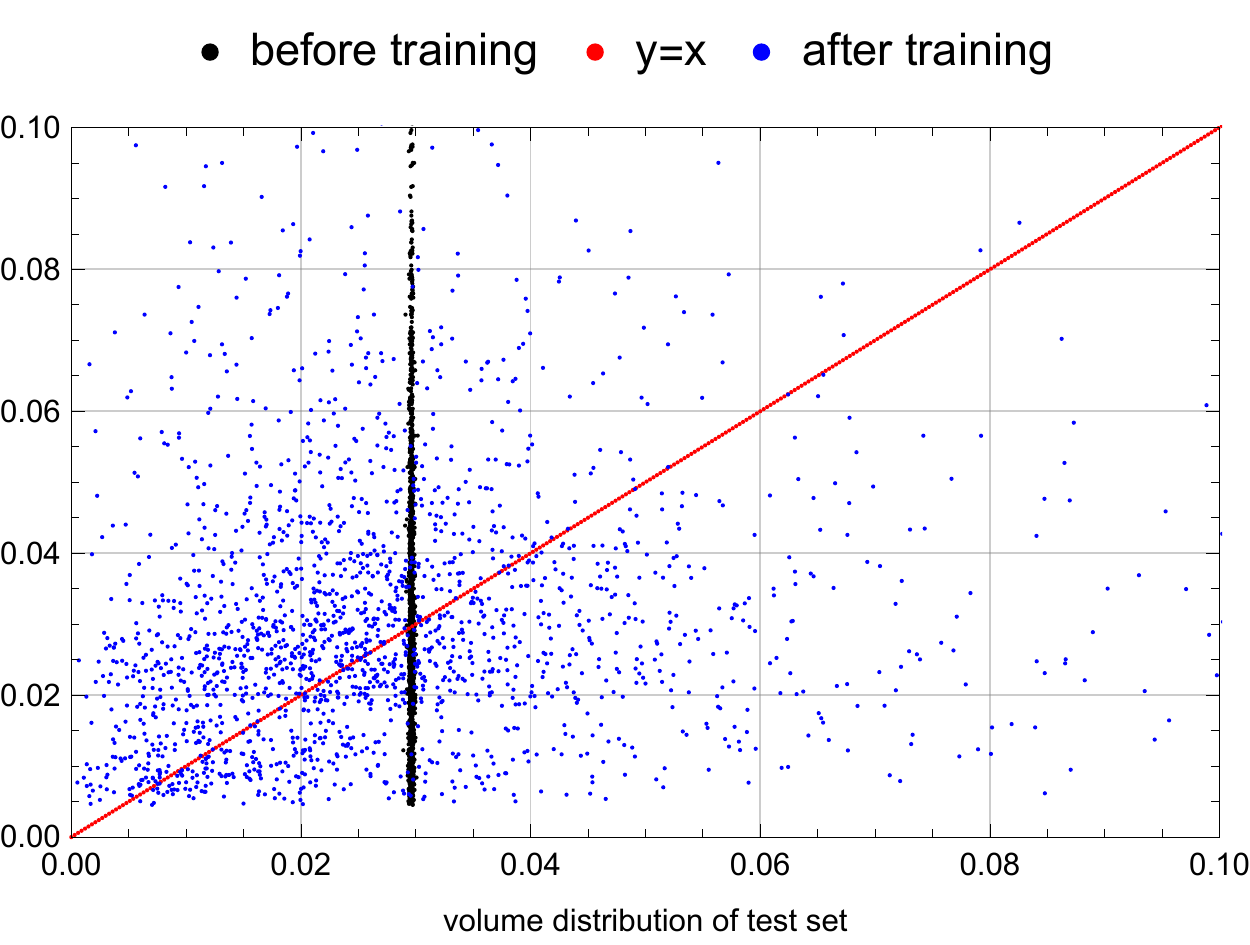}\\[20pt]
\caption{\small\textit{Evolution of $\sigma$-measures and volume distributions with respect to the Fubini--Study metric, for the Tian--Yau manifold ~\eqref{eq:ty}.
Figures in the first column show the evolution of $\sigma$ for both training and test sets, for three activation functions, namely, ReLU, $\tanh$, and logistic sigmoid, which correspond to the three rows, while figures in the second column show the volume distribution for the training points, before and after training.
Figures in the third column show the same for the test points.}}
  \label{fig:sigmasTY}
  \vspace{-10pt}
\end{figure}

In contrast to the previous hypersurface cases where one has permutation symmetries that permit us to infer an identical structure of the metric in all of the patches, for the Tian--Yau manifold one has four different classes of patches. This makes it hard to evaluate the $\mu$-measure. One possibility to follow this approach is to employ a pair of neural networks per class. 

Another example of a complete intersection Calabi--Yau that could be addressed using similar modifications is the Schoen manifold described below:

\be
\text{Schoen}: \quad 
\left[ \ba{c||cc}
\mathbb{P}^2 & \,3 & 0 \cr
\mathbb{P}^2 & \,0 & 3 \cr
\mathbb{P}^1 & \,1   & 1
\ea \right]_{\chi=0} ~.
\label{eq:schoen1}
\ee

The configuration matrix for this manifold is the transpose of the configuration matrix for the Tian--Yau manifold~\eref{eq:ty}.
The manifold is topologically distinct, however.
The Hodge pair for this manifold is $(h^{1,1},h^{1,2}) = (19,19)$, and therefore the Euler characteristic is $\chi = 0$. 
This manifold is interesting in its own right for string phenomenology. It is a split of the bicubic complete intersection Calabi--Yau on which semi-realistic heterotic string models have been realized. 

\newpage
\section{Discrete symmetries}\label{sec:symms}
Discrete symmetries of Calabi--Yau manifolds play an important r\^ole in string model building. In the Standard Model, discrete symmetries are merely hypothesized to explain certain processes or their absence. A phenomenologically interesting example pertains to the absence of proton decay. In this case, a discrete R-symmetry in the minimal supersymmetric Standard Model protects the proton from decay. It also ensures a stable lightest supersymmetric partner, which is then a candidate for dark matter. A second example of the important r\^ole discrete symmetries play in particle phenomenology is given by the discrete structure of mixing matrices such as the CKM and the PMNS matrices in the quark and lepton sectors, respectively. A symmetry group invoked to explain certain such discrete structures is $\Delta(27)$, which is isomorphic to $(Z_3{\times}Z_3){\rtimes}Z_3$. (Curiously, $\Delta(27)$ as an orbifold group gives simple constructions of the Standard Model as a worldvolume theory on D$3$-branes~\cite{Aldazabal:2000sa,Berenstein:2001nk}.) In the language of superstrings, these discrete symmetries cannot merely be hypothesized, but actually descend from isometries of the underlying compactification space. As such, constructing Calabi--Yau spaces with phenomenologically interesting discrete symmetries that remain unbroken after compactification is of paramount importance. 

A number of theoretical investigations have led to the classification of discrete symmetries of Calabi--Yau threefolds. These have resulted in large datasets of discrete symmetries of complete intersection Calabi--Yau threefolds that are freely acting~\cite{Candelas:1987du,Candelas:2008wb,Candelas:2010ve,Braun:2010vc}. This allowed a complete classification of the discrete symmetries of the resulting quotient Calabi--Yau threefolds, on which heterotic models are typically built~\cite{lukas2020discrete, mishra2017calabi}. Further, it was shown that discrete symmetries can be enhanced at special loci in the complex structure moduli space~\cite{candelas2018highly,mishra2017calabi}. Similar attempts have been made on the Kreuzer--Skarke dataset~\cite{braun2018discrete}, although due to the vastness of the dataset, a full classification has not been possible. These investigations point to the fact that discrete symmetries are rather rare in Calabi--Yau spaces, with phenomenologically interesting symmetries even more so. Many semi-realistic string derived standard models are built on Calabi--Yau quotients by freely acting symmetries. The quotient manifolds have topological properties distinct from the original manifold~\cite{Candelas:2008wb,Candelas:2010ve,candelas2016hodge} and are often more suitable for model building. 

With the advent of machine learning in studying the string landscape, one might perceive faster classification of such symmetries on the known Calabi--Yau databases, although there has not been much success thus far~\cite{Bull:2018uow, krippendorf2020detecting}. In this section, we report modest preliminary results in this direction. We hope this stimulates further investigation into a machine motivated discrete symmetry learning of Calabi--Yau spaces. 

We will discuss discrete symmetries for K3 and the quintic threefold. Although we do not inform our neural network architectures of any symmetry explicitly, we hope to understand whether such symmetries are nonetheless being learned, as a consequence of learning the flat metric. As such, we quantify the extent of learned symmetry until epoch $n$, by ${\delta_n}$, defined as,
\begin{align}
\delta_n(\tau):=\frac{1}{N}{\sum_{z}}\text{abs}\left(\frac{g_{NN}(\theta_n; z) - g_{NN}(\theta_n; \tau.z)}{g_{NN}(\theta_n; z)}\right) ~, \label{symmlearn}
\end{align}
where $g_{NN}(\theta_n; z)$ denotes the metric constructed using the neural network(s) in Figure~\ref{fig:arch}, with $\theta_n$ denoting the network's parameters at epoch $n$. Here, $\tau$ denotes a specific instance of a discrete symmetry of the manifold in question, acting linearly on the combined homogeneous co-ordinates of the ambient space. The sum is over a random selection of $N$ points used for training and testing. An overall decreasing behavior of $\delta_n(\tau)$ with increasing number of epochs, $n$, would be indicative of the symmetry $\tau$ being learned. Below, we outline how well certain symmetries were captured by our neural network architectures for the K3 surface and the Fermat quintic. 

\subsection{The quartic K3 surface}
From the quartic K3 surface equation reproduced in~\eqref{K3second}, it is easy to note the symmetry permuting the homogeneous co-ordinates $[z_1: z_2: z_3: z_4]$ of $\mathbb{P}^3$. The permutation symmetry is utilized during the training process to ensure that the metric agrees at the intersection of different patches. We set one of the co-ordinates to $1$ defining a patch, in effect breaking the permutation symmetry $S_4$. There is a second set of symmetries that acts diagonally on the combined homogeneous co-ordinates. The action of this symmetry is given in~\eqref{K3_symm1} and an instance of it, $\tau$, in~\eqref{K3_symm_example}. 

\begin{align}
\text{K3}: ~~~&z_1^4 + z_2^4 + z_3^4 + z_4^4 = 0 \subset \mathbb{P}^3 ~\label{K3second}\\[5pt]
&z_p\rightarrow \ \omega_p\ z_p ~; \text{~~with~~} p \in \{1,2,3,4\}   \text{~~and~~} \omega_p \in \mathbb{Z}_4 ~. \label{K3_symm1}\\
&\tau:~~  z_1 \mapsto ~i\ z_1~,~z_2 \mapsto -\ z_2~,~z_3 \mapsto -i\ z_3 \label{K3_symm_example} ~.
\end{align}

\begin{figure}[t]
 \vspace{0pt}
\centering
\includegraphics[scale=.613]{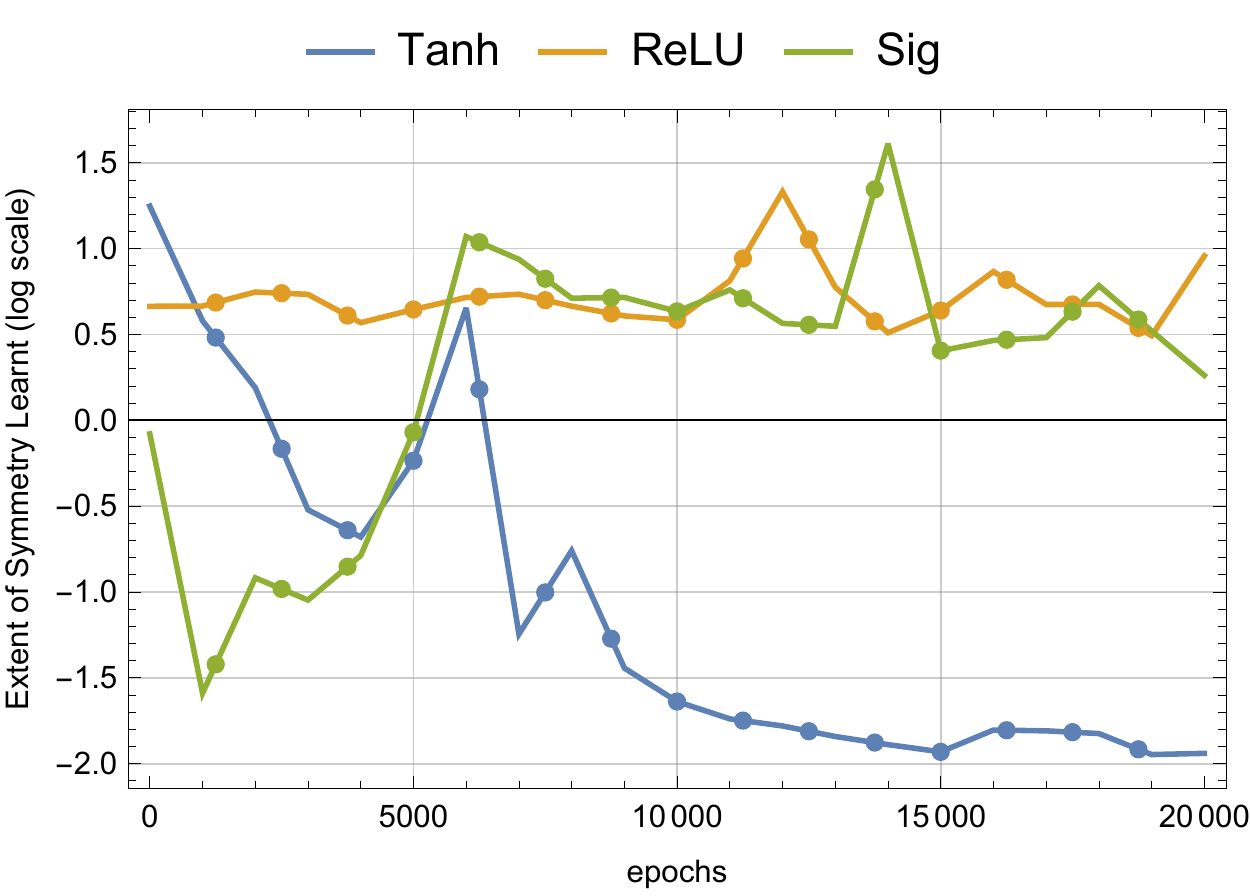}\\[5pt]
\caption{\small\textit{Learning discrete isometries of the quartic K3 surface defined in~\eqref{K3second}. The plot shows the extent of learning given by~\eqref{symmlearn} for the symmetry of the K3 surface described in \eqref{K3_symm_example}.  We observe similar behavior for other symmetries of the class~\eqref{K3_symm1}.}}
  \label{fig:symms_K3}
    \vspace{10pt}
\end{figure}

In Figure~\ref{fig:symms_K3}, we report the extent of symmetry learned by monitoring $\delta_n$ over the epochs. We find that not all architectures capture this symmetry. In fact, the architecture with $\tanh$ activation function is the only one that shows a behavior indicative of the symmetry $\tau$~\eqref{K3_symm_example} being learned.

\subsection{The Fermat quintic}

For the Fermat quintic similar considerations apply, although the discrete symmetries themselves are different.  In~\eqref{Q_symm1} we identify a class of symmetries that could have possibly been learned during the process of training. An example of this symmetry is given in~\eqref{Q_symm_example} and it acts diagonally on the homogeneous co-ordinates $[z_1: z_2 : z_3: z_4: z_5]$. 
\begin{align}
\text{Fermat quintic}:~~ &z_1^5 + z_2^5+ z_3^5+ z_4^5 + z_5^5 = 0 \subset \mathbb{P}^4 ~\label{Qsecond}\\[5pt]
&z_p \rightarrow \omega_p~z_p ~; \text{~~with~~}p \in \{1,2,3,4,5\}  \text{~~and~~} \omega_p \in \mathbb{Z}_5 ~. \label{Q_symm1}\\
& \tau:~~  z_1 \mapsto e^{2\pi \mathrm{i} /5} \ z_1~,~z_2 \mapsto e^{4\pi \mathrm{i}  /5} \ z_2~,~z_3 \mapsto e^{8\pi \mathrm{i} /5}\ z_3,~z_4 \mapsto e^{6\pi \mathrm{i}  /5}\ z_4  \label{Q_symm_example}
\end{align}

\begin{figure}[t]
 \vspace{0pt}
\centering
\includegraphics[scale=.613]{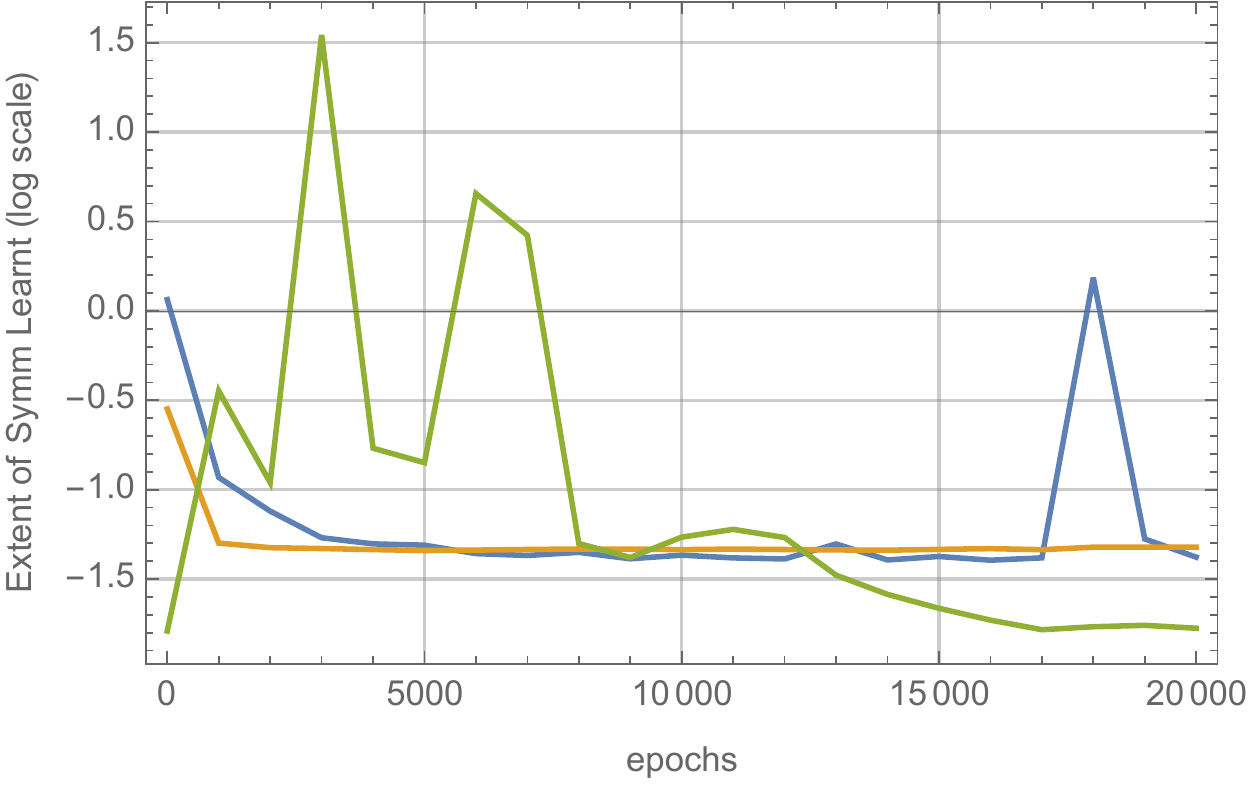}\\[5pt]
\caption{\small\textit{Learning discrete isometries of the Fermat quintic defined in~\eqref{Qsecond}. The plot shows the extent of learning given by~\eqref{symmlearn} for the symmetry of the quintic surface described in \eqref{Q_symm_example}.  We observe similar behavior for other symmetries of the class~\eqref{Q_symm1}.}}
  \label{fig:symms_Q}
    \vspace{10pt}
\end{figure}

In Figure~\ref{fig:symms_Q}, we show the extent of learning for the symmetry described in~\eqref{symmlearn}. In this example, we note that the architecture with the logistic sigmoid activation function has captured the symmetry to the largest extent, while the other architectures show behavior consistent with this learning.  

\newpage

\section{Discussion and outlook}\label{sec:disc}
In this paper, we have reported preliminary investigations on machine learning numerical metrics on Calabi--Yau spaces. We created neural network models for metrics over Calabi--Yau spaces of complex dimensions up to three, choosing examples of the complex torus $T^2$ in dimension one and the quartic K3 surface in dimension two. In dimension three, we chose to work with the Fermat quintic, a second member of the Dwork family, as well as the phenomenologically interesting Tian--Yau manifold. Interestingly, the same neural network architecture described in Figure~\ref{fig:arch} could approximate Ricci flat metrics on all these spaces. In these experiments, the $\tanh$ activation function yielded the best $\sigma$-measures on these spaces. 

In most of our experiments, we have taken a small number of points. This is in contrast with the relatively large number of hyperparameters in the neural networks employed. For the case of the Fermat quintic, where we have increased the number of points for training, we saw improving results on the test set, demonstrating that the neural networks are indeed capable of generalization. We expect this trend to apply in all other geometries considered. Additionally, we would like to employ the full power of the loss function to include the $\kappa$- and $\mu$-measures in all cases, not just for the Fermat quintic. An important issue to address in this direction is to handle complex matrix products directly in order to compute Frobenius norms more efficiently. This was not feasible in the \texttt{PyTorch} implementation as it only allows us to manipulate real quantities. Once these pieces are in order, one would scan for an optimal network architecture that weighs in the capacity for generalization as well as loss minimization. Another avenue to explore is the possibility to constrain network hyperparameters in such a way that  K\"ahlericity is built in to the metric solution \textit{ab initio}.

In Section~\ref{sec:symms}, we commented on discrete symmetries of Calabi--Yau manifolds. While discrete symmetries are very important from a string model building perspective, they are rare. As such, a full classification of them is computationally intensive. There are many ways of incorporating symmetries into a machine learning model involving neural networks. Some of the traditional methods include data augmentation~\cite{beymer1995face,niyogi1998incorporating}, feature averaging, DeepSets~\cite{zaheer2017deep}, and building more complex equivariant architectures~\cite{Cohen2016group}. The inverse problem of discovering symmetries of the dataset itself, or in our case, the manifolds, is not well-studied. An example of such a study is~\cite{van2018learning}, wherein the marginal likelihood was employed to compute symmetries in the MNIST dataset~\cite{deng2012mnist}. In this paper, we lay the ground work for an alternate machine driven approach to discovering symmetries. Through examples of symmetries of the quartic K3 surface and the Fermat quintic, we demonstrate that discrete symmetries can be learned while a network is being trained on an auxiliary task. In this case, the task is to learn the flat metric. We found that our networks were able to learn discrete symmetries over the course of training without any effort on our part to inform the learning process or the network's architecture of these symmetries.

A discrete symmetry may be realized linearly in a given representation of a complete intersection Calabi--Yau, and simultaneously be realized non-linearly in an equivalent representation. An example of this phenomenon involves the Schoen manifold (reproduced below), a split of the bicubic. This manifold admits an equivalent representation which is also noted alongside:
\be
\text{Schoen}: \quad 
\left[ \ba{c||cc}
\mathbb{P}^1 & \,1   & 1 \cr 
\mathbb{P}^2 & \,3 & 0 \cr
\mathbb{P}^2 & \,0 & 3 
\ea \right]_{\chi=0} 
\qquad  \simeq \qquad
\left[ \ba{c||cccccc}
\mathbb{P}^1 &~0   & 0   & 0   & 0   & 1   & 1 \cr 
\mathbb{P}^2 &~1   & 1   & 0   & 0   & 1   & 0 \cr 
\mathbb{P}^2 &~1   & 1   & 0   & 0   & 1   & 0 \cr 
\mathbb{P}^2 &~0   & 0   & 1   & 1   & 0   & 1 \cr
\mathbb{P}^2 &~0   & 0   & 1   & 1   & 0   & 1 \cr
\ea \right]_{\chi=0}  ~.
\label{eq:schoen}
\ee
The manifolds in~\eqref{eq:schoen} are equivalent. There are symmetries of this manifold that are linearly realized in the extended representation but not in the split bicubic representation. Non-linearly realized symmetries are hard to classify using traditional methods. It would be interesting to understand if non-linear symmetries are also picked up by the neural networks during the training process. We hope to return to this analysis in the future. 

\begin{figure}[h]
 \vspace{0pt}
\centering
\includegraphics[scale=.613]{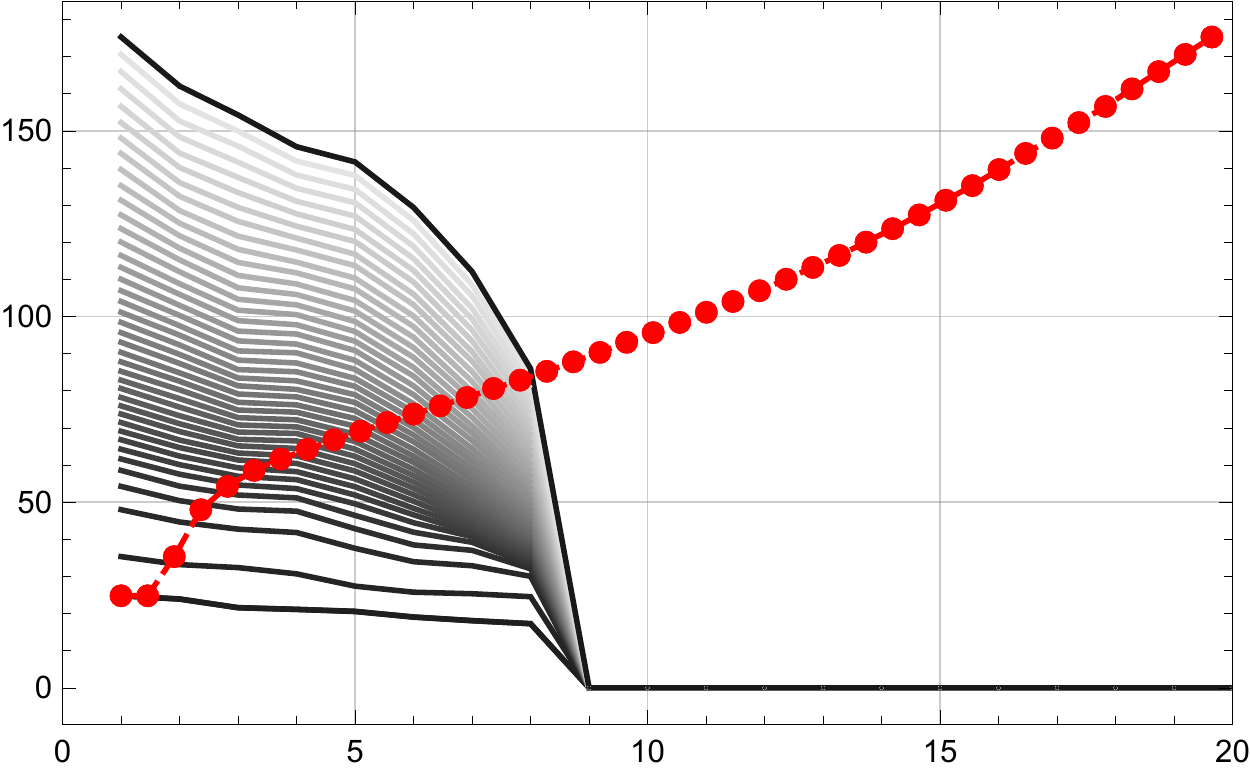}\\[5pt]
\caption{\small\textit{Evolution of the eigenvalues of the weight matrices in the first layer of the neural network approximating the flat metric on the Fermat quintic. The $x$-axis indexes the eigenvalues after sorting. The $y$-axis represents the eigenvalues of the (symmetrized) weight matrix. Different grey curves denote different training epochs, starting with the curve in black at the beginning of the training and finishing with the lightest gray curve at the bottom at the end of training. The curve in red tracks the largest eigenvalue at every epoch. We see the largest eigenvalue as well as the overall spectrum lower as training progresses.}}
  \label{fig:spectra}
    \vspace{10pt}
\end{figure}

The observations above warrant an analysis of the neural networks we have considered. The aim of such an analysis would be to understand how learning of the metric happens, as well as to derive hints for the underlying analytic function for the flat metric. Such an analysis is a necessity in mitigating the black box nature of the resulting neural network models. Although we wish to return to such analysis in a future work, a few comments are in order. A particular approach to analysis is using the recent Neural Tangent Kernel (NTK)  formulation~\cite{jacot2018neural}. Under the NTK regime, we would have an analytic formulation of the function to which our networks would converge. This would allow us to make meaningful statements for the underlying analytic flat metric. Secondly, in order to understand the training dynamics, we would like to understand how the spectra of matrices of the network parameters evolve over the course of learning. In Figure~\ref{fig:spectra}, we show this for the Fermat quintic. We note that during the course of training, the overall spectra lowers. We also note that condition number, which is the ratio of the largest and smallest eigenvalues, decreases over the course of the training. We observe similar behavior for other weight matrices in all the examples we have considered. We would like to understand if this provides hints for the underlying analytic function approximating the metric.

Finally, a Topological Data Analysis (TDA) approach, tracking persistent homology of the distribution of the network parameters might reveal further insights into the learning  process. Persistence has been shown to be a useful tool for interpretability in deep learning~\cite{Bruel2018exposition,Gabella2019topology}. Recently, a new complexity measure of neural networks has been proposed with the aid of topological methods~\cite{rieck2018neural}. This has helped better understand certain machine learning best practices, like dropout and batch–normalization. We expect such topological methods to shed further light on creating more accurate and interpretable neural network models of Ricci flat metrics of Calabi--Yau manifolds.

In importing machine learning methods to high energy theoretical physics, there has so far only been limited success in translating the learned associations into interpretable analytic formulae.
(See~\cite{Klaewer:2018sfl, Brodie:2019dfx, Brodie:2019pnz, Craven:2020bdz} for computations in line bundle cohomology on surfaces and in knot theory.)
An open problem, of course, is to reverse engineer the performance of the neural network to obtain analytic expressions for flat metrics.
So far, we have no results in this direction.

Work in progress details the Ricci flow approach more explicitly.
There are a number of strategies we adopt.
One of these is simply to use $\sum_{a,\bar{b}} |\text{Ric}_{a\bar{b}}|$ as the loss function to update the metric according to the Ricci flow equation.
When this loss is minimized to zero, we have arrived at the flat metric in the same K\"ahler class as the Fubini--Study metric, which is the starting point of the flow.
Another strategy is to solve the partial differential equations~\eref{eq:metricci} and~\eref{eq:bhe} using machine learning techniques.
Indeed, deep learning is particularly good at solving certain systems of equations~\cite{raissi:2017,li2020fourier}.

As approximations for the metric we have considered a system of two neural networks providing the metric entries in a natural matrix factorization. This method proves to work relatively well for all the geometries considered. Because of this we would like to highlight its modularity: It could serve not only to approximate the metric of Calabi--Yau manifolds, but more in general to complex and real manifolds as well. It is also important to compare our architecture and results to the expectations from Donaldson's algorithm. For the quintic threefold, our neural networks have 1,023,525 parameters. Donaldson's balanced metric at level $k$ will have $N_k^2$ real parameters, with $N_k$ the number of monomials up to order $k$: \begin{equation}
N_k=\begin{pmatrix}5+k-1 \\ k\end{pmatrix} - \begin{pmatrix}k-1 \\ k-5\end{pmatrix}\,, \quad k\geq 5
\end{equation}   
 For $k>11$ ($N_k=1155$) the balanced metric will have more parameters than the neural network approximation. For the quintic we obtain $\sigma$ values of order $10^{-3}$ on the training set. Similarly, contemplating a larger dataset we obtain $\sigma$ values of $0.022$. A simple exponential decay extrapolation \cite{Headrick:2009jz} following the results of \cite{Anderson:2010ke,Ashmore:2019wzb} suggests that a $\sigma$ value of $10^{-3}$ can be attained for $k\sim 16$. The  value $\sigma=0.022$ can be reached for $k\sim 11$.
 
\begin{figure}[h!]
\centering
\includegraphics[scale=.612]{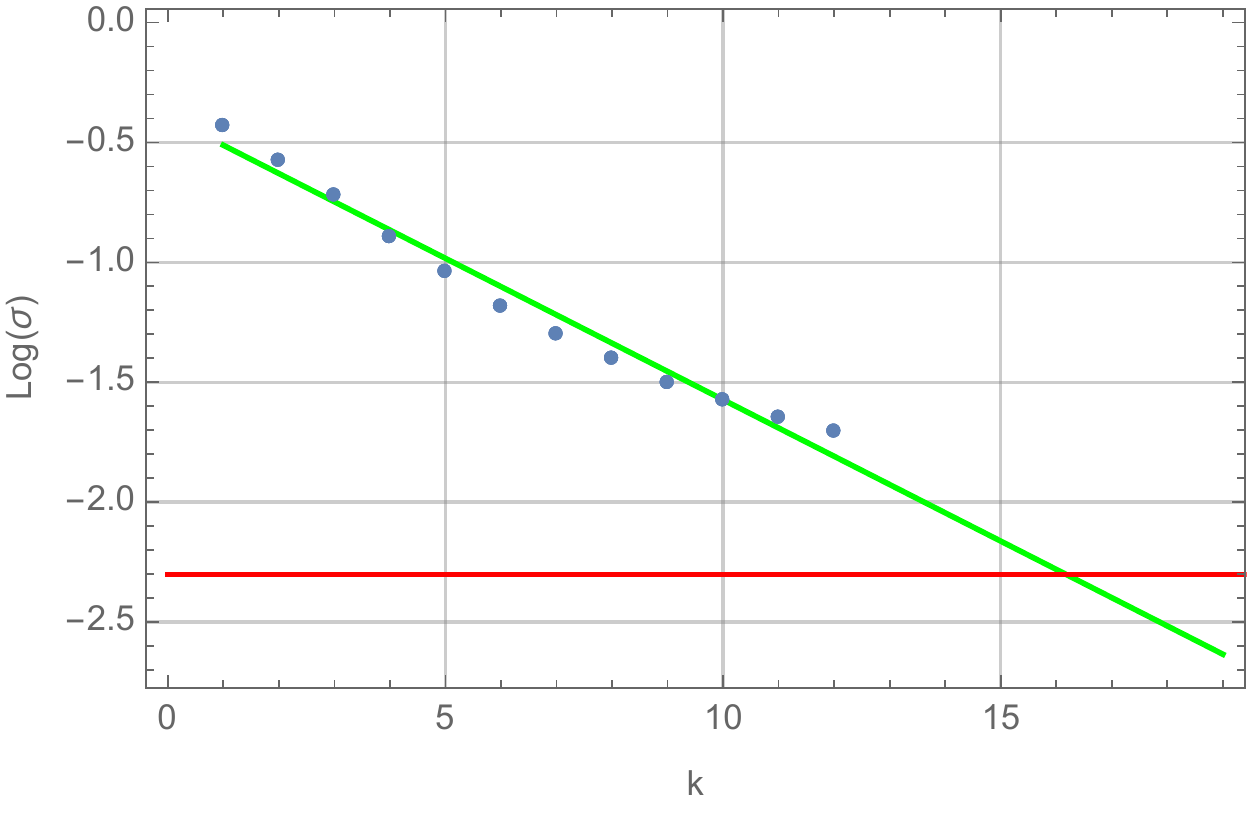}\\[-5pt]
\caption{\small\textit{The value of $\sigma$ reached by the neural network array on the Fermat quintic would be attained at level $k\sim 16$ using Donaldson's algorithm. The exponential extrapolation was constructed using the data of \cite{Ashmore:2019wzb}.}}
  \label{fig:Donaldson}
  \vspace{-10pt}
\end{figure}

These methods are broadly applicable.
In particular, due to the modularity of the architecture, we can envision finding a suitable loss function that enables us to compute metrics on $G_2$ manifolds.\footnote{We thank Anthony Ashmore for a discussion on this point.}
We can as well envision applying machine learning methods to finding new solutions to general relativity or supergravity.
Numerical methods have been applied to construct stationary black hole solutions in various contexts, for example, by using shooting methods to solve ordinary differential equations or employing the Newton--Raphson procedure to solve partial differential equations~\cite{Headrick:2009pv, Wiseman:2011by, Dias:2015nua}; neural networks may supply an additional tool as a supplement to these approaches.

We have so far only studied a handful of Calabi--Yau threefolds, including the Fermat quintic and Tian--Yau Calabi--Yau threefolds.
We aim to find numerical Ricci flat metrics for the base manifolds implicated in top down constructions of the Standard Model.
Interestingly, the configuration matrix in~\cite{Braun:2005ux} corresponding to the split bicubic or Schoen manifold is the transpose of the Tian--Yau configuration matrix~\eref{eq:ty}; its metric is amenable to attack using similar methods.
We hope to create neural network models for Ricci flat metrics on this manifold as well as on the other complete intersection Calabi--Yaus as an extension of this work. In future work, we also hope to report on a first principles theoretical computation of the mass of the electron based on these models. 


\section*{Acknowledgments}
We thank Anthony Ashmore, Andrew Dancer, Yarin Gal, Mario Garcia-Fernandez, James Gray, Nikhil Raghuram, Lewis Smith, Mark van der Wilk, and Francisco Vargas for conversations and the Simons Center for Geometry and Physics for hospitality at the ``Strings, Geometry, and Data Science'' meeting in January 2020.
VJ is supported by the South African Research Chairs Initiative of the Department of Science and Technology and the National Research Foundation.
VJ and DKMP are supported by a Simons Foundation Mathematics and Physical Sciences Targeted Grant, 509116.
CM is supported by a Fellowship by the \text{Accelerate Science} program at the Computer Laboratory, University of Cambridge. CM acknowledges support for part of the work by the UK Government’s Defence \& Security Programme in support of the Alan Turing Institute. 
\appendix

\bibliographystyle{JHEP}
\bibliography{ref}

\end{document}